\documentclass[useAMS,usenatbib]{mn2e}

\usepackage{graphicx,color}
\usepackage{amssymb,amsmath}
\usepackage{amsmath} 
\usepackage{multicol}
\usepackage{latexsym}
\usepackage{amsfonts}
\usepackage{amssymb}
\usepackage{wasysym}
\usepackage{textcomp}
\usepackage{longtable,lscape}
\usepackage{mathtools}

\usepackage{aas_macros}

\usepackage{natbib}
\usepackage{amssymb,amsmath}

\title[Dynamically evolved Galactic OCs]{Catching Galactic open clusters in
advanced stages of dynamical evolution
}
\author[M. S. Angelo et al.]{M. S. Angelo$^{1}$\thanks{E-mail:
mangelo@lna.br}, A. E. Piatti$^{2,3}$, W. S. Dias$^{4}$ and F. F. S. Maia$^{5}$ \\ 
\noindent
$^1$Laborat\'orio Nacional de Astrof\'isica, R. Estados Unidos 154, 37530-000 Itajub\'a, MG, Brazil\\
$^2$Consejo Nacional de Investigaciones Cient\'ificas y T\'ecnicas, Av. Rivadavia 1917, C1033AAJ, Buenos Aires, Argentina \\
$^3$Observatorio Astron\'omico, Universidad Nacional de C\'ordoba, Laprida 854, 5000, C\'ordoba, Argentina\\
$^4$UNIFEI, Instituto de Ci\^encias Exatas, Universidade Federal de Itajub\'a, Av. BPS 1303 Pinheirinho, 37500-903 Itajub\'a, MG, Brazil \\
$^5$Departamento de F\'isica, ICEx, Universidade Federal de Minas Gerais, Av. Ant\^onio Carlos 6627, 31270-901 Belo Horizonte, MG, Brazil
}

\begin{document}

\date{Accepted 2018 April 4. Received 2018 March 29; in original form 2018 March 5}

\pagerange{\pageref{firstpage}--\pageref{lastpage}} \pubyear{2018}

\maketitle

\label{firstpage}

\begin{abstract}


During their dynamical evolution, Galactic open clusters (OCs) gradually lose their stellar content mainly because of internal relaxation and tidal forces. In this context, the study of dynamically evolved OCs is necessary to properly understand such processes. We present a comprehensive Washington $CT_1$ photometric analysis of six sparse OCs, namely: ESO\,518-3, Ruprecht\,121, ESO\,134-12, NGC\,6573, ESO\,260-7 and ESO\,065-7. We employed Markov chain Monte-Carlo simulations to robustly determine the central coordinates and the structural parameters and $T_1\times(C-T_1)$ colour-magnitude diagrams (CMDs) cleaned from field contamination were used to derive the fundamental parameters. ESO\,518-03, Ruprecht\,121, ESO\,134-12 and NGC\,6573 resulted to be of nearly the same young age (8.2\,$\leq\textrm{log}(t\,\textrm{yr}^{-1})\leq$\,8.3); ESO\,260-7 and ESO065-7 are of intermediate age (9.2\,$\leq\textrm{log}(t\,\textrm{yr}^{-1})\leq$\,9.4). All studied OCs are located at similar Galactocentric distances (R$_{G}\sim6-6.9\,$kpc), considering uncertainties, except for ESO\,260-7 ($R_{G}=8.9\,$kpc). These OCs are in a tidally filled regime and are dynamically evolved, since they are much older than their half-mass relaxation times ($t/t_{rh}\gtrsim30$) and present signals of low-mass star depletion.   We distinguished two groups: those dynamically evolving towards final disruptions and those in an advanced dynamical evolutionary stage. Although we do not rule out that the Milky Way potential could have made differentially faster their dynamical evolutions, we speculate here with the possibility that they have been mainly driven by initial formation conditions.

\end{abstract}


\section{Introduction}

Open clusters (OCs) in our Galaxy are subject to disruption effects as they dynamically evolve. Those that survive the early gas-expulsion stage ($t\lesssim3\,$Myr; \citeauthor{Portegies-Zwart:2010}\,\,\citeyear{Portegies-Zwart:2010}) enter in a subsequent phase, during which the stellar population is essentially gas-free. During this phase, different effects take place on the cluster overall dynamics: mass loss due to stellar evolution, preferential evaporation of low-mass stars caused by the internal two-body relaxation, ruled by the Galaxy tidal field, and segregation of higher mass stars and binaries in the cluster centre. As a consequence, the OCs experience structural changes and relations between parameters like core, half-mass and tidal radii together with their present day mass functions can be used as probes of their dynamical states (e.g., \citeauthor{Glatt:2011}\,\,\citeyear{Glatt:2011}). Due to their physical nature, highly dynamically evolved OCs often contain relatively few members and therefore present low contrast with the Galactic field population (\citeauthor{Angelo:2017}\,\,\citeyear{Angelo:2017}; \citeauthor{Pavani:2011}\,\,\citeyear{Pavani:2011}).


Despite these difficulties, it is desirable a uniform characterization of these challenging and usually overlooked objects. Determining the parameters of a large sample of OCs, covering different evolutionary stages and positions within the Galaxy, helps to constrain model parameters (e.g., initial number of stars, initial mass function and fraction of primordial binaries) aimed at investigating the cluster evolution (e.g., \citeauthor{de-La-Fuente-Marcos:1997}\,\,\citeyear{de-La-Fuente-Marcos:1997}; \citeauthor{Portegies-Zwart:2001}\,\,\citeyear{Portegies-Zwart:2001}). According to the most updated version of the OCs catalogue compiled by \citeauthor{Dias:2002}(\citeyear{Dias:2002}, version 3.5 as of 2016 January), a limited number of objects have been characterized in detail. Many works have been developed in order to circumvent this problem by employing large databases and uniform procedures either based on star-by-star analysis methods (e.g., \citeauthor{Kharchenko:2001}\,\,\citeyear{Kharchenko:2001}; \citeauthor{Kharchenko:2005}\,\,\citeyear{Kharchenko:2005}; \citeauthor{Kharchenko:2013}\,\,\citeyear{Kharchenko:2013}, hereafter K13; \citeauthor{Mermilliod:2003}\,\,\citeyear{Mermilliod:2003}) or integrated properties of Galactic OCs (e.g., \citeauthor{Lata:2002}\,\,\citeyear{Lata:2002}).

In this context, we have made use of unpublished Washington photometric system data sets to perform a search for unstudied or poorly studied OCs, which have  sparse appearance and low contrast against the Galactic field, possibly consisting of dynamically evolved stellar aggregates (\citeauthor{Piatti:2016}\,\,\citeyear{Piatti:2016}; \citeauthor{Piatti:2017}\,\,\citeyear{Piatti:2017}; \citeauthor{Piatti:2017a}\,\,\citeyear{Piatti:2017a}). As a result of the search, we found six overlooked OCs, namely: ESO\,518-3, Ruprecht\,121, ESO\,134-12, NGC\,6573, ESO\,260-7 and ESO\,065-7. These objects are older than their half-mass relaxation times (Section \ref{discussion}), therefore presenting signals of being dynamically evolved. We employed Washington $T_{1}\times(C-T_{1})$ colour-magnitude diagrams (CMDs) and a decontamination technique to establish membership likelihoods (Section \ref{CMD_analysis}). Parallaxes measured from the \textit{Gaia} satellite \citep{Gaia-Collaboration:2016}, whenever available, were incorporated into our analysis.  

This paper is organized as follows: in Section \ref{data_collection_reduction}, we describe the collection and reduction of the photometric data. In Sections \ref{center_RDPs_struct_params} and \ref{data_analysis}, we derive cluster structural (core, half-mass and tidal radii) and photometric (reddening, distance, age and mass) parameters from radial density profiles and CMDs, respectively. In Section \ref{discussion} we analyse the dynamical state of the studied OCs. Finally, our conclusions are summarized in Section \ref{conclusions}.


\section{Data collection and reduction}
\label{data_collection_reduction}

\begin{table*}
 \small
  \caption{Observations log of the studied OCs.}
  \label{log_observations}
 \begin{tabular}{lccccccc}
 
  \hline

Cluster & $\rmn{RA}_{J2000}$  & $\rmn{DEC}_{J2000}$ & $\ell$ & $b$ & Filter & Exposure & Airmass  \\   
            &  ($\rmn{h}$:$\rmn{m}$:$\rmn{s}$) & ($\degr$:$\arcmin$:$\arcsec$) & ($^{\circ}$) & ($^{\circ}$) &   & (s)  &    \\
\hline            

ESO 518-3   & 16:47:06  &  -25:48:18   & 355.0710     &  12.4265    & $C$ & 30,300,300   & 1.0,1.0,1.0   \\
                     &                 &                    &                     &                   & $R$ & 5,30,30         & 1.0,1.0,1.0     \\

Ruprecht 121   & 16:41:47 &  -46:09:00   & 338.7241    &  0.0944   & $C$ & 30,30      & 1.1,1.1  \\
                         &                &                    &                    &                & $R$ & 5,5          & 1.1,1.1  \\

ESO 134-12     & 14:44:47   &  -59:08:31   & 317.0236     &  0.5929     & $C$ & 30,300,300   & 1.2,1.2,1.2   \\
                         &                  &                   &                     &                  & $R$ & 5,30,30         & 1.2,1.2,1.2   \\

NGC 6573 & 18:13:39 & -22:07:48    & 9.0490   & -2.0889   & $C$ & 30,300,300    & 1.0,1.0,1.0    \\
                  &                &                   &               &                & $R$  & 5,30,30         & 1.0,1.0,1.0        \\
                  
ESO 260-7  & 08:48:00 &  -47:01:48   & 266.3638     & -2.1920     & $C$ & 30,300,300     & 1.1,1.1,1.1   \\
                    &                &                    &                     &                  & $R$ & 5,30,30            & 1.1,1.1,1.1    \\

ESO 065-7 & 13:29:10 &  -71:15:54   & 305.9831     & -8.6171 	   & $C$ & 45,450,450            & 1.3,1.3,1.3   \\
                   &                &                    &                     &                   & $R$ & 5,10,10,60,60       & 1.3,1.3,1.3,1.3,1.3 \\

\hline
\end{tabular}
\end{table*}

Washington $C$ and Kron-Cousins $r$ images were downloaded from the public website of the National Optical Astronomy Observatory (NOAO) Science Data Management (SDM) Archives\footnote[1]{http://www.noao.edu/sdm/archives.php}. The images were taken using the Tek2K CCD imager (scale of 0.4 arcsec\,pixel$^{-1}$) attached to the 0.9-m telescope at the Cerro Tololo Inter-American Observatory (CTIO, Chile) during the nights 2008 May 08 and May 10 (CTIO programme no. 2008A-0001, PI: Clari\'a). A set of calibration images (bias, dome and sky flat exposures per filter) were also downloaded in order to remove the instrumental signature on the science images. The collection of images retrieved is listed in Table \ref{log_observations}.

The processing of the images was performed with the {\fontfamily{ptm}\selectfont 
QUADRED} package in {\fontfamily{ptm}\selectfont 
IRAF}\footnote[2]{{\fontfamily{ptm}\selectfont 
IRAF} is distributed by the National Optical Astronomy Observatories, which is operated by the Association of Universities for Research in Astronomy, Inc., under contract with the National Science Foundation.}. We followed the usual procedures adopted for CCD photometry: the CCD frames were bias/overscan subtracted, trimmed overscan frame regions and flat-field corrected. The multi-extension files were then transformed into single FITS files.

\begin{table}
 \begin{minipage}{85mm}
  \caption{Mean values of the fitted coefficients and residuals for the presently calibrated $CT_1$ photometric data set.}
  \label{results_FITPARAMS}
 \begin{tabular}{lcccc}
 
\hline

Filter   &   Zero                            &   Extinction                    &  Colour             &  Residual               \\
           &   point                            &   coefficient                   &    term               &   (mag)                  \\

\hline

$C$          &  3.873$\pm$0.008  &   0.282$\pm$0.002  &   -0.165$\pm$0.001   &       0.010                \\
$T_{1}$    &  3.290$\pm$0.030  &   0.089$\pm$0.002  &   -0.026$\pm$0.015   &       0.010                \\

\hline
\end{tabular}
\end{minipage}
\end{table}

The instrumental magnitudes and the position of the stars in each frame were derived using a point spread function (PSF)-fitting algorithm. We used the software STARFINDER \citep{Diolaiti:2000} which was designed for the analysis of crowded stellar fields, adapted to be executed automatically. The code supposes isoplanatism, which is a reasonable assumption since the fields of view are relatively small ($\sim13\times13\,$arcmin$^2$). In order to discriminate and reject unlikely detections, without losing faint stars contaminated by the background noise, only those objects with correlation coefficients between the measured profile and the modeled PSF greater than 0.7 were kept in each image. The astrometric solutions for our frames were computed from the set of positions of the stars in the detector reference system and their equatorial coordinates as catalogued in the UCAC4 \citep{Zacharias:2004}. The transformation between the CCD reference system and equatorial system was made through linear equations. The solutions resulted in precisions typically better than $\sim$0.1\,arcsec. The $c$ and $r$ magnitude data sets were then cross-matched by means of the equatorial coordinates for each detected star and a single master table was constructed by compiling the objects detected in the shorter exposure frames, in order to prevent the inclusion of saturated objects, and adding successively to the list the non-coincident ones found in the larger exposure frames.

To transform instrumental magnitudes to the standard system, we measured nearly 100 magnitudes per filter of stars in the standard fields SA\,101, SA\,107 and SA\,110 (\citeauthor{Landolt:1992}\,\,\citeyear{Landolt:1992}; \citeauthor{Geisler:1996}\,\,\citeyear{Geisler:1996}) using the {\fontfamily{ptm}\selectfont 
APPHOT} package within {\fontfamily{ptm}\selectfont 
IRAF}. These selected areas were observed in a wide range of airmass ($\sim1.1-2.5$). The adopted analytical forms of the transformation equations are as follows:
\begin{align}
    c\, & =\,c_{1} + C + c_{2}\times X_{C} + c_{3}\times(C-T_{1}),  \\
    r\, & =\,t_{11} + T_{1} + t_{12}\times X_{T_{1}} + t_{13}\times(C-T_{1}), 
\end{align}
\noindent
where lowercase and capital letters represent instrumental and standard magnitudes, respectively. $X_{C}$ and $X_{T_{1}} $ represent the effective airmass. The transformation coefficients $c_{i}$ and $t_{1i}$ ($i$\,=\,1,2 and 3), representing the zero-point, extinction and colour-terms for each filter, respectively, were adjusted with the {\fontfamily{ptm}\selectfont 
FITPARAMS} task in {\fontfamily{ptm}\selectfont 
IRAF} for each night. The mean values and residuals for all the nights are listed in Table \ref{results_FITPARAMS}. As proposed by \cite{Geisler:1996}, the $R$ filter is well-known to be an excellent subtitute of the $T_1$ filter, with the advantages of increased transmission at all wavelenghts and the absence of any red-leak problems. Therefore, we used here instrumental $r$ magnitudes to obtain standard 
$T_{1}$ ones.       

\begin{figure}
\begin{center}
 \includegraphics[width=7.0cm]{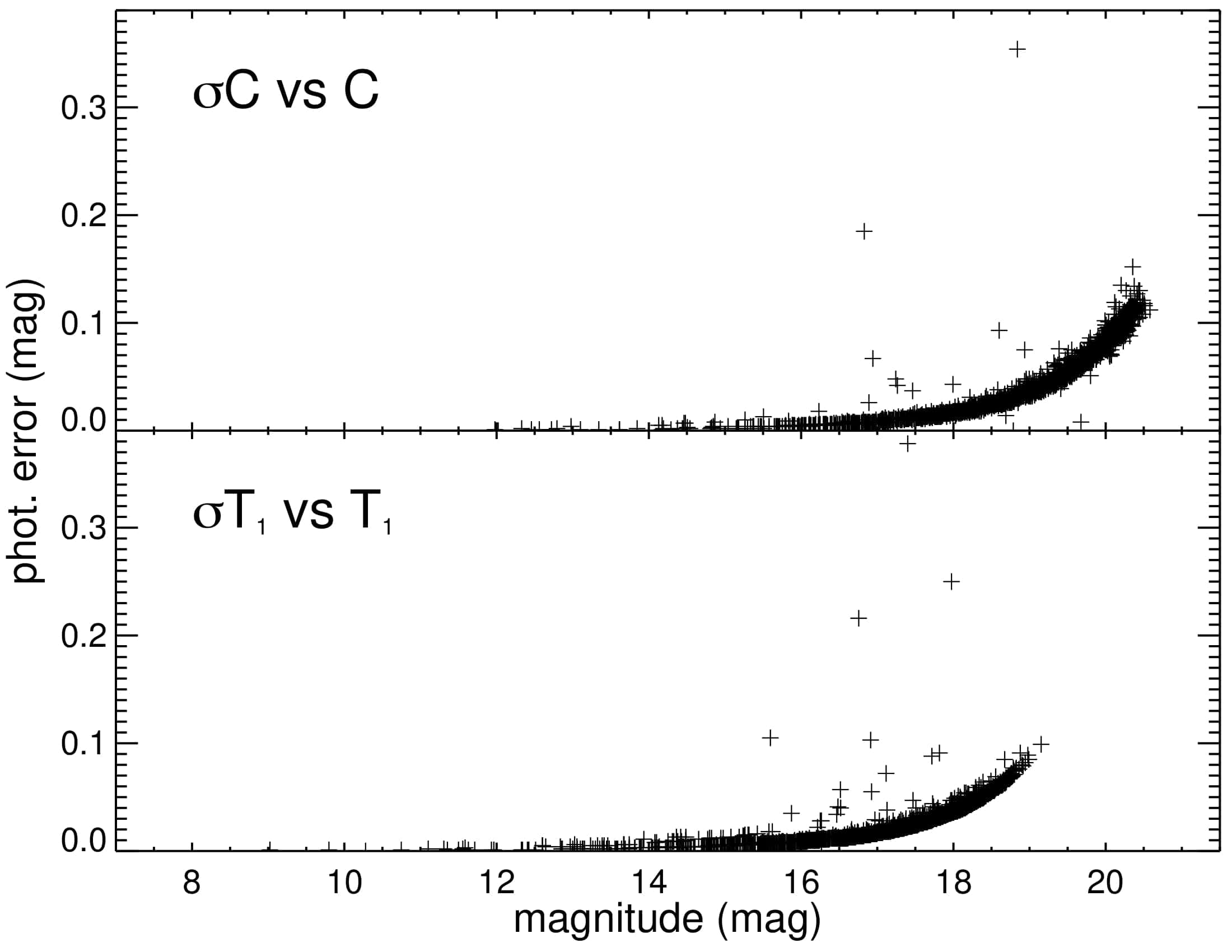}
 \caption{  Photometric uncertainties in $C$ and $T_{1}$ filters for stars in the field of ESO\,065-7. }
   \label{errors_C_T1_ESO065-7}
\end{center}
\end{figure}

We finally transformed the instrumental magnitudes to the $CT_1$ Washington system according to the equations (1) and (2), using the {\fontfamily{ptm}\selectfont 
INVERTFIT} task within {\fontfamily{ptm}\selectfont 
IRAF}. The final table for each OC consists of an internal identifier (ID) per star, its $\rmn{RA}$ and $\rmn{DEC}$ coordinates, the $C$ and $T_{1}$ magnitudes followed by their respective uncertainties. A fragment of this information for ESO\,065-7 is presented in Table \ref{photcatalog_ESO065-7}. Figure \ref{errors_C_T1_ESO065-7} illustrates the typical photometric uncertainties in our photometry, which were derived from the STARFINDER algorithm and then properly propagated into the final magnitudes via the {\fontfamily{ptm}\selectfont IRAF INVERTFIT } task.

\begin{table}
 \tiny
\begin{minipage}{85mm}
  \caption{Identifiers, coordinates and calibrated magnitudes for stars in the field of ESO\,065-7.}
  \label{photcatalog_ESO065-7}
 \begin{tabular}{ccccc}
 
\hline

Star ID   &   $\rmn{RA}$       &   $\rmn{DEC}$       &  $C$           &  $T_{1}$               \\
              &   ($^{\circ}$)         &   ($^{\circ}$)           &   (mag)       &   (mag)                  \\
\hline

 1  & 202.3413696 & -71.2930832 & 13.612\,$\pm\,$0.001 & 10.748\,$\pm\,$0.001  \\       
 2  & 202.6452332 & -71.1639481 & 13.616\,$\pm\,$0.001 & 12.326\,$\pm\,$0.001  \\       
 3  & 202.2594604 & -71.2158890 & 14.172\,$\pm\,$0.001 & 12.967\,$\pm\,$0.004  \\        
4  & 202.0158386 & -71.3307495 & 15.681\,$\pm\,$0.004 & 13.055\,$\pm\,$0.004  \\       
5 & 202.1492004 & -71.3342743 & 17.398\,$\pm\,$0.011 & 15.480\,$\pm\,$0.006  \\        
$-$   &        $-$         &           $-$      &            $-$              &            $-$                \\
\hline 
\end{tabular}
\end{minipage}
\end{table}

\begin{figure}
\begin{center}
 \includegraphics[width=8.0cm]{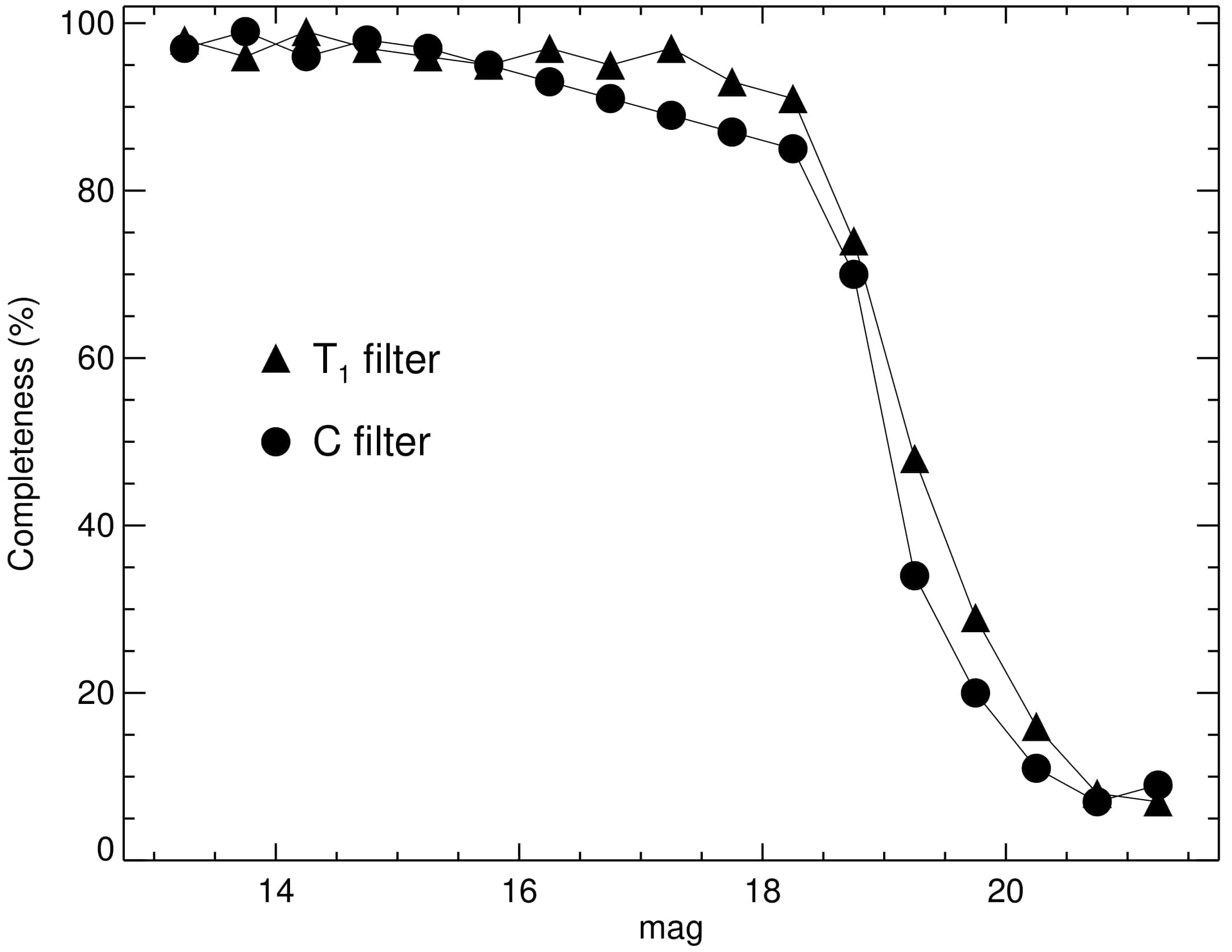}
 \caption{  Completeness as a function of $C$ and $T_{1}$ magnitudes. }
   \label{completeness_versus_mag}
\end{center}
\end{figure}

We performed artificial star tests to derive the completeness level at different magnitudes. We used the stand-alone {\fontfamily{ptm}\selectfont ADDSTAR } program in the {\fontfamily{ptm}\selectfont DAOPHOT } package \citep{Stetson:1990a} to add synthetic stars. We added a number of stars equivalent to $\sim$5 per cent of the measured stars in order to avoid in the synthetic images significantly more crowding than in the original images. Since the fields are no crowded as to consider a dependence of the photometry completeness with the distance to the cluster centre (see, e.g. \citeauthor{Piatti:2014}\,\,\citeyear{Piatti:2014}; \citeauthor{Piatti:2016a}\,\,\citeyear{Piatti:2016a}; \citeauthor{Piatti:2017b}\,\,\citeyear{Piatti:2017b}), the artificial stars were added randomly in space and in magnitude in the $C$ and $R$ images. On the other hand, to avoid small number statistics in the artificial-star analysis, we created a thousand different images for each original one. We used the option of entering the number of photons per ADU in order to properly add the Poisson noise to the star images. We then repeated the same steps to obtain the photometry of the synthetic images as described above. The star-finding efficiency was estimated by comparing the output and the input data for these stars using the DAOMATCH and DAOMASTER tasks. The results are shown in Figure \ref{completeness_versus_mag}. The completeness level for both $C$ and $T_1$ filters is $\gtrsim90\%$ for magnitudes brighter than $\sim$18\,mag and falls to nearly 0 for magnitudes $\gtrsim20\,$mag.

\section{Structural parameters}
\label{center_RDPs_struct_params}

\begin{figure*}
\begin{center}

\parbox[c]{1.0\textwidth}
  {
   
    \includegraphics[width=0.495\textwidth]{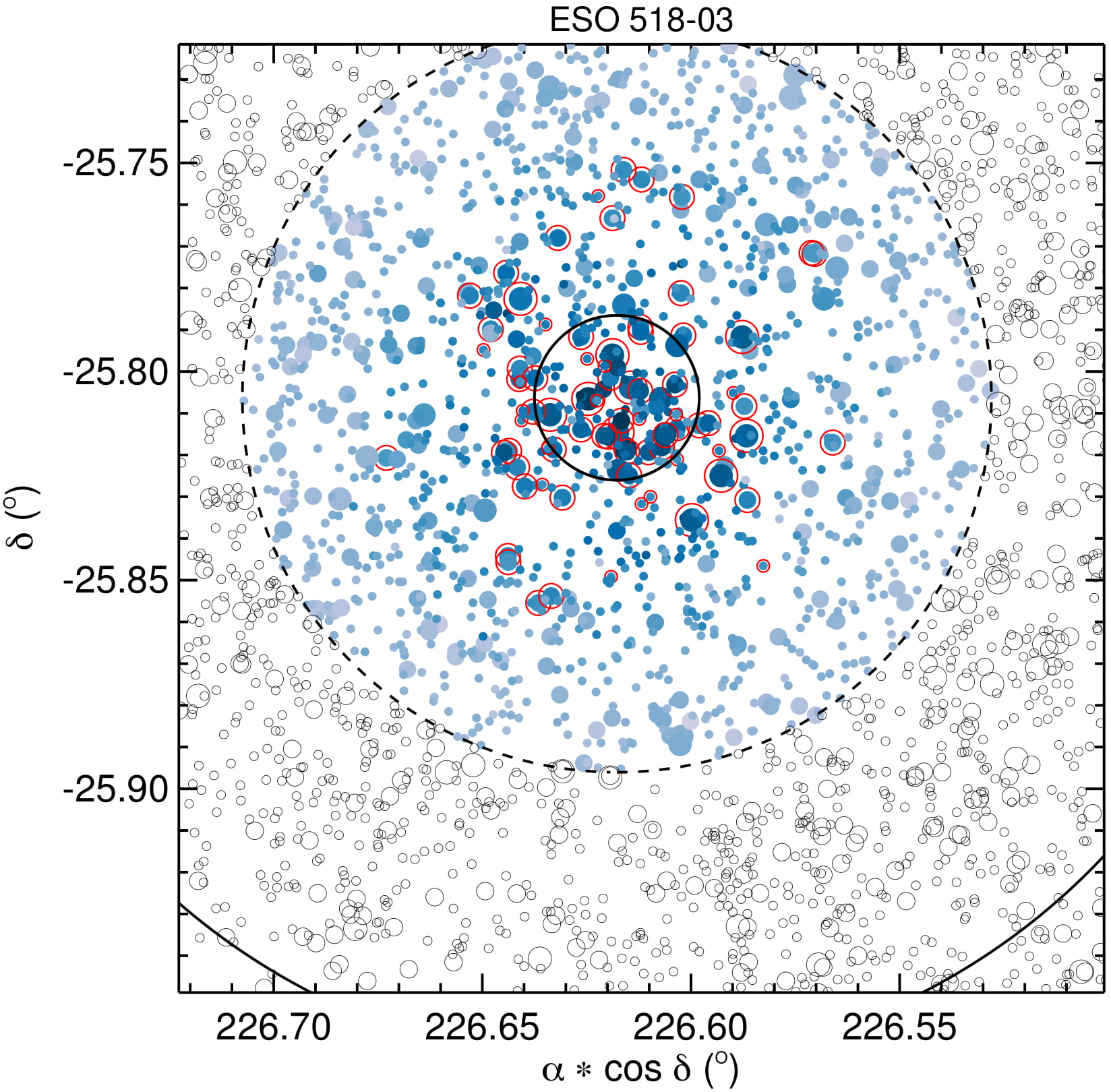}
     \includegraphics[width=0.502\textwidth]{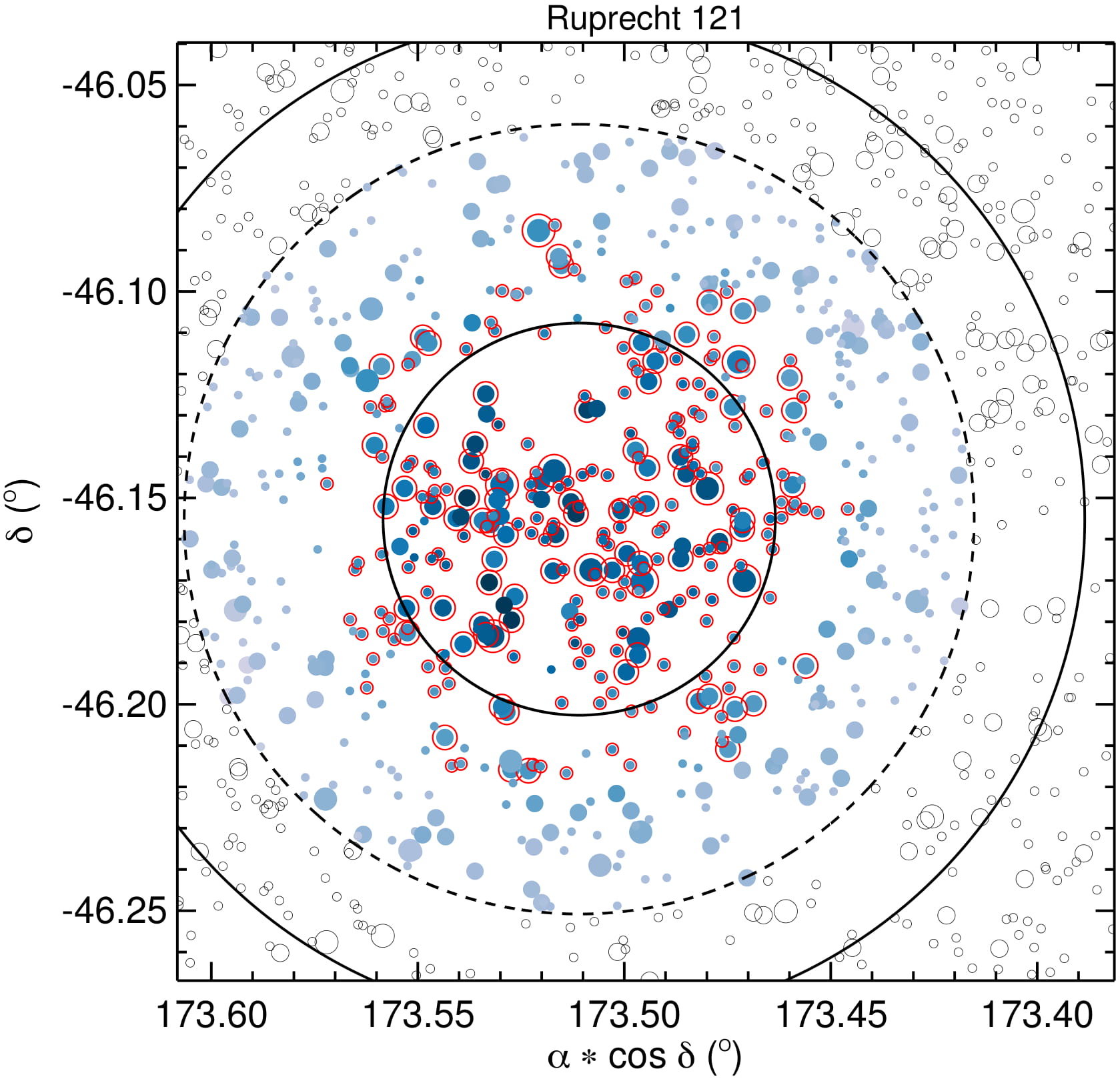}
    \includegraphics[width=0.50\textwidth]{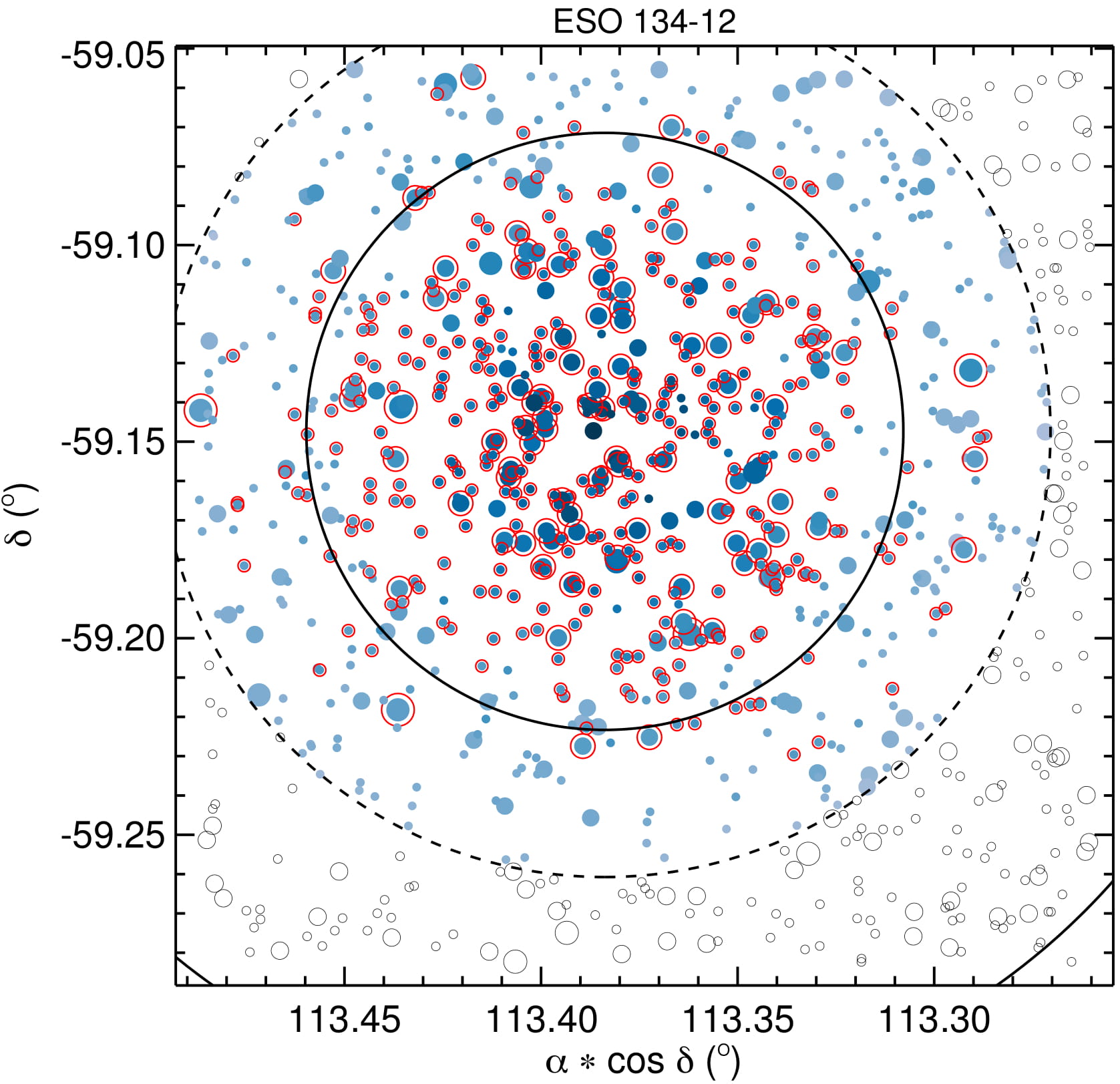}
   \includegraphics[width=0.50\textwidth]{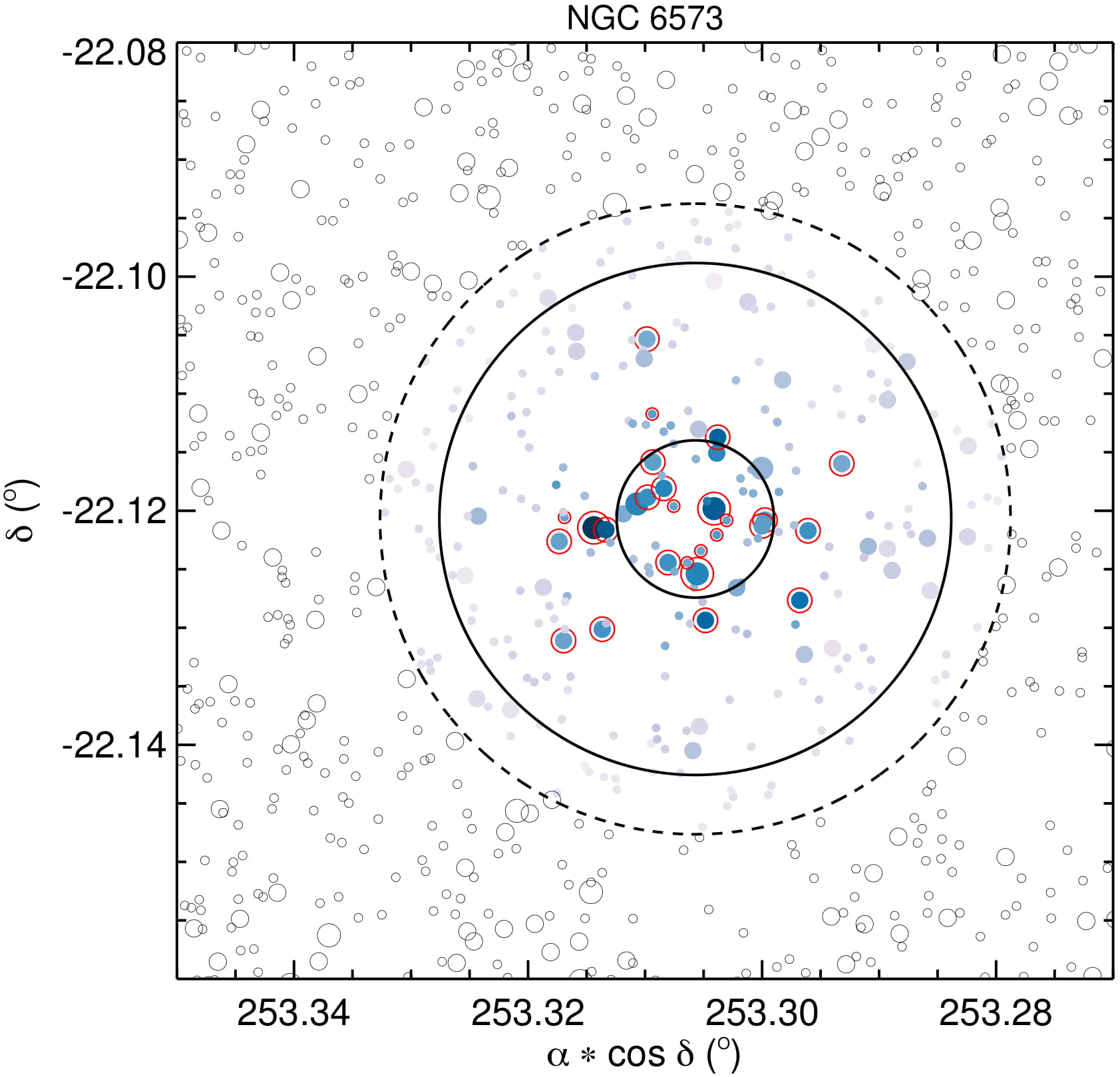}
  }
\caption{ Schematic finding charts of ESO\,518-3 (top-left panel), Ruprecht\,121 (top-right panel), ESO\,134-12 (bottom-left panel) and NGC\,6573 (bottom-right panel). Colours indicate membership likelihoods (Section \ref{CMD_analysis}) and the symbols' sizes are proportional to the stars' brightnesses. Photometric members are highlighted with open red circles. The solid lines indicate the core (inner circle) and tidal radii (outer circle). The dashed circle indicates the limiting radius for the decontamination procedure, beyond which stars are selected for the control field (open circles). North is up and East to the left. }

\label{finding_charts_part1}
\end{center}
\end{figure*}

\begin{figure*}
\begin{center}

\parbox[c]{1.0\textwidth}
  {
   \includegraphics[width=0.495\textwidth]{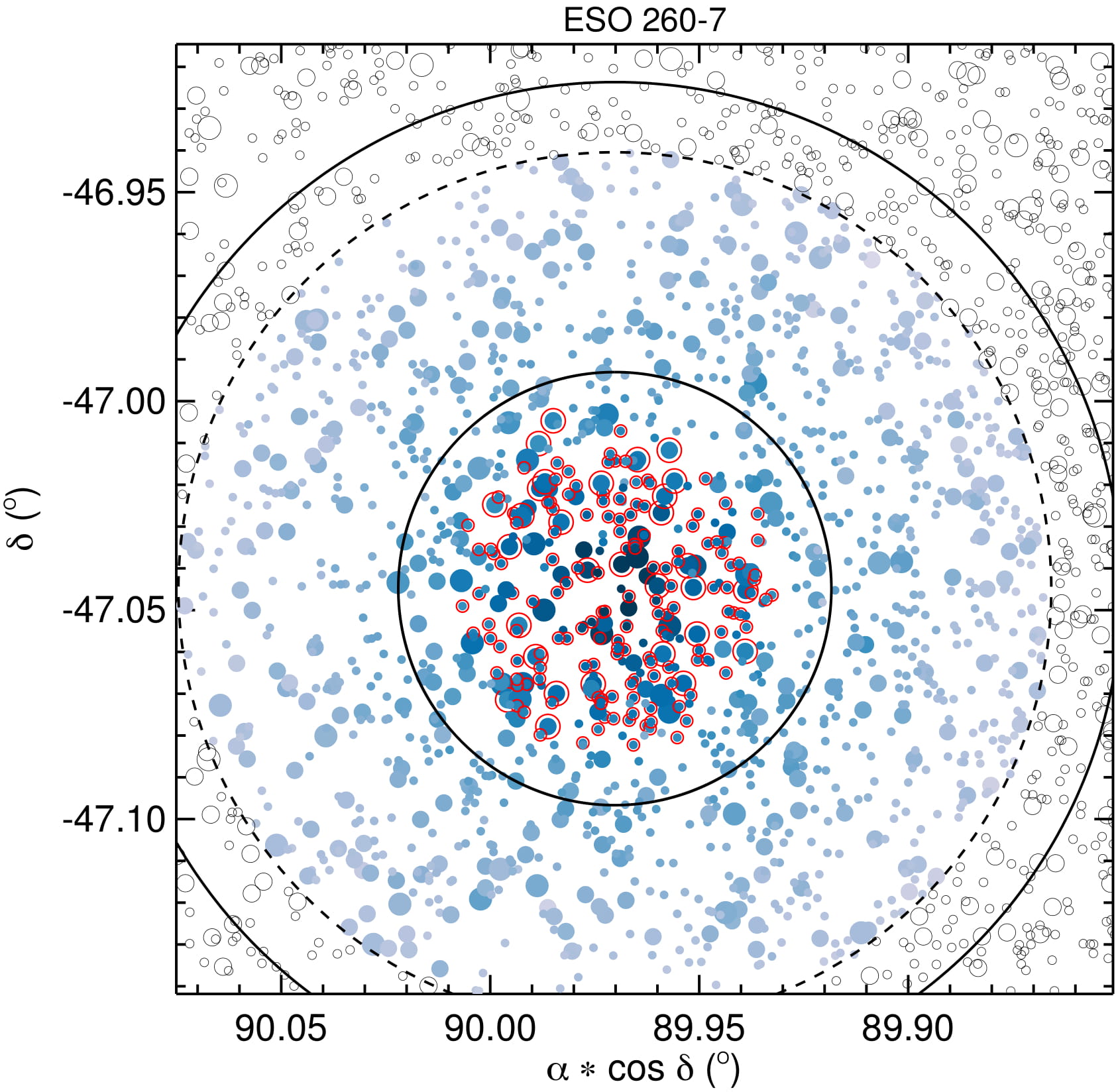}
  \includegraphics[width=0.505\textwidth]{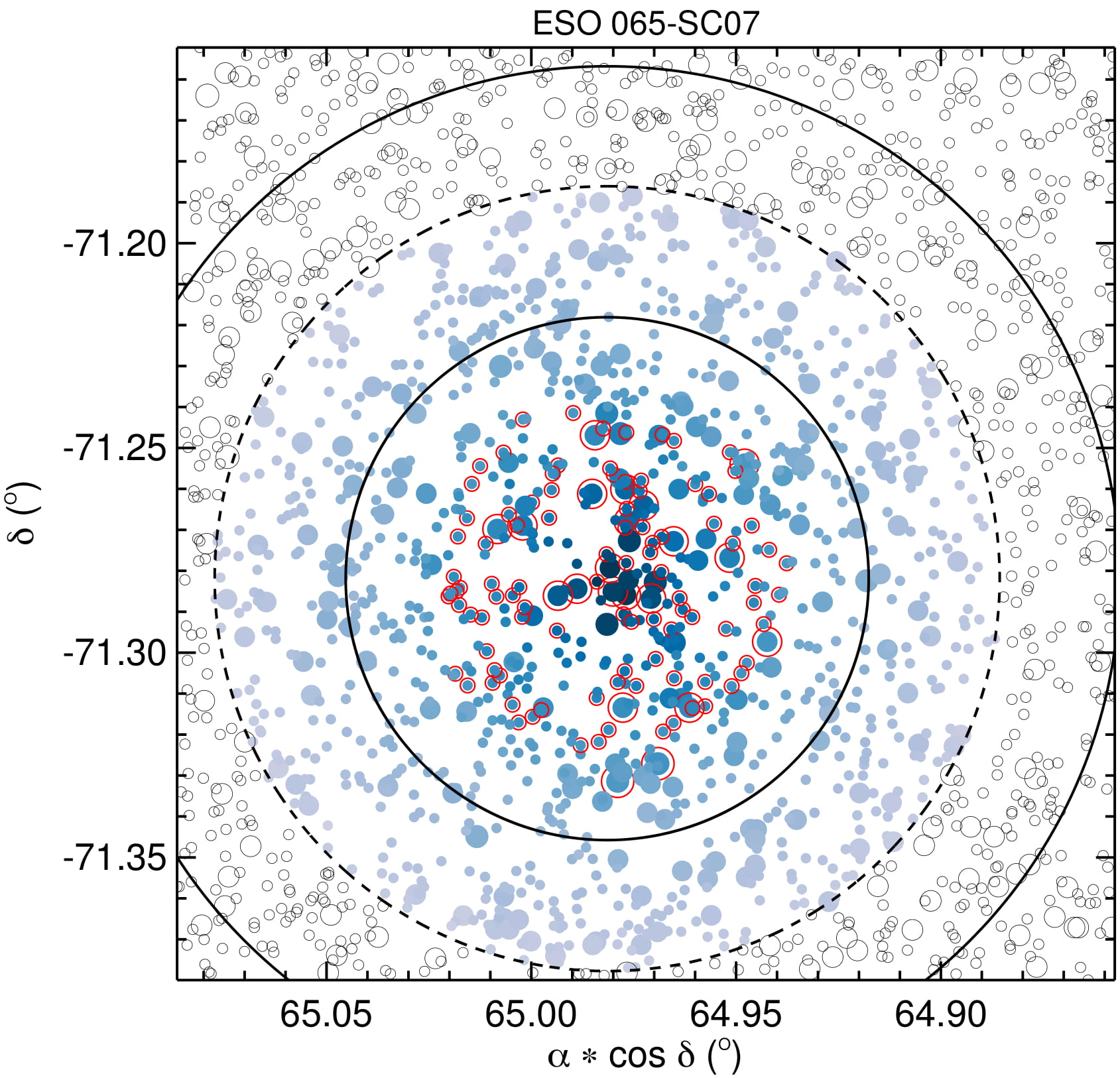}
  }
\caption{Same of Figure \ref{finding_charts_part1}, but for ESO\,260-7 (left panel) and ESO\,065-7 (right panel).}

\label{finding_charts_part2}
\end{center}
\end{figure*}

Figures \ref{finding_charts_part1} and \ref{finding_charts_part2} show the finding charts for the OCs studied in this paper. Colours indicate membership likelihoods (Section \ref{CMD_analysis}); their sizes being proportional to the stars' brightnesses. Those considered photometric members are highlighted with red open circles. For each OC, stars in the selected control field are represented by open circles.

Deriving structural parameters of stellar clusters usually involves finding the central coordinates of the population 
followed by the fitting of an analytical profile to the stars radial density profile, which is usually built by binning the 
data into annulus around the derived centre and computing their mean stellar density. There are, however, several
drawbacks to this approach. For instance, it is not easy to quantify and often overlooked, how a small change on 
the central coordinates impact the final structural parameters. Moreover, the choice of the bin size and uniformity 
are also an issue when creating the radial density profile. Finally, sampling near the centre of the cluster is always a 
problem when using these methods. They require small radial bins which in turn will produce very small projected 
areas and poor statistics, leading to artificially high density values in the central bins.

To circumvent these problems, we devised a bayesian method to derive the structural parameters of the clusters 
by employing a Metropolis-Hastings algorithm \citep{Hastings:1970} to conduct Markov chain Monte-Carlo
(MCMC) simulations on the data set. Sampling of the parameter space was done using the method described in 
\citet{Goodman:2010}, which has been largely used in astronomy, having implementations in \emph{R}\footnote[3]{
https://www.rdocumentation.org/packages/LaplacesDemon} and \emph{Pyton}\footnote[4]{http://dfm.io/emcee/}
\citep{Foreman-Mackey:2013} programming languages.

\begin{figure}
\includegraphics[width=0.48\linewidth]{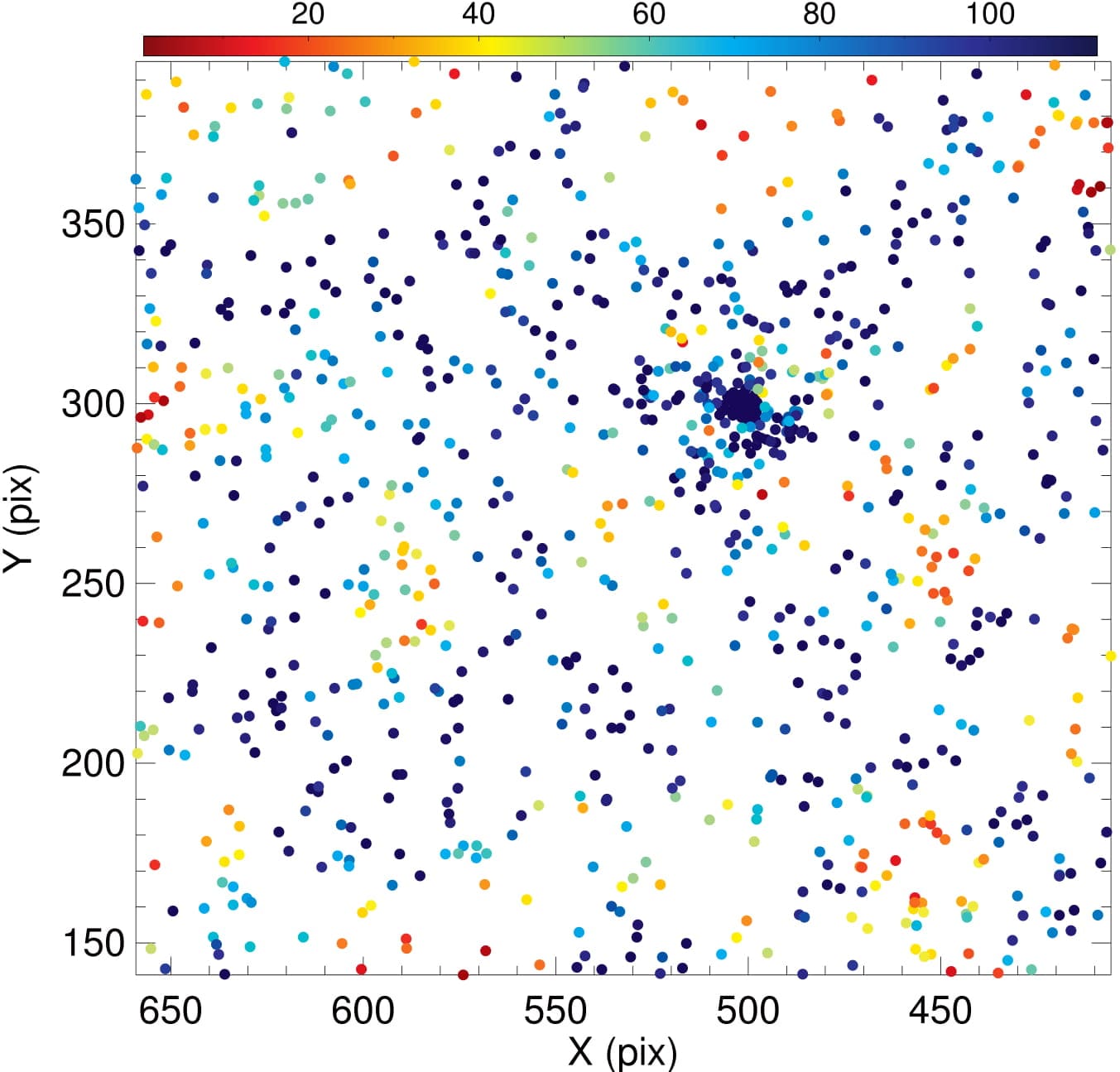}
\includegraphics[width=0.49\linewidth]{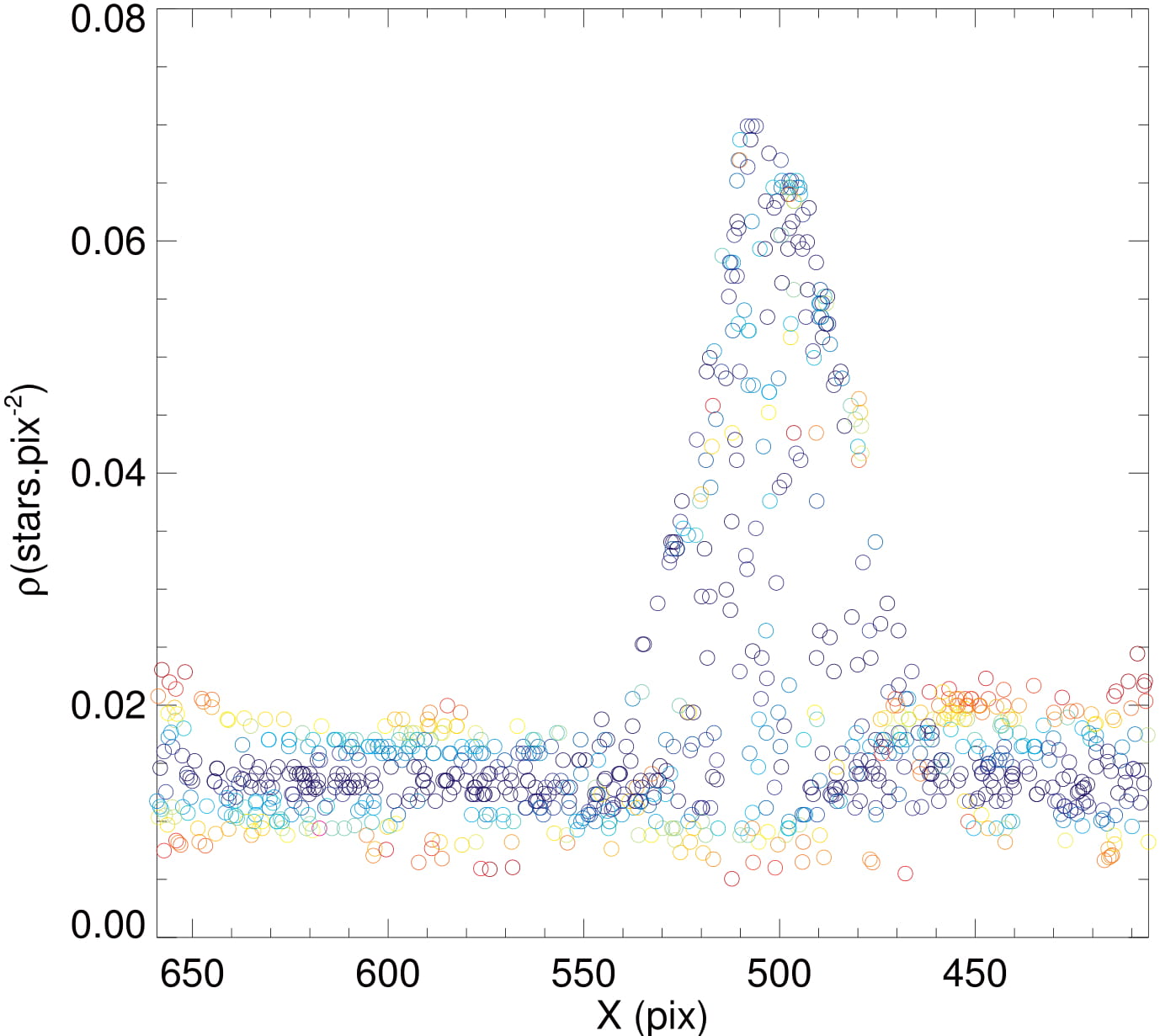}
\caption{Synthetic cluster chart (left) and its marginal density profile across the x-axis (right). The off-centre 
positioning of the cluster was intentional. The colourbar represents the individual likelihood of each star derived 
from the best K62 profile model.}
\label{fig:mcchart}
\end{figure}

Input data included the stellar coordinates and the local stellar density ($\rho$), calculated at each star 
position using a large circular kernel with a fixed size of 6 times the average nearest-neighbour distance of the 
sample. (see Figure \ref{fig:mcchart}). Once an analytical model has been chosen and one 
set of parameter values is proposed, a model stellar density ($m$) is computed at each datum coordinate and a 
likelihood value ($\mathcal{L}$) is calculated for this particular set of parameters through the equation:

\begin{equation}
\ln \mathcal{L} = \sum_i^{N_{star}}\ln\left\{{\frac{1}{2\pi\sigma}\exp \left[ -\frac{(\rho_i - m_i)^2}{2\sigma^2} \right]}\right\},
\end{equation}

\noindent
where $\sigma$ is the standard deviation of $(\boldsymbol{\rho} - \mathbf{m})$. In this paper, we employed the \citeauthor{King:1962}'s\,\,(\citeyear{King:1962}, hereafter K62) model, expressed by

\begin{equation}
  f(r)=f_{0}\left[ \frac{1}{\sqrt{1+(r/r_c)^2}} - \frac{1}{\sqrt{1+(r_t/r_c)^2}} \right]^{2}, 
\end{equation}

\noindent where $r_c$ and $r_t$ are the core and tidal radii, respectively.

The initial guess and prior distribution of the parameters were defined based on the input data. Centre coordinates 
and central density were estimated as the median values of the 5\% highest density points, while background 
density was estimated as the median of the 20\% lowest density points. Core radius and tidal radius were estimated 
from the marginalised density profile as the full width at half-maximum and the full width at background level, 
respectively. The priors were defined as wide normal distributions around these values.

\begin{figure}
\includegraphics[width=0.999\linewidth]{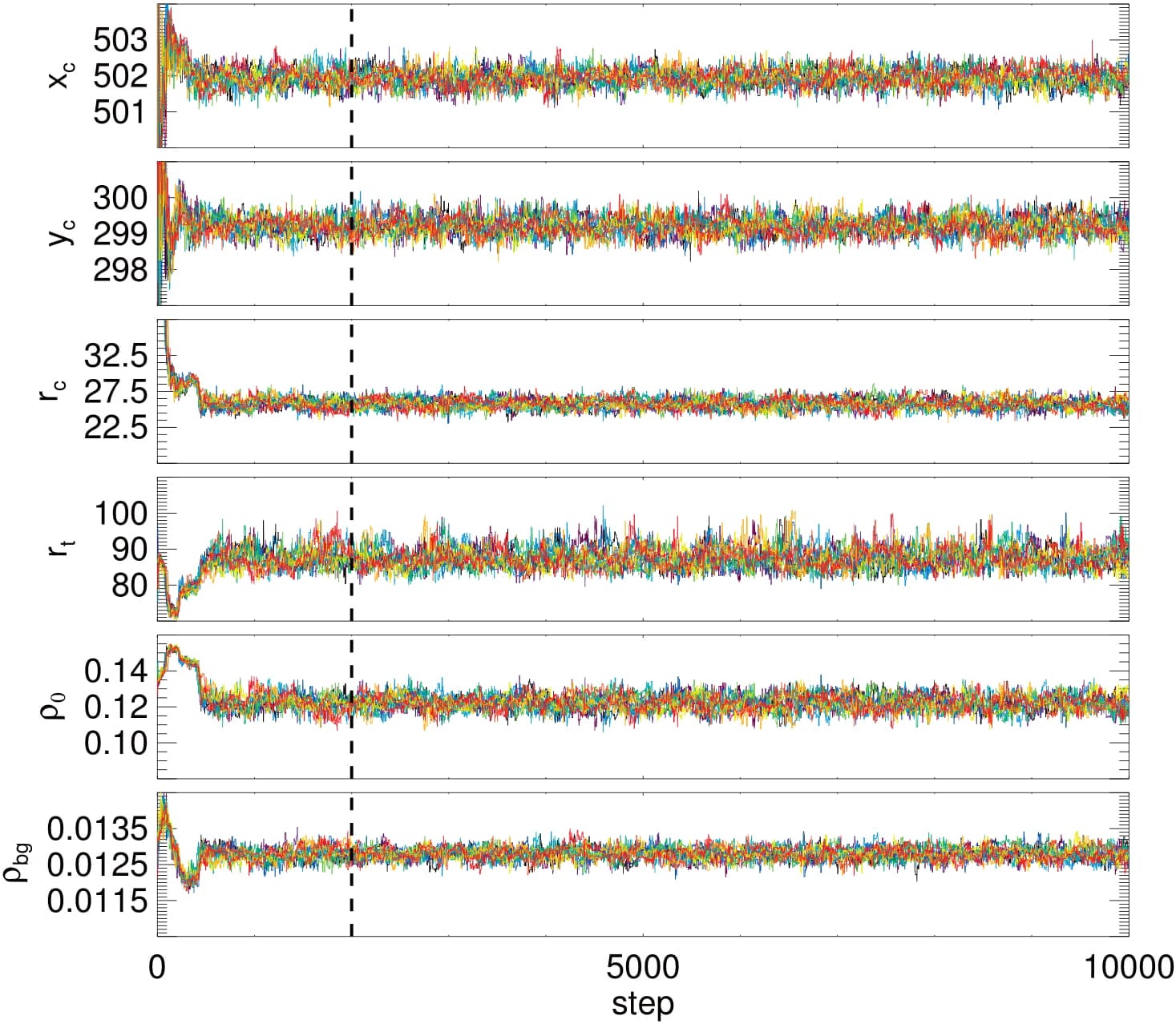}
\caption{Time series of all parameters in the MCMC chain. The twelve threads are represented by different coloured 
lines. The vertical dashed line marks the adopted transition from the transient states to the stationary distribution.}
\label{fig:mctime}
\end{figure}

The cluster central coordinates and the K62 model structural parameters constitute a 6D parameter 
space that was randomly sampled by twelve parallel threads (usually called 'walkers') in a 10000 steps simulation. 
We have found out that this number of threads provided a sufficient sampling of the parameter space as different 
realisations of the simulation showed no noticeable variations in the parameters posterior distributions. The chain 
length was defined after the verification that a stationary distribution is typically achieved after 1000 steps, ensuring 
that the rejection (or 'burning') of the first 20\% of the steps is sufficient to remove the transient states (see 
Figure \ref{fig:mctime}).

As required by the Markov process, a newly proposed parameter set is aways accepted if its likelihood is larger than 
the likelihood of the current parameter set and is accepted with a probability equal to the ratio of the likelihoods 
otherwise ($P_{t\,\rightarrow\, t+1} = \mathcal{L}_{t+1}/\mathcal{L}_t$). We have also adopted a commonly used 
scale function to control the length of the random steps in the parameter space, ensuring that the global acceptance 
ratio of the proposed states stays in the optimal range between 20\% and 30\%, as suggested for the 
Metropolis-Hastings algorithm \citep{Sherlock:2013}. 

To test the method, a 200 star synthetic cluster was generated following the probability density function of the 
K62 profile function, immersed in background twice as dense as the cluster mean density, covering an area 
twice as large as that of the cluster. Figure \ref{fig:mcchart} shows the chart of the simulated cluster and its marginal 
density profile across the x-axis, from which the initial guess and prior distribution of the parameters were calculated.

\begin{figure}
\centering
\includegraphics[width=0.8\linewidth]{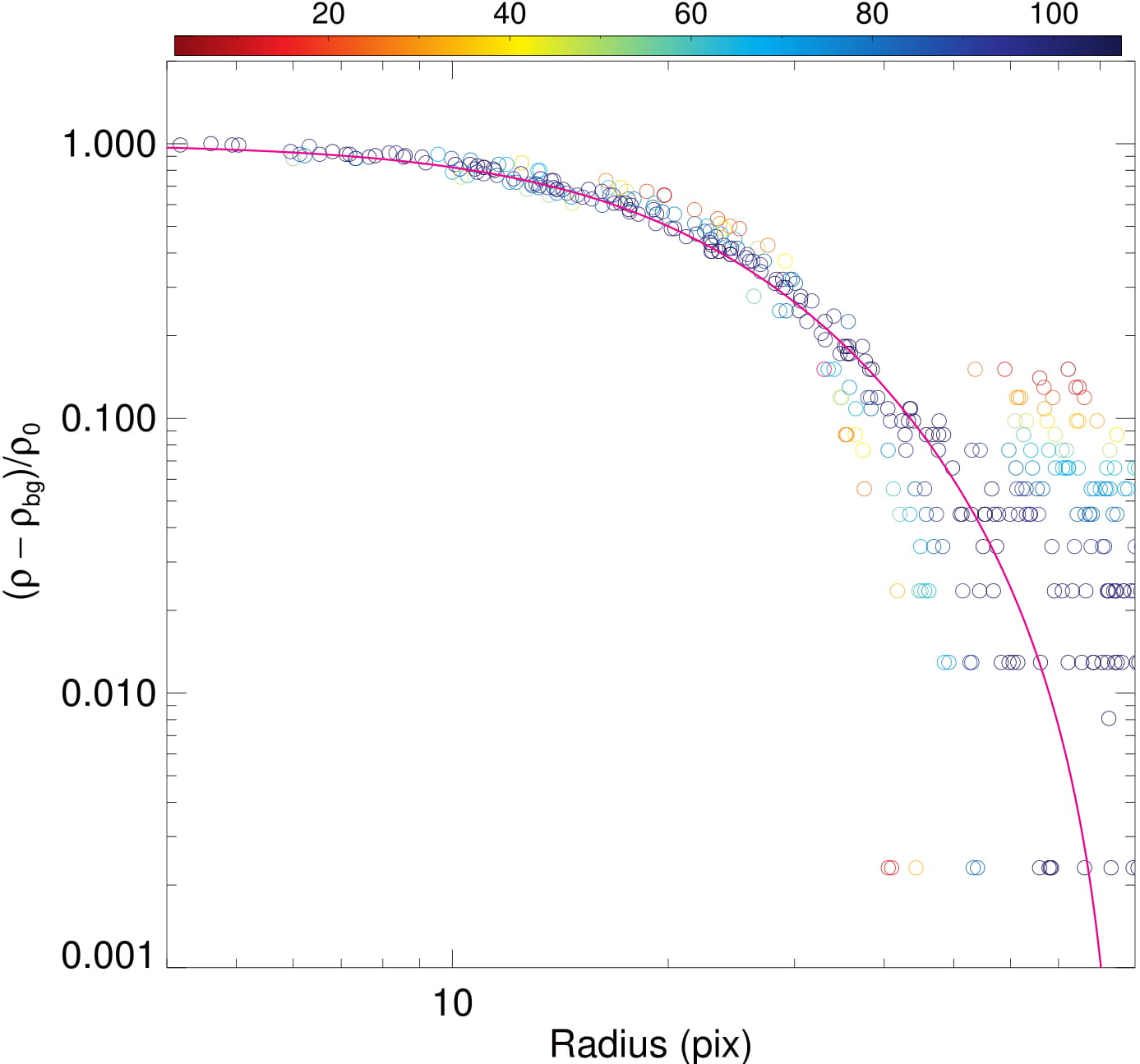}
\caption{\textbf{Normalised background subtracted} radial density profile of the cluster around the determined centre coordinates, showing a K62 profile model function (solid line) with the determined structural parameters overlaid over the data. The colourbar represents the individual likelihood of each star as derived from this model.}
\label{fig:mcfit}
\end{figure}

Figure \ref{fig:mctime} shows the time series of all parameters. It can be seen that after a transient period of about
1000 steps, all treads converge to a stationary posterior distribution from which the final parameters are drawn by 
calculating their marginalised mean and standard deviation. Table \ref{tab:mcpar} compares the simulated cluster
true parameters with the ones recovered from the MCMC simulation. Given the large size of the adopted density 
estimator kernel (similar to the cluster core radius) resulted in an artificial flattening of the density points near the 
centre of the cluster. To prevent this from affecting our simulations, we removed the central points with radius inferior 
to half the kernel size from the simulations, and therefore we could not relate the recovered central stellar density 
with the true (arbitrary) value used in the simulation. Despite that, all other recovered parameters have shown 
excellent agreement with their true values and also 'fits' very well with the data as it can be seem from 
Figure \ref{fig:mcfit}.

\begin{table}
\caption{Simulated cluster parameters}
\centering
\begin{tabular}{l c r@{$\pm$}l } \hline
parameter  & real & \multicolumn{2}{c}{recovered} \\ \hline
$x_c$ (pix) & 500 & 501.9 & 0.2 \\ 
$y_c$ (pix) & 300 & 299.2 & 0.2 \\
$r_c$ (pix)  & 25 & 25.8 & 0.7 \\
$r_t$  (pix) & 90 & 87 & 3 \\
$\rho_{bg}$ (pix)$^{-2}$ & 0.0128 & 0.0128 & 0.0002 \\ \hline
\end{tabular}
\label{tab:mcpar}
\end{table}

\begin{figure*}
\centering
\includegraphics[width=11.0cm]{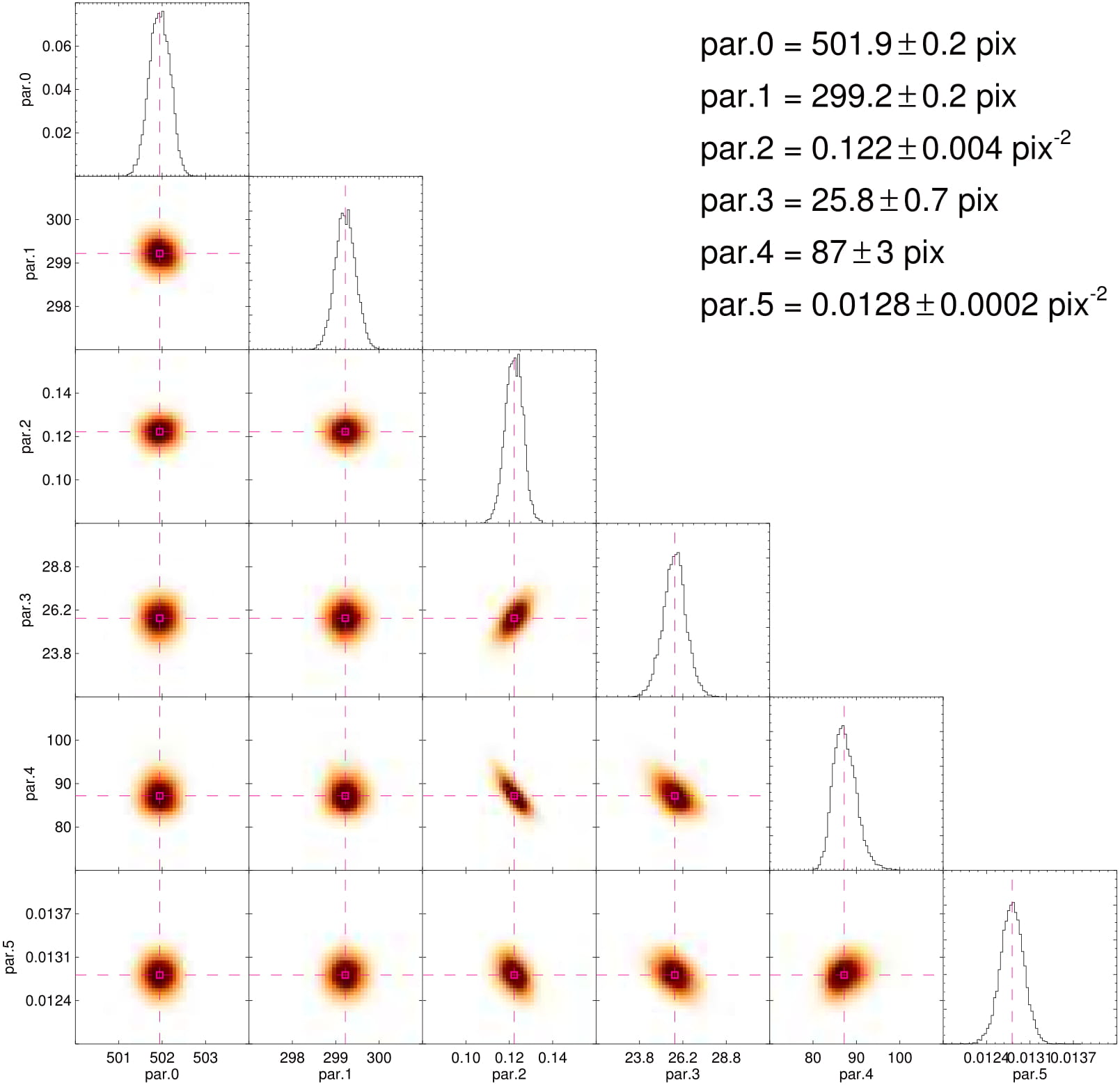}
\caption{Corner plot showing the density of the posterior states around the most likely parameters (dashed lines). Parameters 
$0-5$ represented are, respectively, the central coordinates ($x_c$, $y_c$), the core stellar density ($\rho_0$), core radius ($r_c$), 
tidal radius ($r_t$) and background density ($\rho_{bg}$).}
\label{fig:cornpl}
\end{figure*}

Another tool that is commonly used to infer the quality of a MCMC simulation is the corner plot, shown in Figure \ref{fig:cornpl}. 
It shows, for each parameter, the marginalised distribution of the posterior states (after discarding the initial burnt states) against 
all other parameters. It is useful for checking, at a glance the uniqueness of the solution found (single peak) and the possible 
existence of correlation between any two parameters (deviation from the round form). It can be seen, for example, that in this 
simulation the parameter 2 ($\rho_0$ - third column) shows a strong negative correlation with the parameter 4 ($r_t$) and a 
weaker one with parameter 5 ($\rho_{bg}$) while also showing a positive correlation with parameter 3 ($r_t$). Being correlated 
with all other structural parameters would mean that $\rho_0$ is not really a free parameter in the simulation, but a quantity that
could be derived through a functional form from the other structural parameters.

After applying the procedure outlined in this section to the studied OCs, we obtained the central coordinates and the best-fitting parameters listed in Table \ref{tab_params_OCs}. The maximum distance between the new centres (columns 2 and 3 of Table \ref{tab_params_OCs}) and the literature ones (columns 2 and 3 of Table \ref{log_observations}) resulted 1.4$\arcmin$ (for ESO\,065-7); the minum difference resulted 0.14$\arcmin$ (for ESO\,518-SC03). The analogous of Figures \ref{fig:mctime} and \ref{fig:mcfit} for other clusters are shown in the Appendix.


\begin{table*}
 \scriptsize
 \hskip-2.0cm
  \caption{  Redetermined central coordinates, Galactocentric distances and structural parameters of the studied OCs.  }
  \label{tab_params_OCs}
 \begin{tabular}{lccccccc}
 
\hline

 Cluster         & $\rmn{RA}_{J2000}$                 & $\rmn{DEC}_{J2000}$               & R$_{\textrm{GC}}$            &    $r_c$                    &    $r_h^{\dag}$             & $r_t$                       &  R$_J^{\dag\dag}$          \\
                 & ($\rmn{h}$:$\rmn{m}$:$\rmn{s}$)    & ($\degr$:$\arcmin$:$\arcsec$)     &  (kpc)                       &    (pc)                     &     (pc)                    &  (pc)                       &  (pc)                      \\                                                

\hline

ESO\,518-3       &  16:47:06                          & -25:48:23                         &6.0\,$\pm$\,0.6               &0.72\,$\pm$\,0.03            &0.61\,$\pm$\,0.06            &5.84\,$\pm$\,0.64            &2.52\,$\pm$\,0.27            \\
Ruprecht\,121    &  16:41:46                          & -46:09:19                         &6.5\,$\pm$\,0.5               &1.31\,$\pm$\,0.04            &0.87\,$\pm$\,0.04            &3.39\,$\pm$\,0.19            &4.23\,$\pm$\,0.37            \\
ESO\,134-12      &  14:44:18                          & -59:08:51                         &6.7\,$\pm$\,0.5               &2.64\,$\pm$\,0.11            &1.27\,$\pm$\,0.06            &6.01\,$\pm$\,0.20            &4.63\,$\pm$\,0.38            \\
NGC\,6573        &  18:13:41                          & -22:07:15                         &6.4\,$\pm$\,0.5               &0.19\,$\pm$\,0.02            &0.12\,$\pm$\,0.03            &0.60\,$\pm$\,0.04            &1.84\,$\pm$\,0.22            \\
ESO\,260-7       &  08:47:58                          & -47:02:41                         &8.9\,$\pm$\,0.5               &3.06\,$\pm$\,0.19            &1.32\,$\pm$\,0.07            &7.17\,$\pm$\,0.32            &4.14\,$\pm$\,0.27            \\
ESO\,065-7       &  13:29:22                          & -71:16:55                         &6.9\,$\pm$\,0.5               &2.67\,$\pm$\,0.12            &0.96\,$\pm$\,0.04            &5.24\,$\pm$\,0.31            &2.65\,$\pm$\,0.23            \\

\hline
\multicolumn{8}{l}{ \textit{Note}: The angular separations (in arcmin) where converted to physical distances (in pc) using the expression:  } \\
\multicolumn{8}{l}{ $(\pi\,\theta/10800)\times10^{[(m-M)_{0}+5]/5}$, where $(m-M)_{0}$ is the cluster distance modulus (see Table \ref{clusters_total_masses}). } \\
\multicolumn{8}{l}{ $^{\dag}$ Half-mass radius (Section \ref{mass_functions}). } \\
\multicolumn{8}{l}{ $^{\dag\dag}$ Jacobi radius (Section \ref{discussion}). }


\end{tabular}
\end{table*}

\section{CMD analysis}
\label{data_analysis}

\subsection{Membership and fundamental parameters determination}
\label{CMD_analysis}

\begin{figure*}
\begin{center}

\begin{minipage}{185mm}

\parbox[c]{1.0\textwidth}
  {
   
    \includegraphics[width=0.5\textwidth]{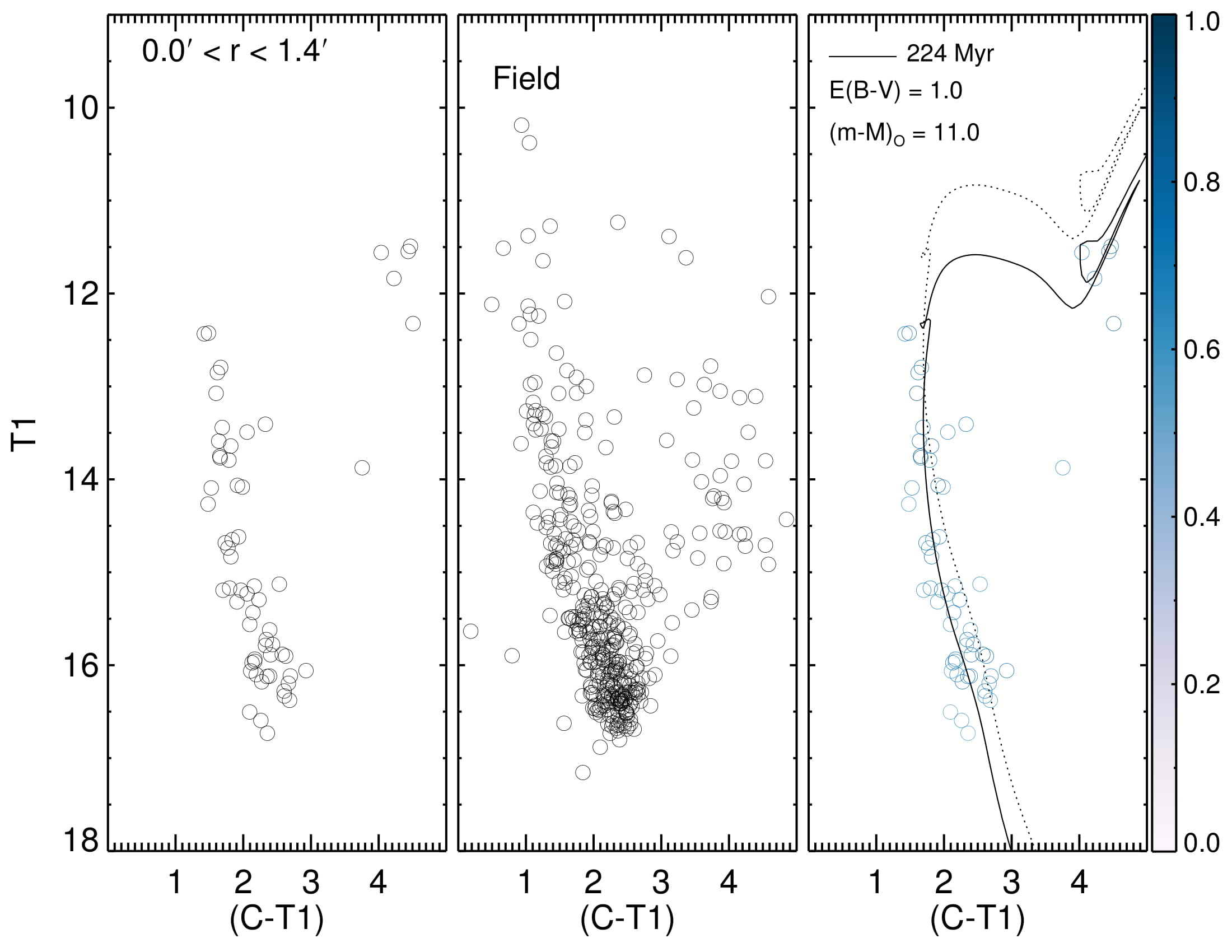}
    \includegraphics[width=0.5\textwidth]{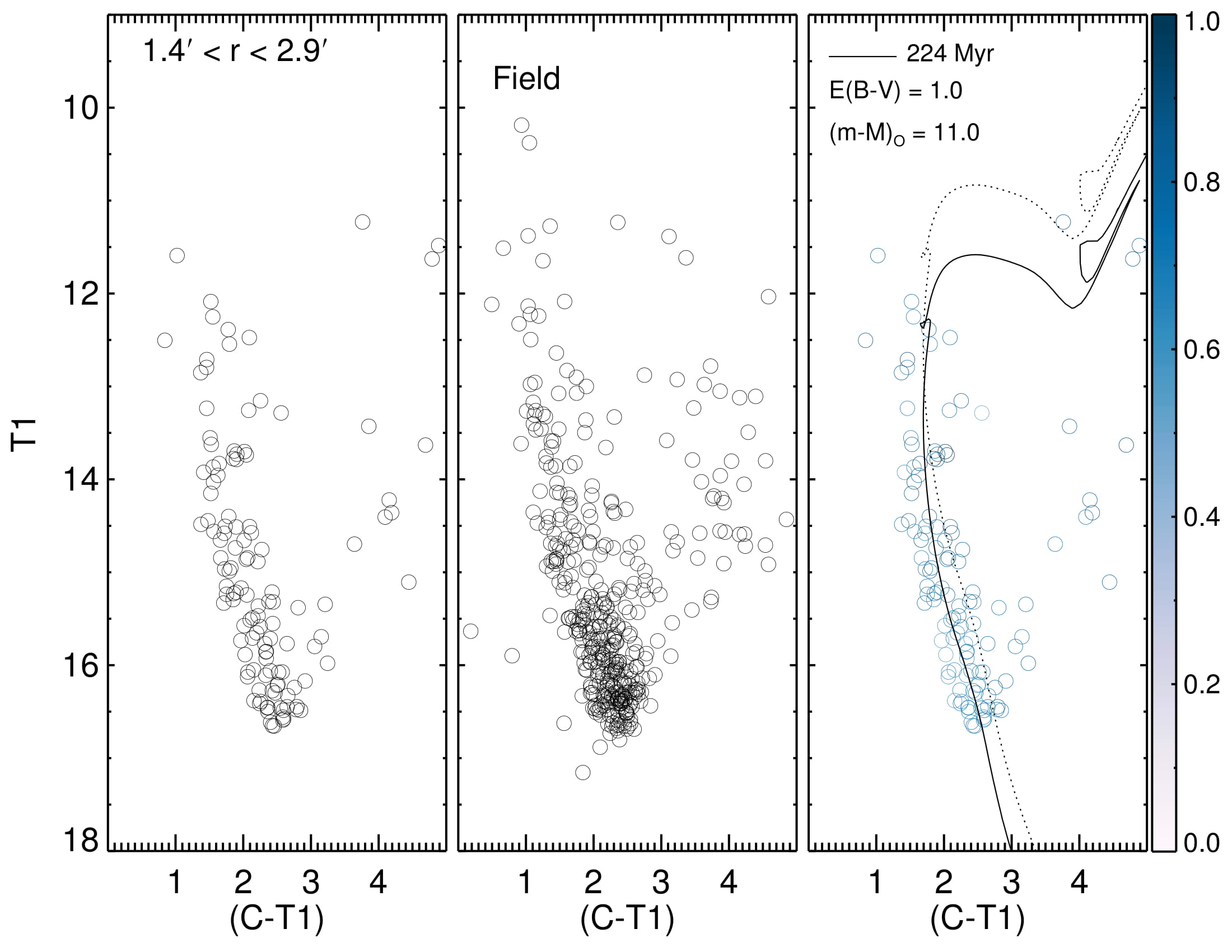}
     \includegraphics[width=0.5\textwidth]{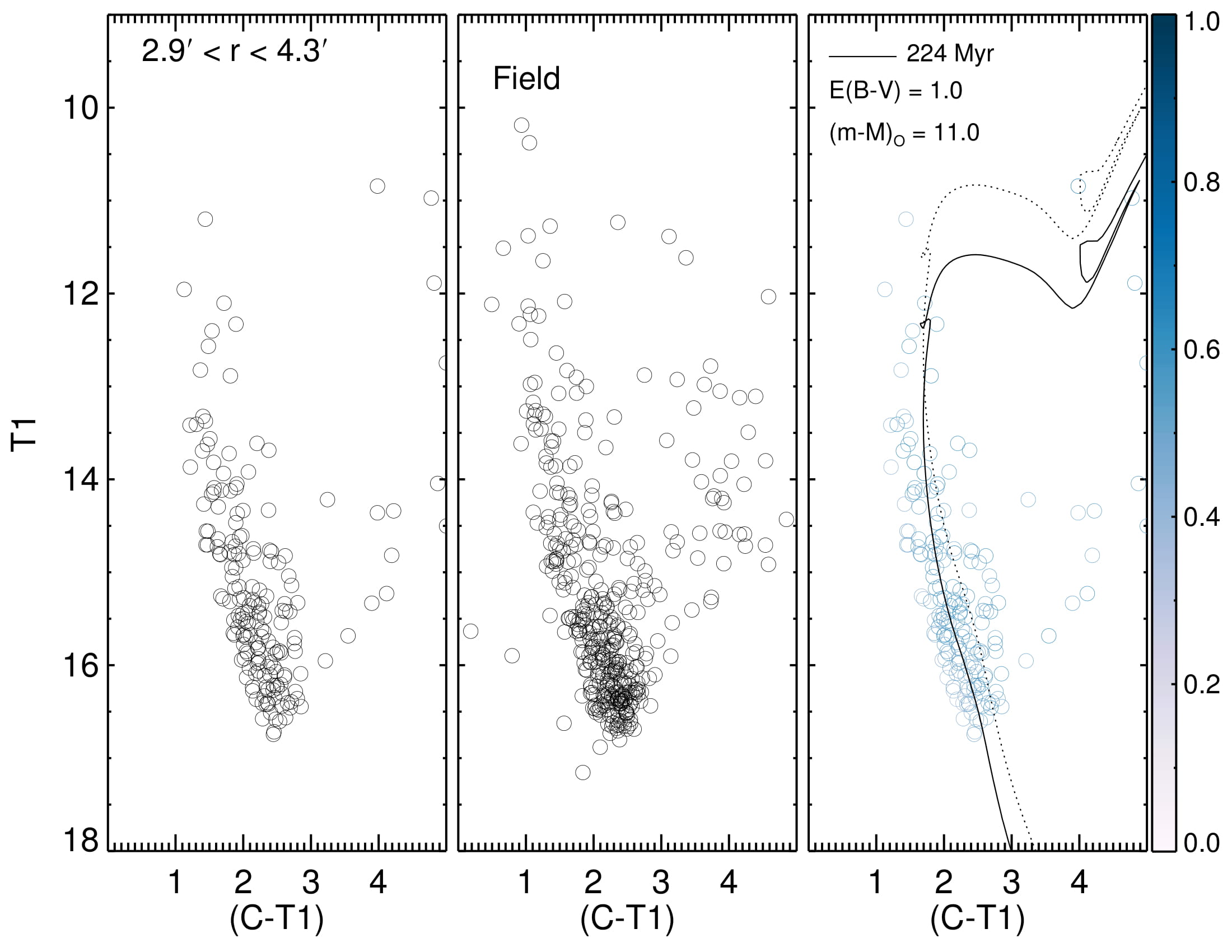}
    \includegraphics[width=0.5\textwidth]{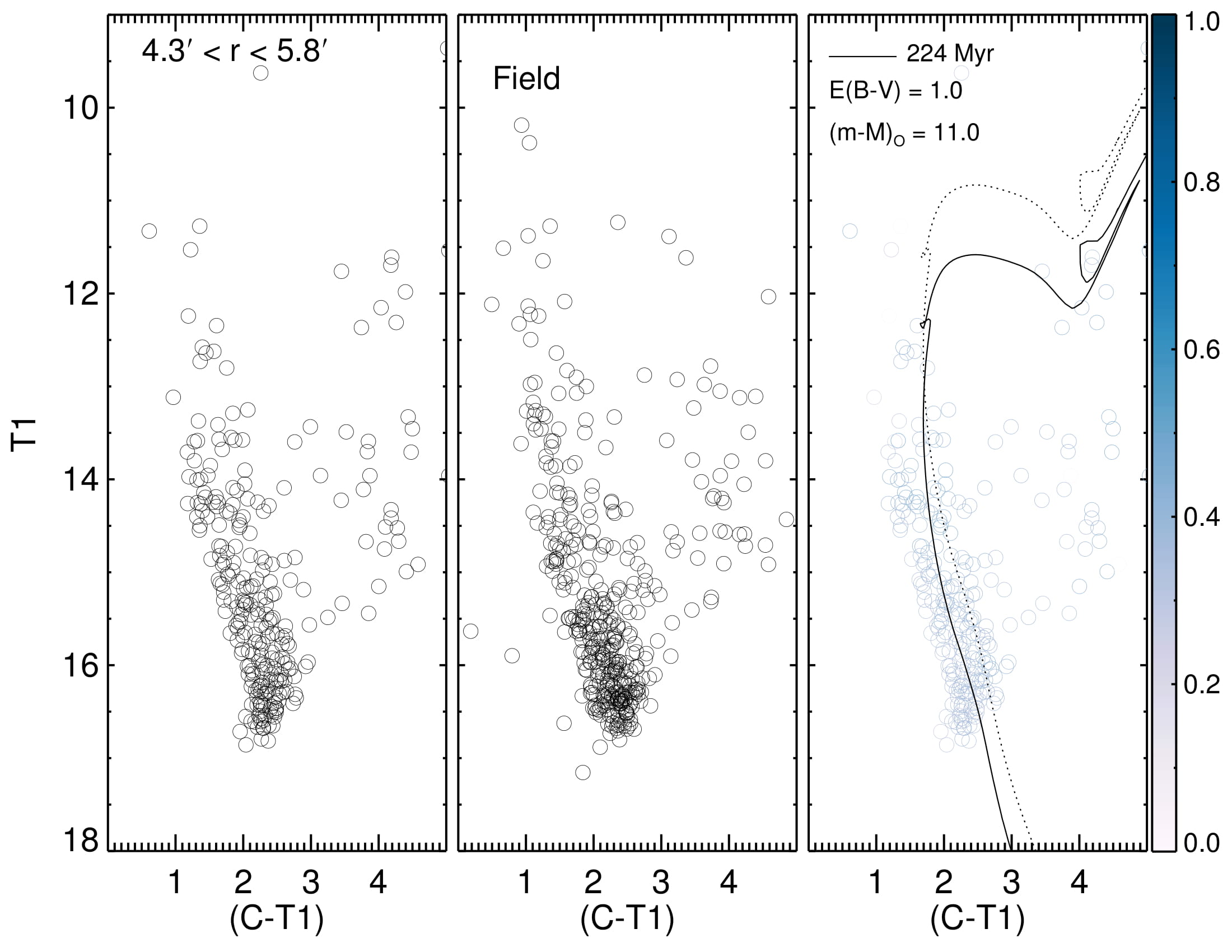}    
 
  }
\caption{   Decontamination procedure applied to stars in four annular regions centered on Ruprecht\,121. In each case, cluster (left) and control field (middle) $T_1\times(C-T_1)$ CMDs were represented. The rightmost CMD shows the derived membership likelihoods, as indicated in the colour bars.   }

\label{CMDs_decontam_cascas_exemplo}
\end{minipage}

\end{center}

\end{figure*}

We extracted $C$ and $T_1$ magnitudes of stars located in concentric annular regions (widths proportional to $r_c$ and centered on the studied OCs; Figure \ref{CMDs_decontam_cascas_exemplo}), as well as of those distributed in a control field. Control field stars are represented with open circles in Figures \ref{finding_charts_part1} and \ref{finding_charts_part2}. Firstly, we used this information to build $T_{1}\times(C-T_{1})$ CMDs for stars inside each selected annular region. Then we executed a routine that evaluates the overdensity of stars in the cluster CMD relative to the control field CMD in order to establish photometric membership likelihoods. The algorithm is fully described in \citeauthor{Maia:2010}\,\,(\citeyear{Maia:2010}). Briefly, the cluster and field CMDs are divided into small cells with sizes that are proportional to the mean uncertainties in magnitude and colour index. Here we used cell sizes ($\Delta{T_{1}}$ and $\Delta(C-T_{1})$) that are, respectively, equal to 20 and 10 times the sample mean uncertainties in $T_{1}$ and $(C-T_{1})$. Taking into account our whole sample, both $\Delta{T_{1}}$ and $\Delta(C-T_{1})$ translate into  $\sim$0.5\,mag. These cells are small enough to detect local variations of field-star contamination along the sequences in the cluster CMD, and large enough to accommodate a significant number of stars. 

Membership likelihoods ($L$) are assigned to stars in each cell according to the weight function

\begin{equation}
  L = e^{  -\left(\frac{N_{\textrm{fld}}*A_{\textrm{clu}}}{N_{\textrm{clu}}*A_{\textrm{fld}}}\right)^2} \times e^{-\left(\frac{r}{r_t}\right)^2},
\end{equation}

\noindent
where $r_t$ is the tidal radius, $N_{\textrm{clu}}$ is the number of stars counted within a cell in the cluster CMD and $N_{\textrm{fld}}$ is the number of stars counted in the corresponding cell within the control field CMD. The multiplicative factors $A_{\textrm{clu}}$ and $A_{\textrm{fld}}$ properly take into account differences in the areas corresponding to the cluster and field stars, respectively, in each decontamination procedure run. This formulation has the advantage of considering both the photometric and structural informations.

The cell positions are changed by shifting the entire grid by one-third of the cell size in each direction. Also, the cell sizes are increased and decreased by one-third of the average sizes in each of the CMD axes. Considering all possible configurations, 81 different grid sets are used. For each pair cluster-control field, the star membership likelihood is derived from the average of the memberships obtained over the whole grid configurations. This procedure is illustrated in Figure \ref{CMDs_decontam_cascas_exemplo}, which shows the results of the decontamination algorithm applied to Ruprecht\,121. Four annular regions with width $\sim0.5\,r_c$ were employed in this case. The colourbars indicate membership likelihoods. Analogous figures for the other studied OCs are showed in the Appendix.

\begin{figure*}
\begin{center}

\begin{minipage}{130mm}

\parbox[c]{1.0\textwidth}
  {
   
    \includegraphics[width=0.5\textwidth]{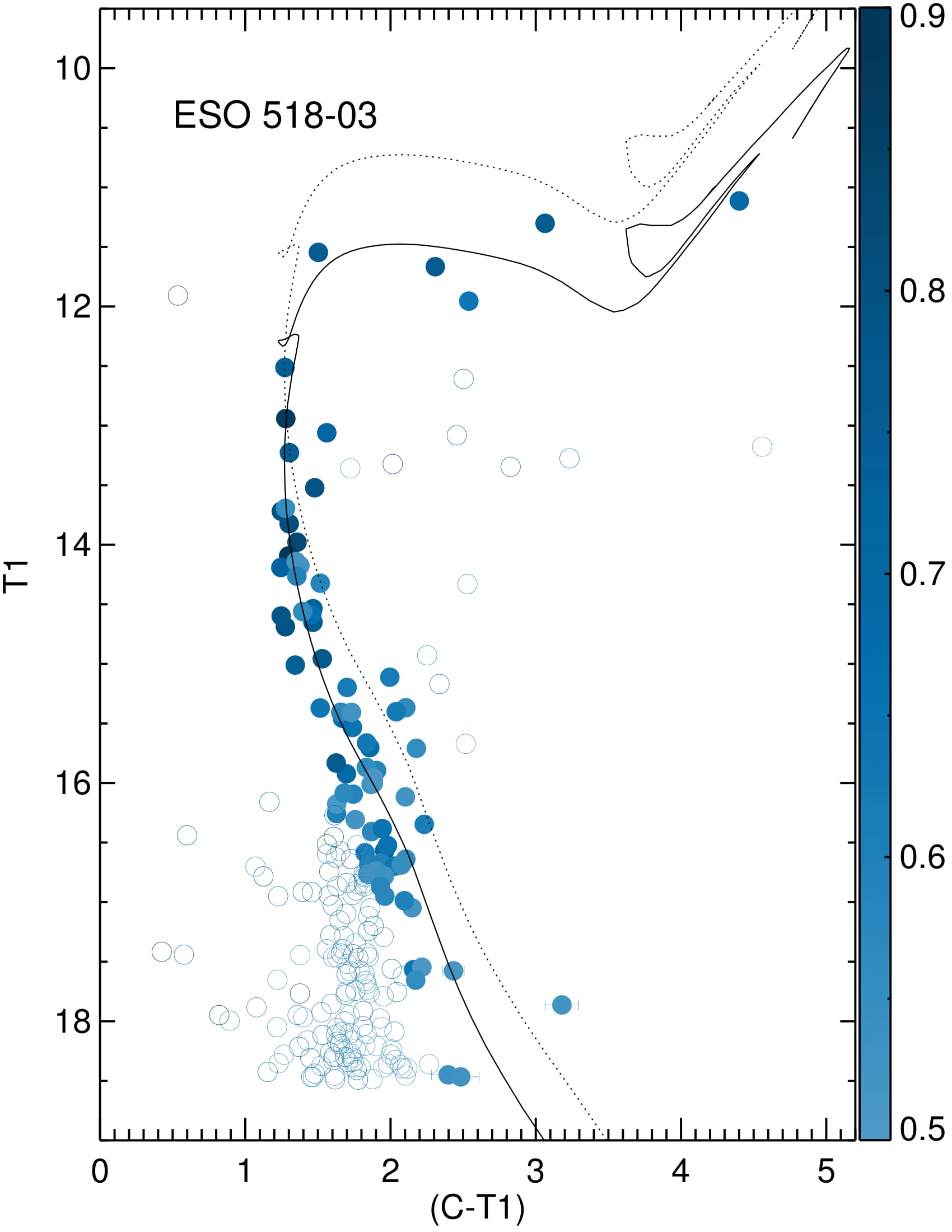}
    \includegraphics[width=0.5\textwidth]{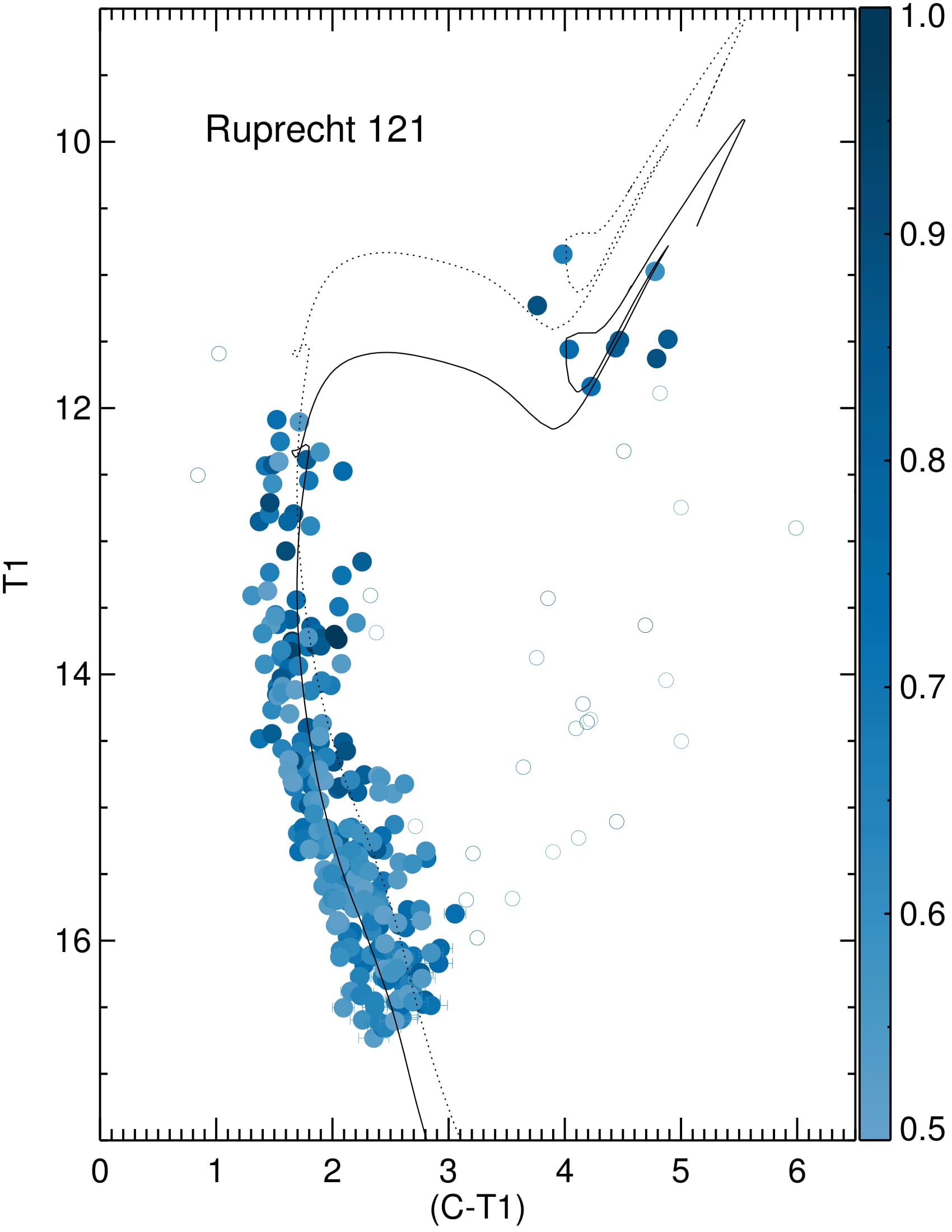}
     \includegraphics[width=0.5\textwidth]{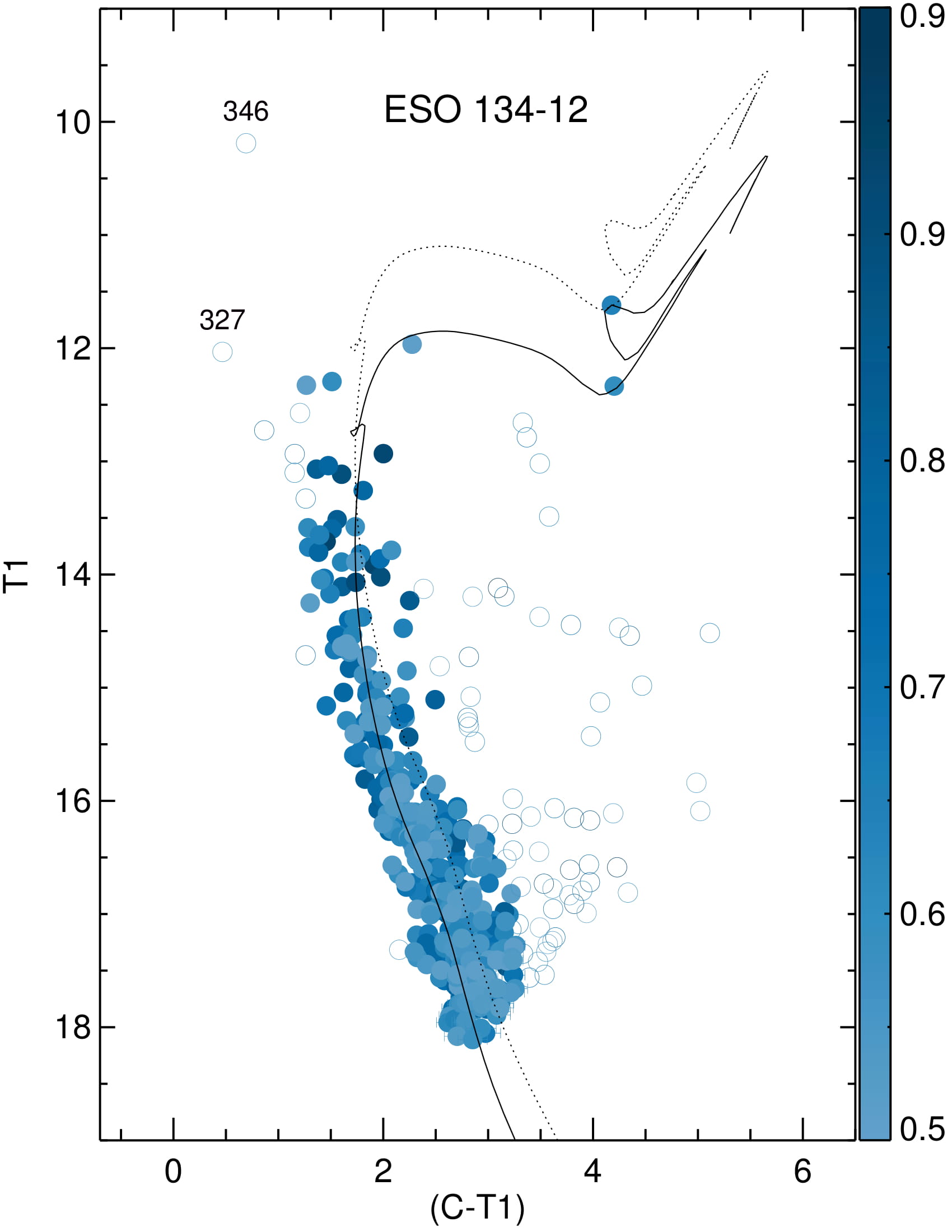}
    \includegraphics[width=0.5\textwidth]{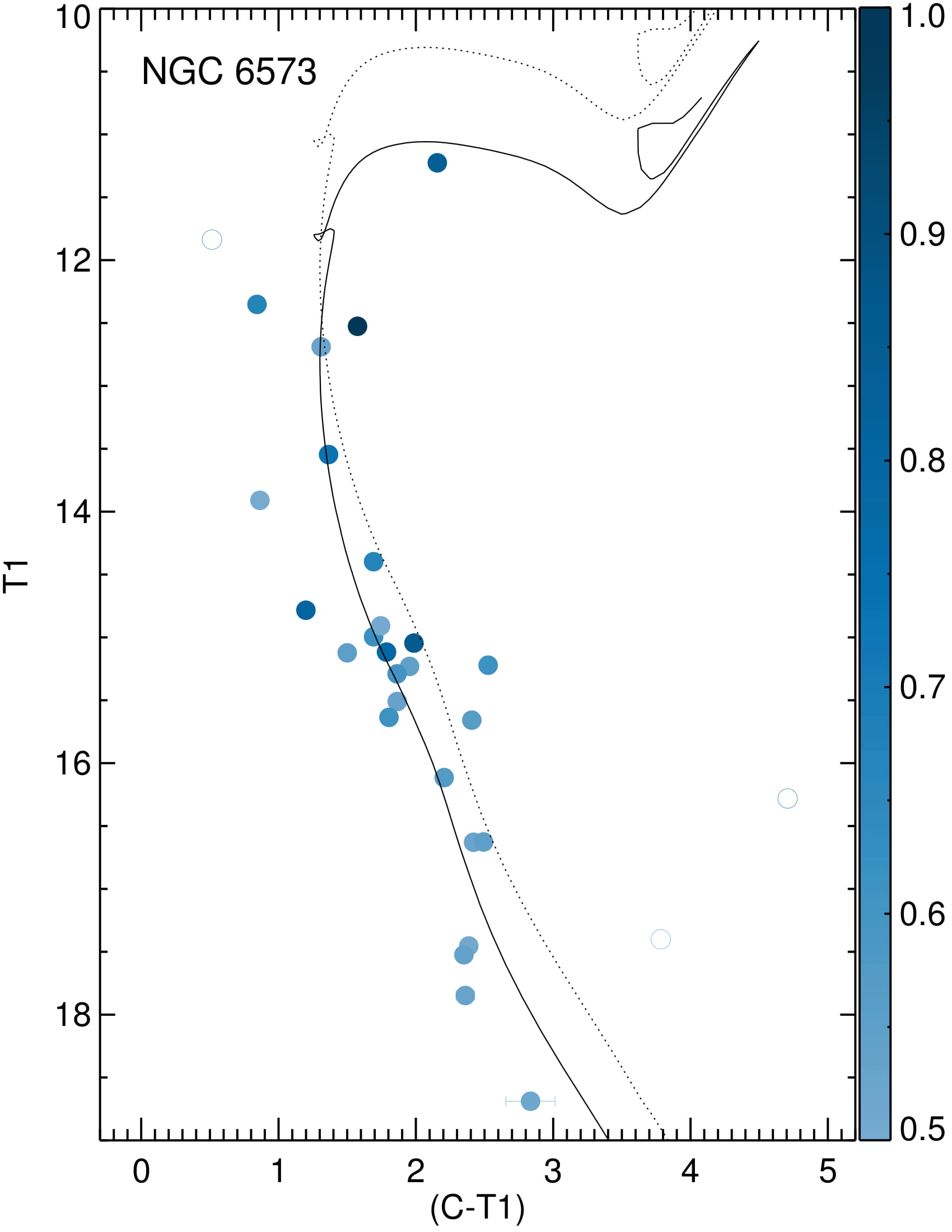}    
 
  }
\caption{Decontaminated $T_{1}\times(C-T_{1})$ CMDs for ESO\,518-3 (top-left panel), Ruprecht\,121 (top-right panel), ESO\,134-12 (bottom-left panel) and NGC\,6573 (bottom-right panel) after running Maia et al.'s (2010) algorithm. The colourbars indicate membership likelihoods ($L$). Only stars with $L\ge50\%$ were plotted. Filled circles represent photometric members and open ones are non-member stars. Continuous lines are PARSEC-COLIBRI isochrones matched to our data, while dotted ones represent the locus of unresolved binaries of equal mass. Stars with available Gaia parallaxes ($\pi$) and proper motions are numbered (internal identifiers).}

\label{CMDs_part1}
\end{minipage}

\end{center}

\end{figure*}

\begin{figure*}
\begin{center}

\begin{minipage}{130mm}

\parbox[c]{1.0\textwidth}
  {
   
    \includegraphics[width=0.5\textwidth]{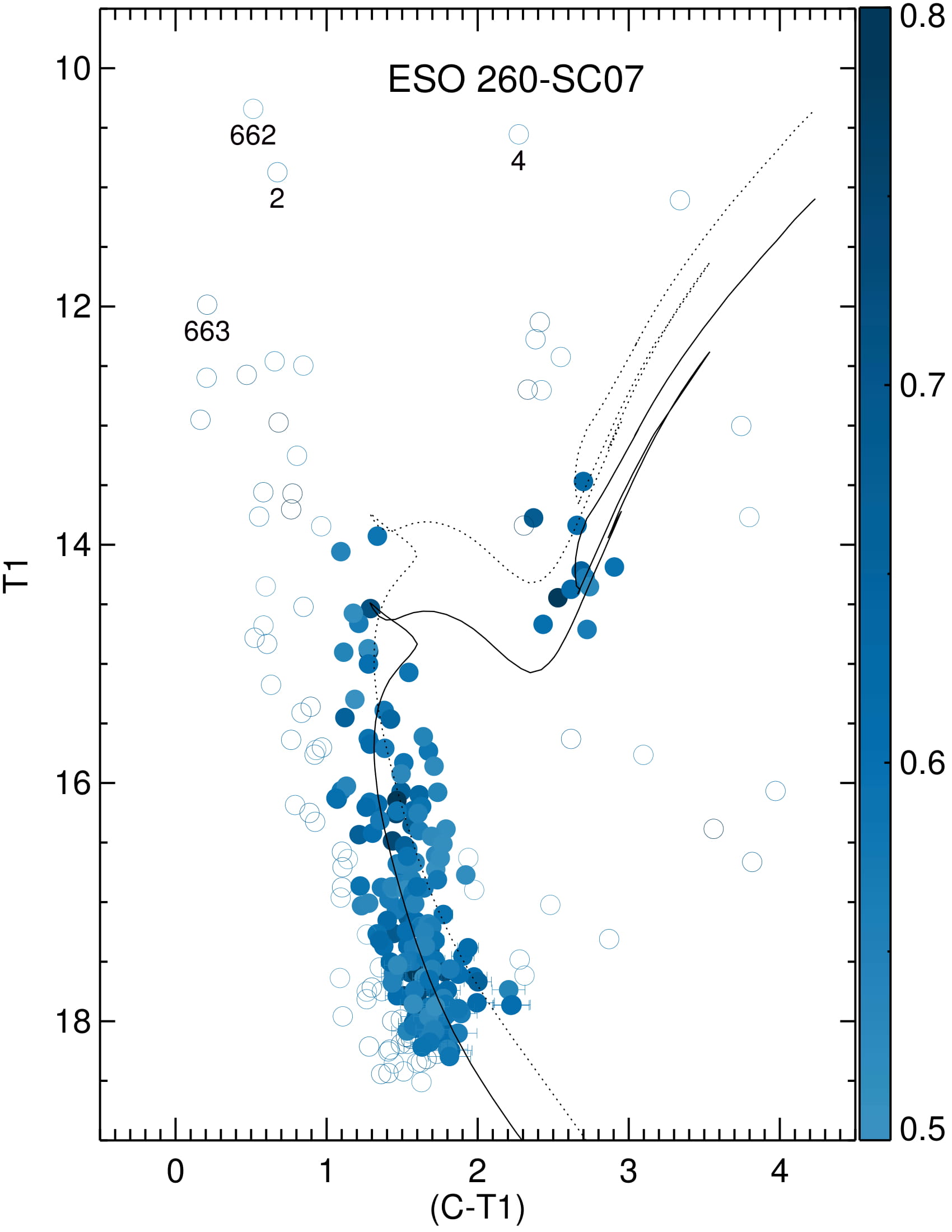}
    \includegraphics[width=0.5\textwidth]{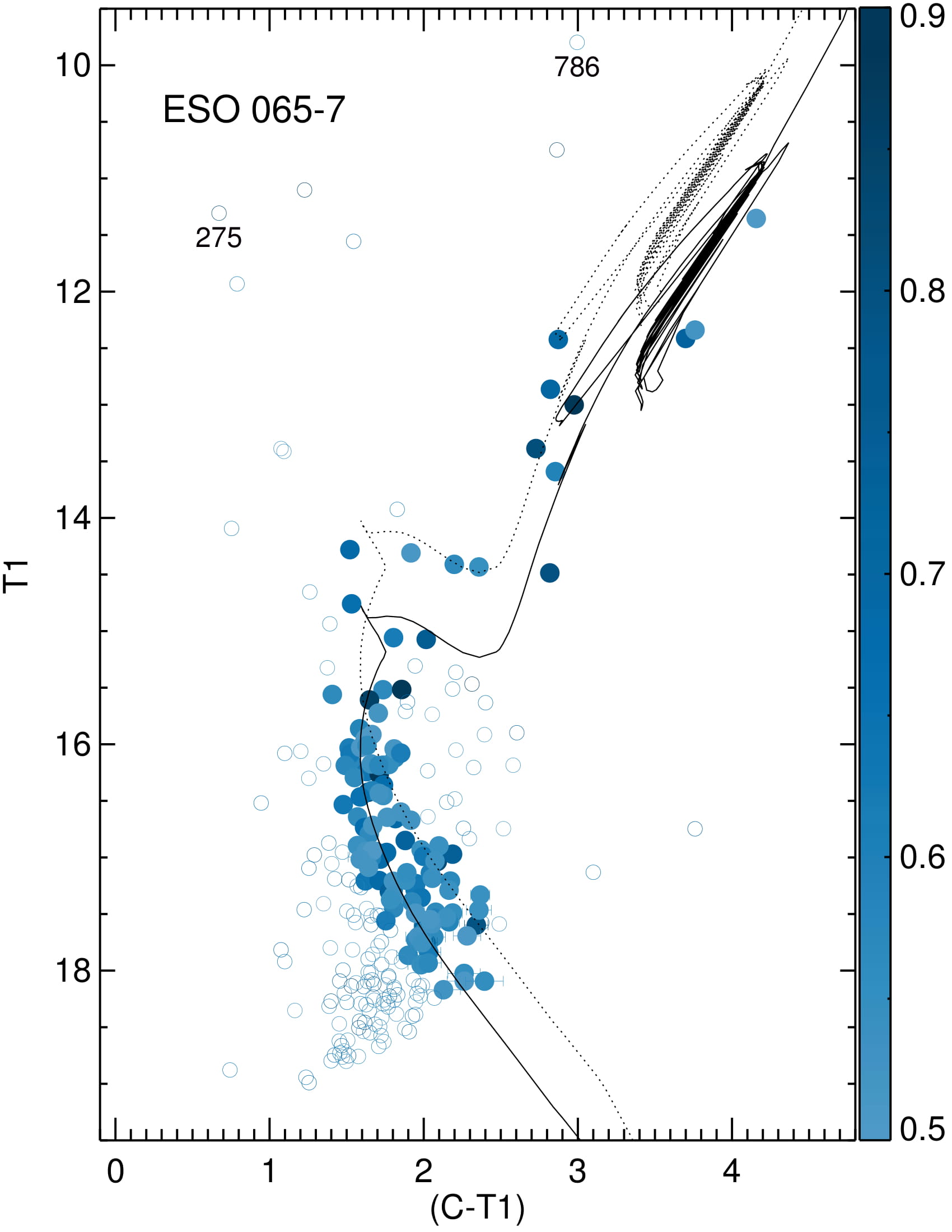}    
 
  }
\caption{Same as Figure \ref{CMDs_part1}, but for the OCs ESO\,260-7 (left panel) and ESO\,065-7 (right panel).}

\label{CMDs_part2}
\end{minipage}

\end{center}

\end{figure*}

Groups of stars with high $L$ values are significantly contrasting with the field and provide useful constraints for isochrone fitting, although some interlopers remain. The presence of residuals in the CMDs, specially red stars (Figures \ref{CMDs_part1} and \ref{CMDs_part2}), is an expected issue, since these clusters are located at relatively low Galactic latitudes ($\vert b\vert<13^{\circ}$, Table \ref{log_observations}) and thus the statistical analysis is subject to a certain degree of variability of the stellar density, reddening, luminosity function and colour distribution of the disc field stars. 

A new generation of PARSEC-COLIBRI isochrones (\citeauthor{Marigo:2017}\,\,\citeyear{Marigo:2017})
was used to estimate the OC astrophysical properties by matching them to the distribution of stars with $L\ge0.5$ (Figures \ref{CMDs_part1} and \ref{CMDs_part2}). We made use of a set of solar metallicity isochrones 
ranging from 7.50 up to 8.60 in steps of 0.05 in log($t$), for the younger OCs, and ranging from 9.00 up to 9.60 for the older OCs. Each isochrone was reddened and vertically shifted by using the following relations: $E(C-T_{1})/E(B-V)=1.97$; $A_{T_{1}}/E(B-V)=2.62$  \citep{Geisler:1996} until matching the observed OC CMD features. Reddening values from dust maps ($E(B-V)_{\textrm{dust}}$, \citeauthor{Schlafly:2011}\,\,\citeyear{Schlafly:2011}; \citeauthor{Green:2015}\,\,\citeyear{Green:2015}), whenever available, were used as initial guesses and then we tested values in the range $\sim0.5-2\times E(B-V)_{\textrm{dust}}$.


We also considered for each isochrone matching the effect of unresolved binaries. To do this we shifted the isochrone in steps of 0.01\,mag in direction of decreasing $T_{1}$ down to 0.75\,mag, which is the limit corresponding to 
unresolved binaries of equal mass. Finally, we chose the isochrone which best resembles the sequences traced by the stars with $L\ge0.5$ (see Table \ref{clusters_total_masses}). Stars with available Gaia DR1 parallaxes ($\pi$, see Table \ref{Gaia_stars}) were labelled  in the CMDs of ESO\,134-12, ESO\,260-7 and ESO\,065-7. As these stars were not fitted by the chosen isochrones (Figures \ref{CMDs_part1} and \ref{CMDs_part2}), they were dismissed as possible clusters members.

\begin{table}
 \begin{minipage}{85mm}
  \caption{Gaia DR1 parallaxes and proper motions for stars in the OC observed fields.}
  \label{Gaia_stars}
 \begin{tabular}{lccrr}
 
\hline

Cluster             &   ID      &   $\pi$                     &  $\mu\rmn{RA}$          &  $\mu\rmn{DEC}$               \\
                        &             &  (mas)                     &  (mas yr$^{-1}$)          &   (mas yr$^{-1}$)                 \\

\hline
ESO\,134-12   &  327     & 0.82\,$\pm$\,0.36   & -7.43\,$\pm$\,1.56       & -8.05\,$\pm$\,0.49     \\
                        &  346     & 2.44\,$\pm$\,0.33   & 9.41\,$\pm$\,1.37        &  4.64\,$\pm$\,0.44    \\
                        &             &                                &                                     &                                    \\

ESO\,260-7   &  2         & 1.81\,$\pm$\,0.34   &  -12.32\,$\pm$\,1.52    & -1.40\,$\pm$\,0.86       \\
                      &  4         & 0.83\,$\pm$\,0.31   &  -4.71\,$\pm$\,1.41      & 10.35\,$\pm$\,1.02      \\
                      &  662     & 1.42\,$\pm$\,0.35   &  -11.28\,$\pm$\,1.00    & 12.03\,$\pm$\,0.86       \\
                      &  663     & 0.26\,$\pm$\,0.52   &  -5.52\,$\pm$\,2.30      & 5.54\,$\pm$\,1.57        \\                   
                      &             &                                &                                     &                                    \\

ESO\,065-7   &  275     & 0.39\,$\pm$\,0.34   & -10.12\,$\pm$\,0.57     & -3.62\,$\pm$\,0.82     \\
                      &  786     & 0.60\,$\pm$\,0.41   & -20.07\,$\pm$\,0.70     & -14.20\,$\pm$\,0.92   \\
                      &             &                                &                                     &                                    \\
\hline

\end{tabular}
\end{minipage}
\end{table}

\begin{table*}
 \begin{center}
  \caption{Fundamental parameters, relaxation times and photometric masses ($M_{\textrm{phot}}$) for the studied clusters. $M_{\textrm{Kroupa}}$ and $N_{\textrm{Kroupa}}$ are upper limits for cluster mass and number of stars, respectively.}
  \label{clusters_total_masses}
 \begin{tabular}{lcccccccc}
 
\hline

Cluster          &  (m-M)$_{0}$          & $d$                 &   $E(B-V)$           &   log($t$/yr)          &   $t_{rh}$                   & $M_{\textrm{phot}}$             & $M_{\textrm{Kroupa}}$   &  $N_{\textrm{Kroupa}}$    \\
                 &   (mag)               & (kpc)               &    (mag)             &                        &   (Myr)                      & ($M_{\odot}$)                   & ($M_{\odot}$)                    &                  \\                                                                                                                        
\hline                                                                                                                                                                                                                                                      

ESO\,518-3       &11.60\,$\pm$\,0.30     &2.09\,$\pm$\,0.29    &0.80\,$\pm$\,0.10     &8.30\,$\pm$\,0.10       & 1.43\,$\pm$\,0.25            & 111\,$\pm$\,17                  & 376                         &  835    \\ 
Ruprecht\,121    &11.00\,$\pm$\,0.30     &1.58\,$\pm$\,0.22    &1.00\,$\pm$\,0.05     &8.35\,$\pm$\,0.10       & 3.18\,$\pm$\,0.24            & 405\,$\pm$\,32                 & 1592                       &  3565   \\
ESO\,134-12      &11.50\,$\pm$\,0.30     &2.00\,$\pm$\,0.28    &1.05\,$\pm$\,0.10     &8.25\,$\pm$\,0.10       & 6.82\,$\pm$\,0.53            & 500\,$\pm$\,33                 & 1520                       &  3360   \\ 
NGC\,6573        &11.00\,$\pm$\,0.30     &1.58\,$\pm$\,0.22    &0.80\,$\pm$\,0.10     &8.35\,$\pm$\,0.10       & 0.12\,$\pm$\,0.04            & 35\,$\pm$\,9                     & 124                         &  279    \\ 
ESO\,260-7       &12.65\,$\pm$\,0.30     &3.39\,$\pm$\,0.47    &0.40\,$\pm$\,0.05     &9.15\,$\pm$\,0.10       & 6.77\,$\pm$\,0.61            & 152\,$\pm$\,14                  & 531                        &  1380   \\
ESO\,065-7       &11.90\,$\pm$\,0.30     &2.40\,$\pm$\,0.33    &0.40\,$\pm$\,0.05     &9.45\,$\pm$\,0.05       & 4.02\,$\pm$\,0.34            & 86\,$\pm$\,10                   & 386                        &  1077   \\

\hline

\end{tabular}
\end{center}
\end{table*}

Fundamental parameters for these objects were also derived by K13. In what follows, we summarize the main discrepancies with our results. ($i$) Regarding $E(B-V)$, the greatest difference is found for ESO\,518-03; relatively to our value, K13 found a value 50\% smaller; for the other clusters, differences vary from 4\% to 30\% ; ($ii$) for log\,($t$), we found good agreement with the literature values within 0.2\,dex in the cases of Ruprecht\,121, ESO\,260-7 and ESO\,065-7; for the other clusters, the differences in log\,($t$) are 0.87 (ESO\,518-3), 0.65 (ESO\,134-12) and 0.45 (NGC\,6573), with K13 presenting older ages; (iii) regarding the derived distances ($d$), our results reproduce the literature ones, within the uncertainties quoted in Table \ref{clusters_total_masses} (uncertainties in $d$ not informed in K13 tables) only in the cases of Ruprecht\,121 and ESO\,134-12; for the other clusters, differences are of the order of $\sim1\,$kpc (smaller distances found by K13 in the cases of ESO\,518-3 and ESO\,260-7; greater distances in K13 for NGC\,6573 and ESO\,065-7). 

We speculate that, at least partially, these differences may be due to the photometric completeness in the 2MASS $JHK_s$-bands employed by K13; since the 6 studied OCs are located at low Galactic latitudes ($b<13^{\circ}$) and/or project in the direction of the Galactic centre, the $K_s\times(J-K_s)$ CMDs analysed by K13 go down $\sim$2\,mag or less below the main sequence turnoff. Particularly in the cases of NGC\,6573 and ESO\,065-7, the isochrone fitting procedure was performed basically on stars in the subgiant and red giant branches, which makes the derived parameters more uncertain.

\subsection{Clusters masses}
\label{mass_functions}

\begin{figure*}


\parbox[c]{0.7\textwidth}
  {
   
    \includegraphics[width=0.7\textwidth]{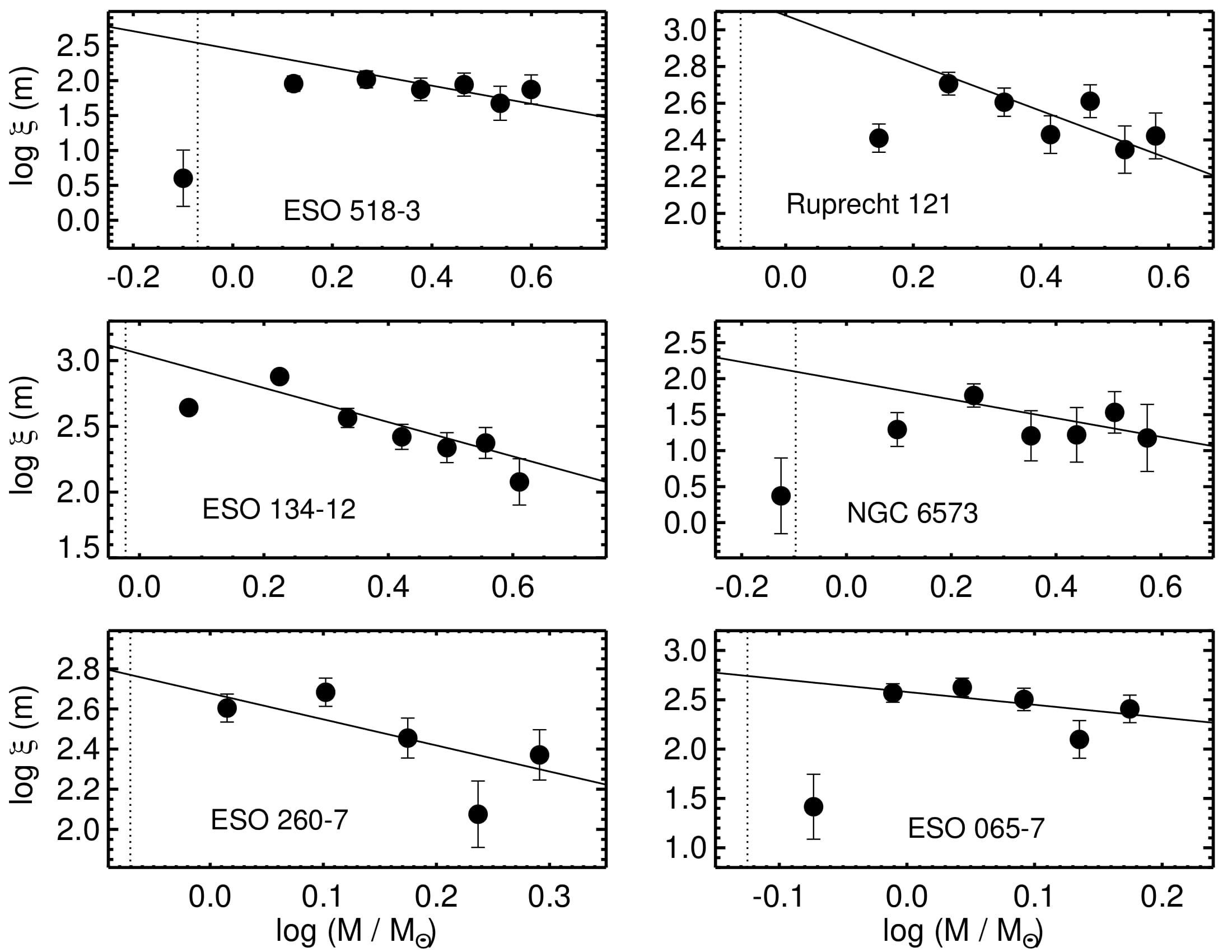}

  }
\caption{Observed mass functions for ESO\,518-3 (top-left panel), Ruprecht\,121 (top-right panel), ESO\,134-12 (middle-left panel), NGC\,6573 (middle-right panel), ESO\,260-7 (bottom-left panel) and ESO\,065-7 (bottom-right panel). The continous lines show the \citeauthor{Kroupa:2001}\,\,(\citeyear{Kroupa:2001}) IMF, for comparison. For reference, the dotted vertical lines correspond to the 50\% completeness limit of our photometry.}

\label{phi_m}


\end{figure*}

We derived individual masses of photometric cluster members by interpolation in the theoretical isochrone, properly corrected by reddening and distance modulus (Figures \ref{CMDs_part1} and \ref{CMDs_part2}), from the star $T_1$ magnitude. The half-mass radius of each OC ($r_h$, Table \ref{tab_params_OCs}) was derived directly from the list of members, by taking their individual masses and distances to the cluster centre. Then, the clusters' mass functions (MF, Figure \ref{phi_m}) were constructed by counting the number of stars in linear mass bins (i.e., $\xi\,(m)=dN/dm$), which were then converted to the logarithmic scale and plotted in Figure \ref{phi_m}. Poisson statistics was assumed for uncertainties determination. The dotted vertical lines correspond to the 50\% completeness limit of our photometry, for reference (see also Figure \ref{completeness_versus_mag}). Star counts inside each bin were weighted by the photometric membership likelihoods (Figures \ref{CMDs_part1} and \ref{CMDs_part2}) and corrected for completeness (Figure \ref{completeness_versus_mag}). Finally, the total observed cluster masses ($M_{\textrm{phot}}$) were determined by discrete sum of the mass bins in Figure \ref{phi_m}. They are listed in Table \ref{clusters_total_masses}, where the uncertainties come from error propagation. 

\begin{figure}
\begin{center}
 \includegraphics[width=8.5cm]{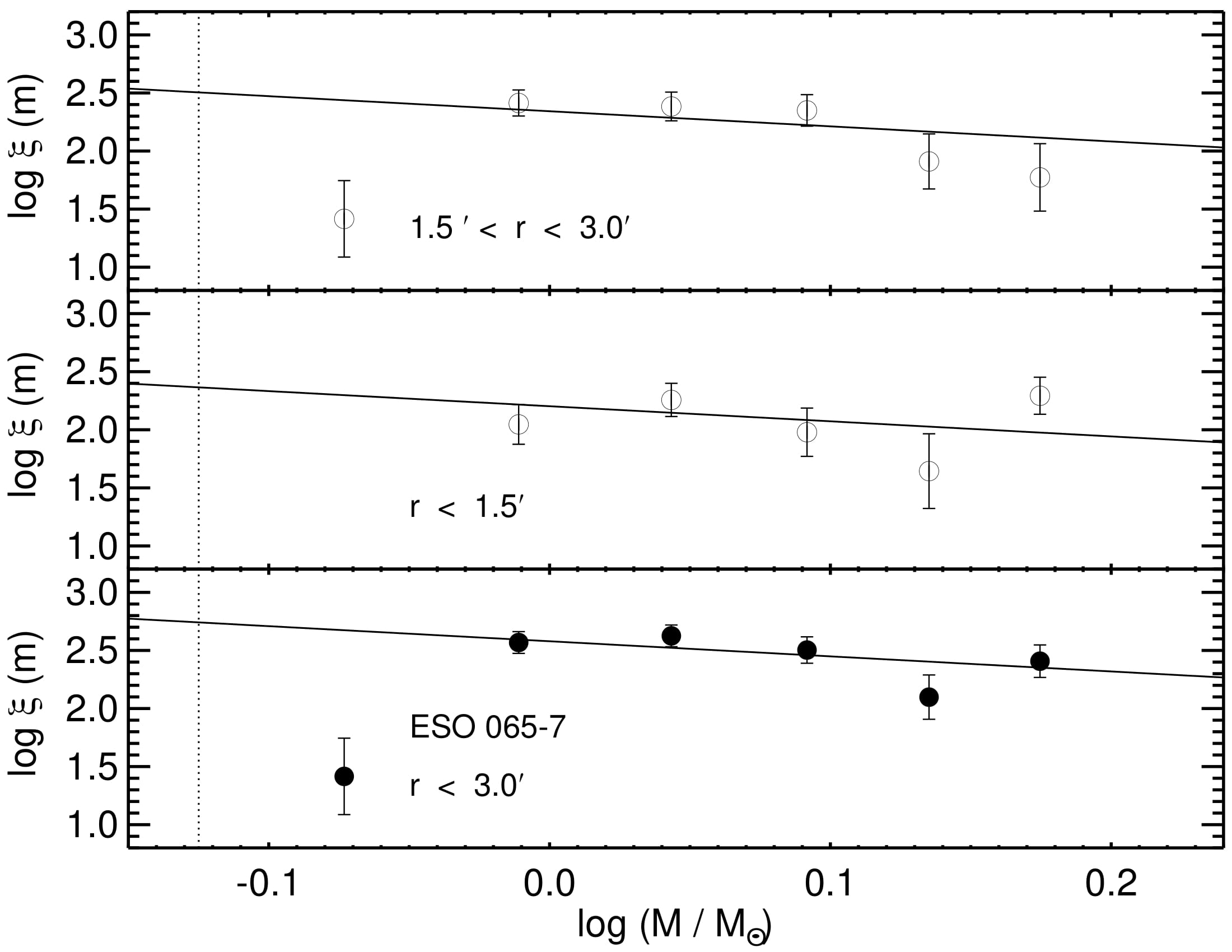}
 \caption{   The plots show the observed mass function of stars in the outer region (top panel), in the inner region (middle panel) and the overall mass function (bottom panel) of each ESO\,065-7, as indicated. The continuous line is the Kroupa (2001) IMF. }
   \label{phi_m_allregions_ESO065-7}
\end{center}
\end{figure}

For comparison purposes, the initial mass function (IMF) of \cite{Kroupa:2001} was scaled according to each photometric cluster mass and superimposed to the observed mass functions (solid lines in Figure \ref{phi_m}). Except for ESO\,260-7, all studied OCs present signals of low-mass stars depletion, which is a signature of dynamically evolved OCs (\citeauthor{de-La-Fuente-Marcos:1997}\,\,\citeyear{de-La-Fuente-Marcos:1997}; \citeauthor{Portegies-Zwart:2001}\,\,\citeyear{Portegies-Zwart:2001}; \citeauthor{Baumgardt:2003}\,\,\citeyear{Baumgardt:2003}). In all cases, the more massive bins are consistent with the IMF steepness, although with some scatter. We also checked for possible variations in the steepness of the derived MF across the studied OCs. For each object, we built two additional MFs: one for stars located in the range $0<r<0.5\,r_{\textrm{max}}$ and another one for the region $0.5\,r_{\textrm{max}}<r<r_{\textrm{max}}$, where $r_{\textrm{max}}$ is the maximum distance of photometric members to the cluster centre. The results are show in Figure \ref{phi_m_allregions_ESO065-7} for ESO\,065-7 and in Figure \ref{phi_m_allregions} for the other 5 OCs. In all cases the derived MFs do not deviate significatively from the Kroupa IMF, except in the lower mass bins, which are affected by preferential evaporation of low-mass stars.

Since there may be member stars below the detection limits, we performed rough estimates of upper mass limits ($M_{\textrm{Kroupa}}$) and number of stars ($N_{\textrm{Kroupa}}$) by integrating the scaled Kroupa's law down to a stellar mass of 0.1$M_{\odot}$. The resulting $M_{\textrm{Kroupa}}$ and $N_{\textrm{Kroupa}}$ values are showed in Table \ref{clusters_total_masses}. As can be seen, the $M_{\textrm{phot}}$ values resulted to be $\sim30$ per cent of $M_{\textrm{Kroupa}}$ for all studied OCs.

\section{Discussion}
\label{discussion}


In order to study the evolutionary stages of the OC sample, we considered parameters closely associated with their dynamical evolution, i.e., age, stellar mass, core, half-mass and tidal radii.  We in addition computed the half-mass relaxation times according to the equation \citep{Spitzer:1987}

\begin{equation}
     t_{rh}=0.138\frac{M^{1/2}\,r_{h}^{3/2}}{G^{1/2}\,\langle m\rangle\,\textrm{ln}(0.4\,N)},
     \label{definicao_trh}
\end{equation}  

\noindent
where $G$ is the gravitational constant, $M$ is the cluster mass ($M_{\textrm{phot}}$, Table \ref{clusters_total_masses}), $r_{h}$ is the half-mass radius (Table \ref{tab_params_OCs}), and $\langle m\rangle=M/N$ is the global mean stellar mass, where $N$ is the number of photometric members (Figures \ref{CMDs_part1} and \ref{CMDs_part2}). 

\begin{figure*}
\begin{center}


\parbox[c]{0.8\textwidth}
  {
   
    \includegraphics[width=0.8\textwidth]{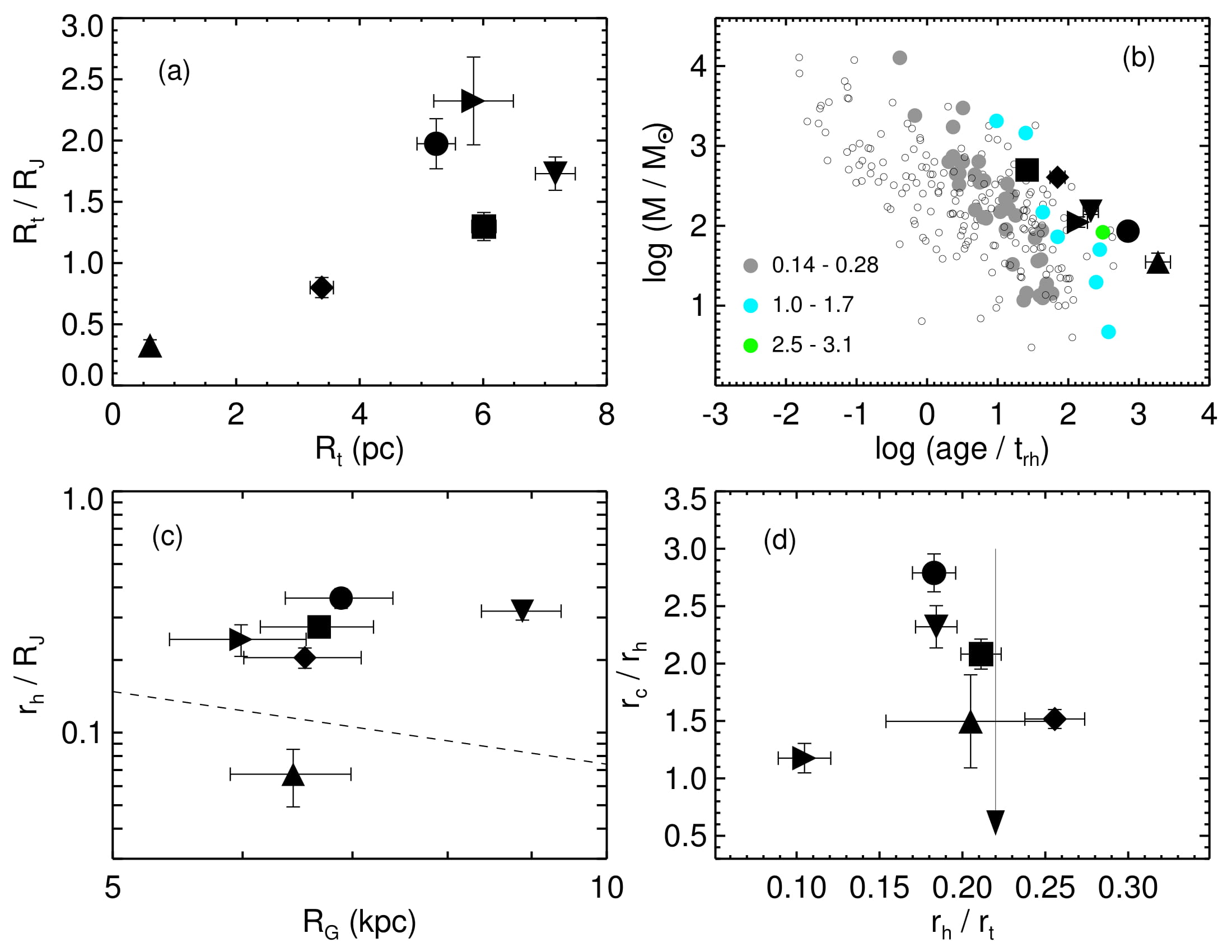}
 
  }
\caption{   Plots (a) $r_t/R_J$ versus $R_t$, (b) logarithm of mass versus age/$t_{rh}$ ratio, (c) $r_h/R_J$ versus $R_G$ and (d) $r_c/r_h$ versus $r_h/t_t$ ratio for the studied clusters. In panel (b), open symbols and filled coloured ones represent 236 OCs from the sample of Piskunov et al. (2007, see text for details). Black filled symbols are: $\blacktriangleright$ (ESO\,518-3), $\blacklozenge$ (Ruprecht\,121), $\blacksquare$ (ESO\,134-12), $\blacktriangle$ (NGC\,6573), $\blacktriangledown$ (ESO\,260-7) and $\bullet$ (ESO\,065-7).  }

\label{plots_radius_age_mphot}


\end{center}

\end{figure*}

The $t_{rh}$ tells us about the typical timescale in which a system reaches thermal equilibrium \citep{Portegies-Zwart:2010}. From a theoretical point of view, it is the timescale on which stars tend to establish a Maxwellian velocity distribution, continously repopulating its high-velocity tail and thus losing stars by evaporation. The $t_{rh}$ values derived for the studied OCs are listed in Table \ref{clusters_total_masses}. The derived ages resulted at least one order of magnitude larger than $t_{rh}$ (Figure \ref{plots_radius_age_mphot}, panel (b)), which suggests that these OCs have had enough time to evolve dynamically. This is also true even if we considered $M_{\textrm{Kroupa}}$ and $N_{\textrm{Kroupa}}$ from Table \ref{clusters_total_masses} to compute $t_{rh}$, still supporting that the OCs have lived many times their $t_{rh}$.

The Jacobi radii (Table \ref{tab_params_OCs}) were also determined from the equation \citep{von-Hoerner:1957}:

\begin{equation}
   R_{J} = \left(\frac{M_{\textrm{clu}}}{3\,M_{\textrm{G}}}\right)^{1/3}\,\times\,R_{\textrm{G}},
   \label{definicao_rJ}
\end{equation}

\noindent
where $M_{\textrm{clu}}$ is the cluster mass and $M_{\textrm{G}}$ is the Milky Way (MW) mass inside the cluster Galactocentric distance ($R_{\textrm{G}}$). This formulation assumes a circular orbit around a point mass galaxy. The value of $M_{\textrm{G}}$ (=$5\times10^{11}M_{\odot}$) was obtained from the MW mass profile of \cite{Taylor:2016}. The $R_{\textrm{G}}$ were obtained from the distances in Table \ref{tab_params_OCs}, assuming that the Sun is located at 8.0\,$\pm$\,0.5\,kpc from the Galactic centre \citep{Reid:1993a}.

\begin{figure}
\begin{center}

\begin{minipage}{85mm}

    \includegraphics[width=1.0\textwidth]{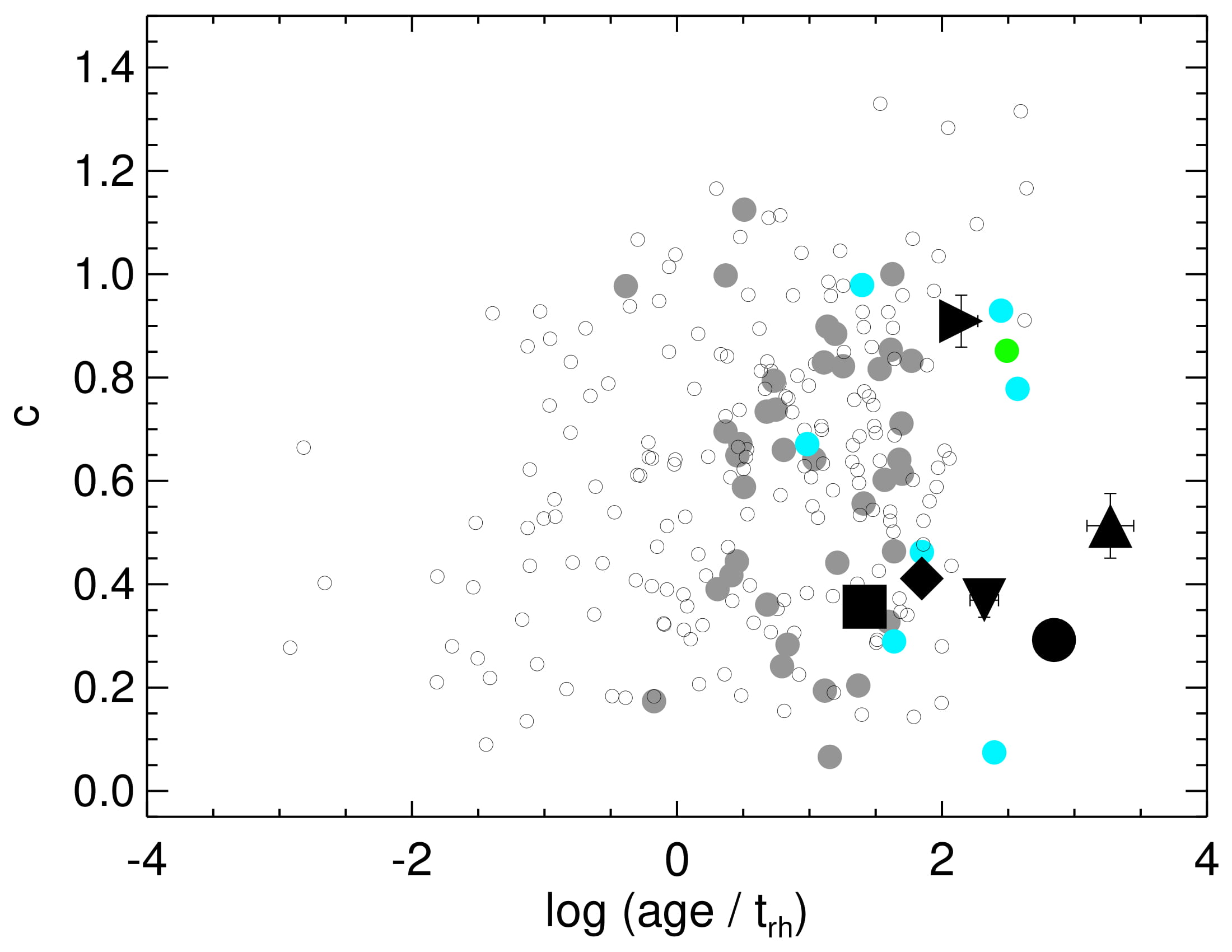}

\caption{ Relationship between the concentration parameter $c$ and log (age/$t_{rh}$). Symbols are the same as those used in Figure \ref{plots_radius_age_mphot}. }

\label{c_versus_age_trh}
\end{minipage}

\end{center}

\end{figure}

The analysis of the derived structural parameters can give some indications of the OCs' dynamical stage. In this context, it is important to highlight that quantities such as mass, $r_t/r_J$, the concentration parameter $c=\log(r_t/r_c)$ (Figure \ref{c_versus_age_trh}) and $r_c/r_h$ depend on the unknown initial formation conditions. What follows are considerations made on the basis of the observed trend between various structural parameters in a large sample of OCs and can give an indication of the dynamical stage of these systems in the simplified assumption that all OCs were born with the same mass, density profile and stellar population.

Figure \ref{plots_radius_age_mphot}, panel (a), shows the ratio $r_t/R_{J}$ for the studied OCs. As can be seen, ESO\,518-3, ESO\,134-12, ESO\,260-7 and ESO\,065-7 present tidal radius larger than their respective Jacobi radii, which suggests that these OCs might be experiencing relatively important mass loss processes, just towards final disruption. On the other hand, Ruprecht\,121 and NGC\,6573 have most of their stellar contents within their Jacobi volume, and therefore they are under relatively less notable mass loss. Ruprecht\,121 may be a transition object entering in its final disruption stage, with its $r_t$ comparable to $R_J$. It is noticeable that NGC\,6573 presents the poorest CMD and, at the same time, it is the most compact OC within our sample. Although underpopulated, its compact structure apparently allowed this OC to keep its remnant stellar content gravitationally bounded throughout its dynamical evolution.

Since the studied OCs have lived many times their typical relaxation times, it is expected some trend of the present-day cluster mass with the age/$t_{rh}$ ratio. Figure \ref{plots_radius_age_mphot}, panel (b), depicts such a relationship for them (black symbols) and for 236 OCs located in the solar neighbourhood analysed by \cite{Piskunov:2007}. Filled grey symbols represent Piskunov's OCs within the age range ($\sim0.14-0.28\,$Gyr) of 4 of our studied OCs, namely: ESO\,518-3 ($t=0.20\,\pm0.05\,$Gyr), Ruprecht\,121 ($t=0.22\pm0.05\,$Gyr), ESO\,134-12 ($t=0.18\pm0.04\,$Gyr) and NGC\,6573 ($t=0.22\pm0.05\,$Gyr); blue symbols represent clusters within the age range $\sim1.0-1.7\,$Gyr, which corresponds to ESO\,260-7 ($t=1.4\pm0.3\,$Gyr); the green symbol corresponds to the age range $\sim2.5-3.1\,$Gyr, which complies with ESO\,065-7 ($t=2.8\pm0.3\,$Gyr).

 \cite{Piskunov:2007} derived $r_c$ and $r_t$ by fitting K62 profiles to the cluster stellar density distributions and masses from their $r_t$ values. We computed here $r_{h}$ values assuming that the cluster stellar density profiles are indistinguishably reproduced by K62 and \cite{Plummer:1911} models. Hence, $t_{rh}$ values were derived from equation \ref{definicao_trh}. The six studied OCs span the most evolved half part of the age/$t_{rh}$ ratio range and tend to present higher masses compared to their solar neighborhood counterparts with compatible age/$t_{rh}$ values. This is an expected result, since all studied OCs (except for ESO\,260-7) are located inside the solar ring ($R_{G}<8\,$kpc) and thus subject to a stronger Galactic tidal field compared to clusters in the solar neighborhood. Their more massive nature may have allowed them to live for many times their age/$t_{rh}$ ratios.  

Figure \ref{plots_radius_age_mphot} (c) shows the $r_h/R_J$ versus $R_G$ plot. The distribution of clusters in this plot gives some indication of the strength of the Galaxy tidal field and it was employed by \cite{Baumgardt:2010} to identify two distinct groups of globular clusters in the Milky Way. We can notice the absence of clusters with $r_h/R_J>0.5$. Such clusters should be quickly destroyed, since they would be subject to strong tidal forces. Following \cite{Baumgardt:2010}, the dashed line was plotted here as a reference and it depicts the position that a cluster with $M=10^5\,$M$_{\odot}$ and $r_h=3\,$pc would have at different Galactocentric distances. Except for NGC\,6573, all studied clusters present $r_h/R_J$ in the range $\sim0.20<r_h/R_J< \sim0.35$ and therefore they are more tidally influenced, which is consistent with the overall disruption scenario. On the other hand, the dynamical evolution of the compact cluster NGC\,6573 ($r_h/R_J\sim0.07$) seem to be mainly driven by its internal relaxation.

In the panel (d) of Figure \ref{plots_radius_age_mphot}, we plotted the positions of the studied OCs in the $r_c/r_h$ versus $r_h/r_t$ plane. We included the evolutionary path \cite[solid arrow,][]{Heggie:2003} a cluster would follow
if it were tidally filled at the beginning of its evolution. In such a case, $r_h/r_t$ remains constant during the cluster evolution and $r_c$ and $r_h$ decrease due to violent relaxation in the cluster core region followed by two-body relaxation, mass segregation and finally core-collapse. This is consistent with the results of direct $N-$body simulations of \cite{Gieles:2008}, for which clusters that begin their evolution in their tidal regime dissolve completely in this regime. The derived $r_h/r_t$ ratio for the present cluster sample is $0.19\pm0.05$, which is consistent with the expected value for a tidally filled cluster.


Figure \ref{c_versus_age_trh} shows the concentration parameter $c$ (=log($r_t/r_c$)) as function of 
log(age/$t_{rh}$) for the studied OCs and for Piskunov et al.' sample. The $c$ parameter of the studied OCs vary from moderate ($\sim$0.3) to relatively high ($\sim$0.9) values. The ensemble of points show a general trend, although with considerable scatter, in which the dynamical evolution conducts to greater $c$ values and that star clusters tend to initially start their dynamical evolution with relatively small-concentration parameters (\citeauthor{Piatti:2016}\,\,\citeyear{Piatti:2016} and references therein). In this context, highly dynamically evolved star clusters, which suffered from severe mass loss, can have relatively low $c$ values.

It is enlightening to note that part of our sample is composed by clusters (namely, ESO\,518-3, Ruprecht\,121, ESO\,134-12 and NGC\,6573) with compatible $R_G$ and similar young ages (Tables \ref{tab_params_OCs} and \ref{clusters_total_masses}, respectively), considering uncertainties. Despite this, we distinghish two groups: one near to disruption (ESO\,518-3 and ESO\,134-12) and those in advanced dynamical stage (Ruprecht\,121 and NGC\,6573). We speculate with the possibility that these differences might be traced back to their progenitor OCs, although we should not rule out that the Galactic gravitational field could play a differential role. The older studied OCs (ESO\,260-7 and ESO\,065-7) are also near to disruption at different $R_G$. Differences in the Galaxy tidal field could, at least partially, explain the difference in the $r_h$ values of these both clusters. Nevertheless, considering their ages and $R_G$, differences in the potential lead to a small variation ($\lesssim10\%$) in $r_h$ (figure 1 of \citeauthor{Miholics:2014}\,\,\citeyear{Miholics:2014}). This suggests that the internal interactions have relaxed the mass distribution internally in such a way that these clusters strongly feel the tidal field of the Galaxy (in fact, ESO\,260-7 and ESO\,065-7 present the highest $r_h/R_J$ ratios among our sample).

\section{Summary and concluding remarks}
\label{conclusions}

In this paper we analysed the evolutionary stages of six sparse OCs, namely: ESO\,518-3, Ruprecht\,121, ESO\,134-12, NGC\,6573, ESO\,260-7 and ESO\,065-7. We employed photometric data in the Washington $C$ and $T_{1}$ bands from publicly available CTIO images. Markov chain Monte-Carlo simulations were performed in order to determine the central coordinates and the structural parameters by fitting K62 model to the stellar profiles. In order to check the method efficiency, it was previously employed on data of a synthetic star cluster.

We employed a decontamination technique that performs statistical comparisons between OC and control field CMDs ($T_1\times(C-T_1)$, in our case), attributing membership likelihoods ($L$) to each star. The more probable members provide useful constraints for isochrone fitting, from which we derived the OCs' fundamental parameters age, distance modulus and reddening. Solar metallicity isochrones provided adequate fits for our OCs. We found that four of our  OCs are relatively young and within a relatively narrow age range (8.2\,$\leq\textrm{log}(t\,\textrm{yr}^{-1})\leq$\,8.3); the  older OCs present reasonably compatible ages (log ($t$\,yr$^{-1}$) in the range 9.2$-$9.4). Except for ESO\,260-7 ($R_G$ = 8.9\,kpc), all the studied OCs are located at compatible Galactocentric distances ($R_{G}$ in the range $\sim6.0-6.9\,$kpc), considering uncertainties.

Individual masses for the member stars were estimated from their positions relatively to the fitted isochrone, 
from which the OC mass functions and present-day observed masses ($M_{\textrm{phot}}$) were obtained. The $M_{\textrm{phot}}$ of the six OCs span a wide range (log\,$M$ = 1.5$-$2.7) and we also estimated upper mass limits assuming a Universal IMF described by a \cite{Kroupa:2001} law, which resulted in the mass range log($M_{\textrm{Kroupa}}$) = 2.1$-$3.2. We found evidence of low-mass stellar depletion, which is expected for dynamically evolved stellar systems.

From the analysis of the derived parameters we found that:

$\bullet$ Two groups of OCs can be identified: a group of OCs towards final
disruption ($r_t/R_J$ $>$ 1) and another in an advanced dynamical stage ($r_t/R_J$ $<$ 1).

$\bullet$ The positions of the studied OCs in the mass versus log (age/$t_{rh}$) and $c$ versus log (age/$t_{rh}$) planes are compatible with the hypothesis of dynamically evolved stellar systems; the derived $r_h/r_t$ ratios are consistent with the expected value for tidally filled clusters. 
 
 $\bullet$ The dynamical evolution of the compact cluster NGC\,6573 seems to mainly driven by internal interactions, while the other OCs are more strongly tidally influenced, as can be seen in the $r_h/R_J$ versus $R_G$ plane. 

$\bullet$ Their distinct dynamical stages could be due to different initial formation conditions, although the Milky Way tidal field could have played also a role.

\section{Acknowledgments}

We thank the anonymous referee for very useful suggestions and discussions. We also thank the financial support of the brazilian agency CNPq (grant No. 304654/2017-5). This work has made use of data from the European Space Agency (ESA) mission Gaia (http://www.cosmos.esa.int/gaia), processed by the Gaia Data Processing and Analysis Consortium (DPAC, http://www.cosmos.esa.int/web/gaia/dpac/consortium). Funding for the DPAC has been provided by national institutions, in particular the institutions participating in the Gaia Multilateral Agreement.


{\footnotesize
\bibliographystyle{mn2e}
\bibliography{referencias}}


\appendix
\section[]{Supplementary material}

\begin{figure*}

\parbox[c]{1.0\textwidth}
{

   \includegraphics[width=0.5\textwidth]{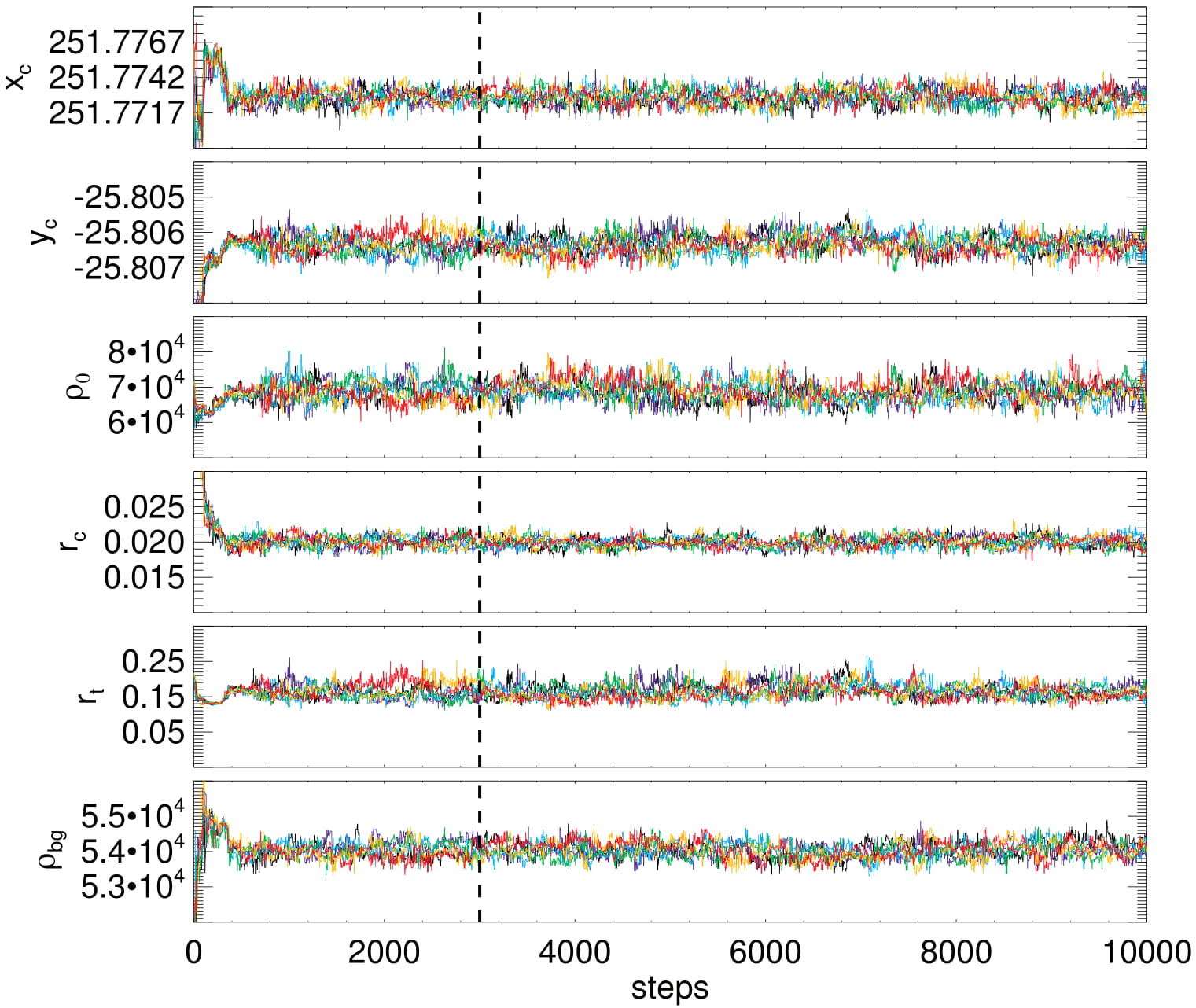}
   \includegraphics[width=0.5\linewidth]{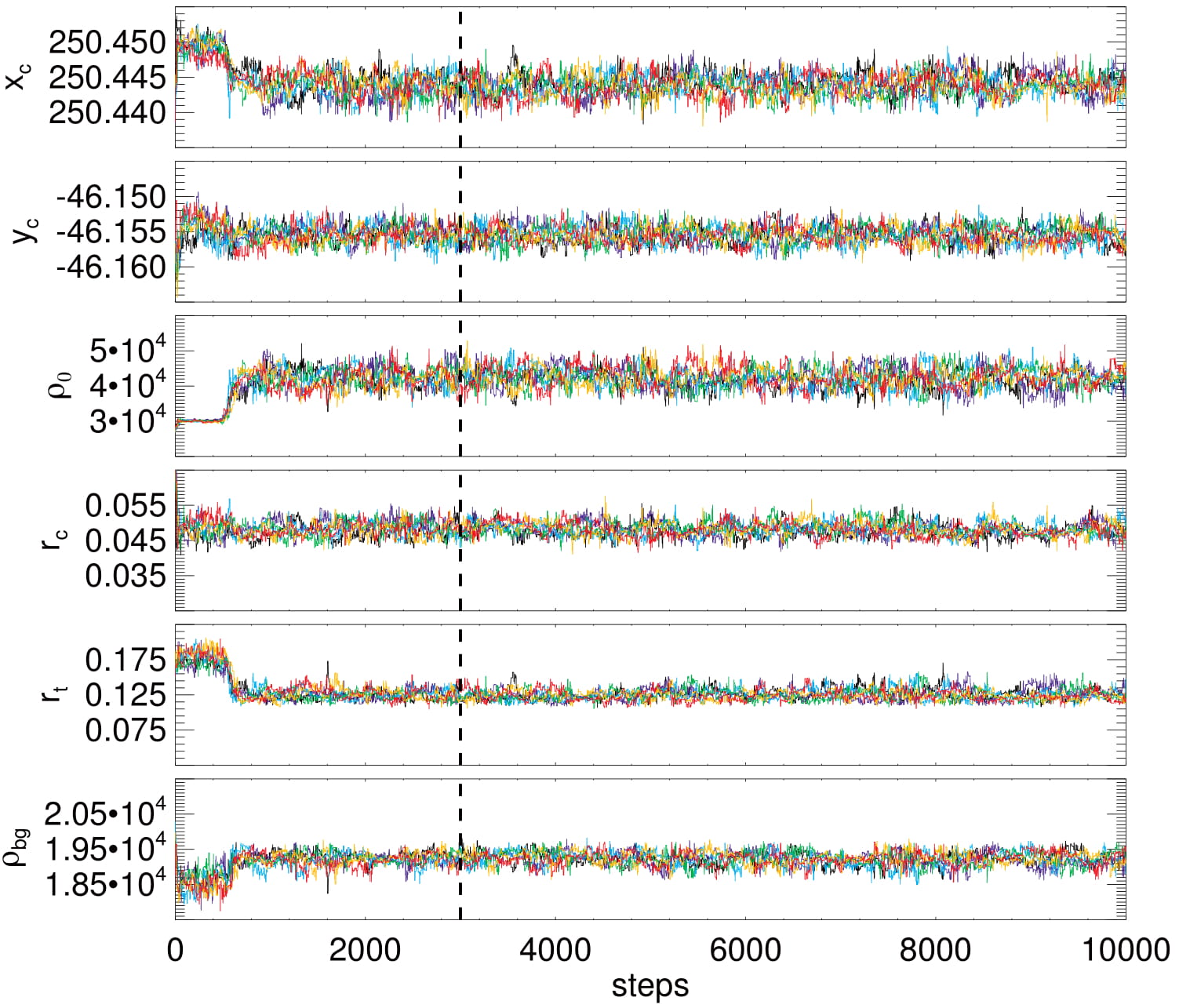}
   \includegraphics[width=0.5\linewidth]{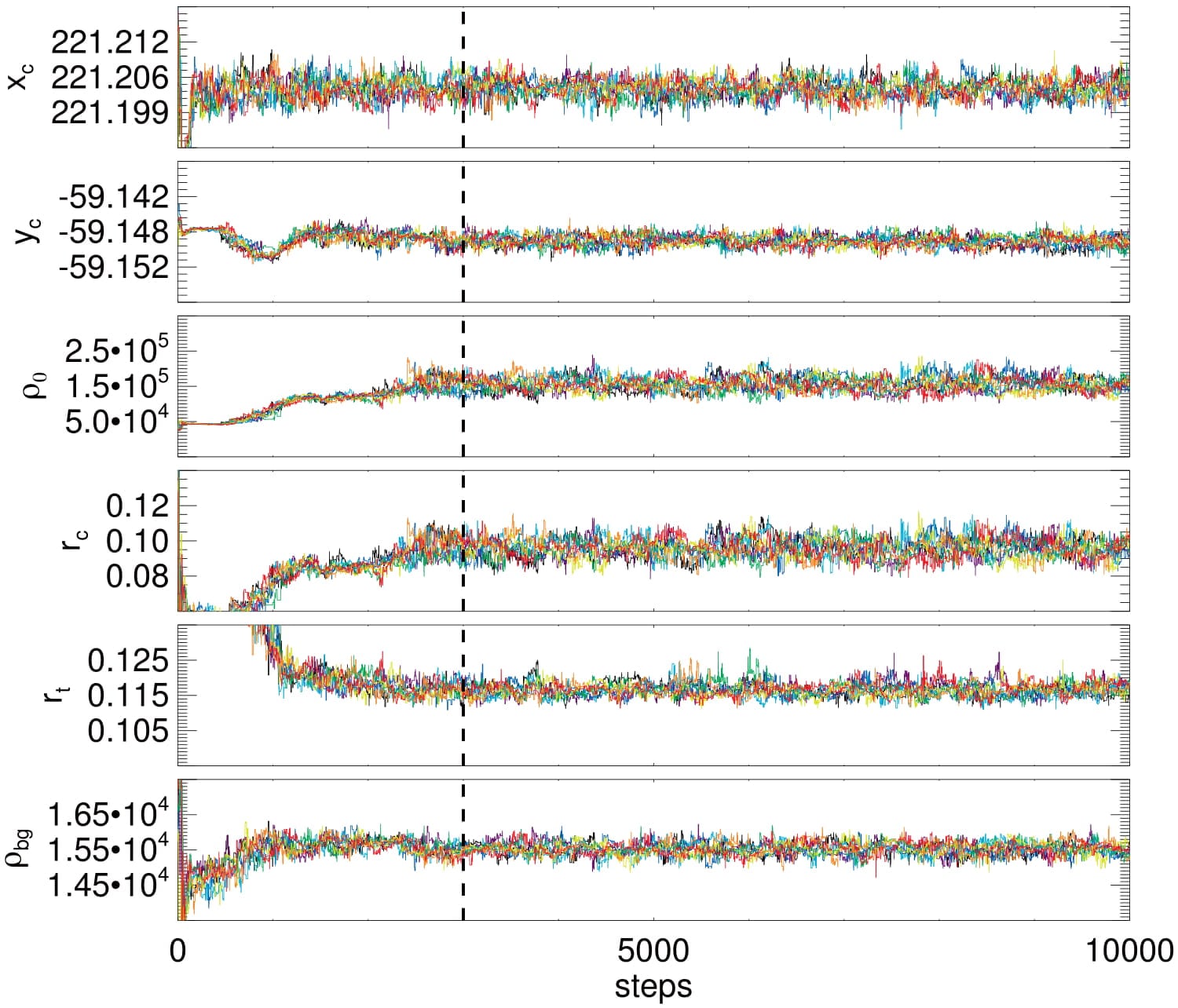}
   \includegraphics[width=0.5\linewidth]{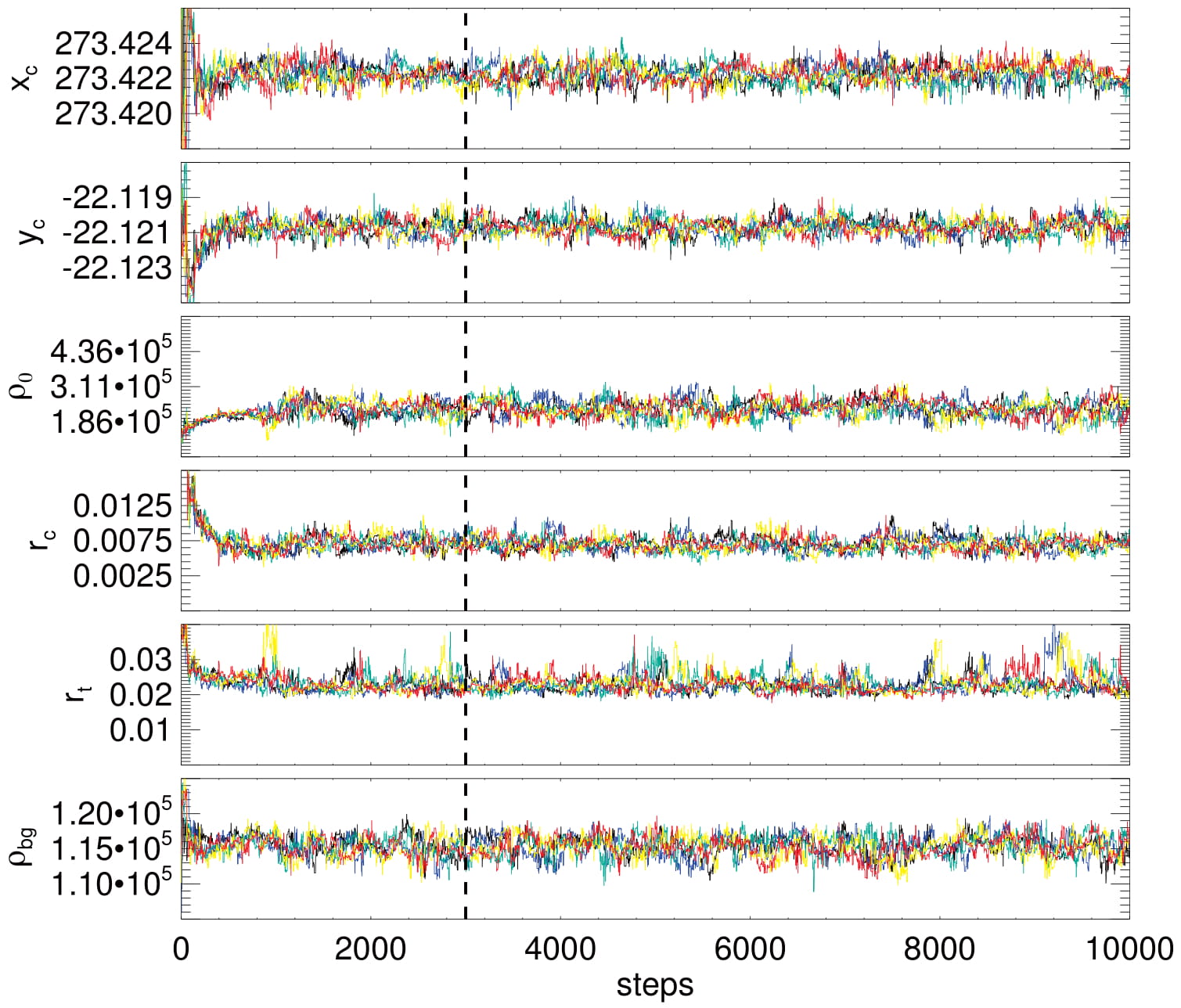}
   \includegraphics[width=0.5\linewidth]{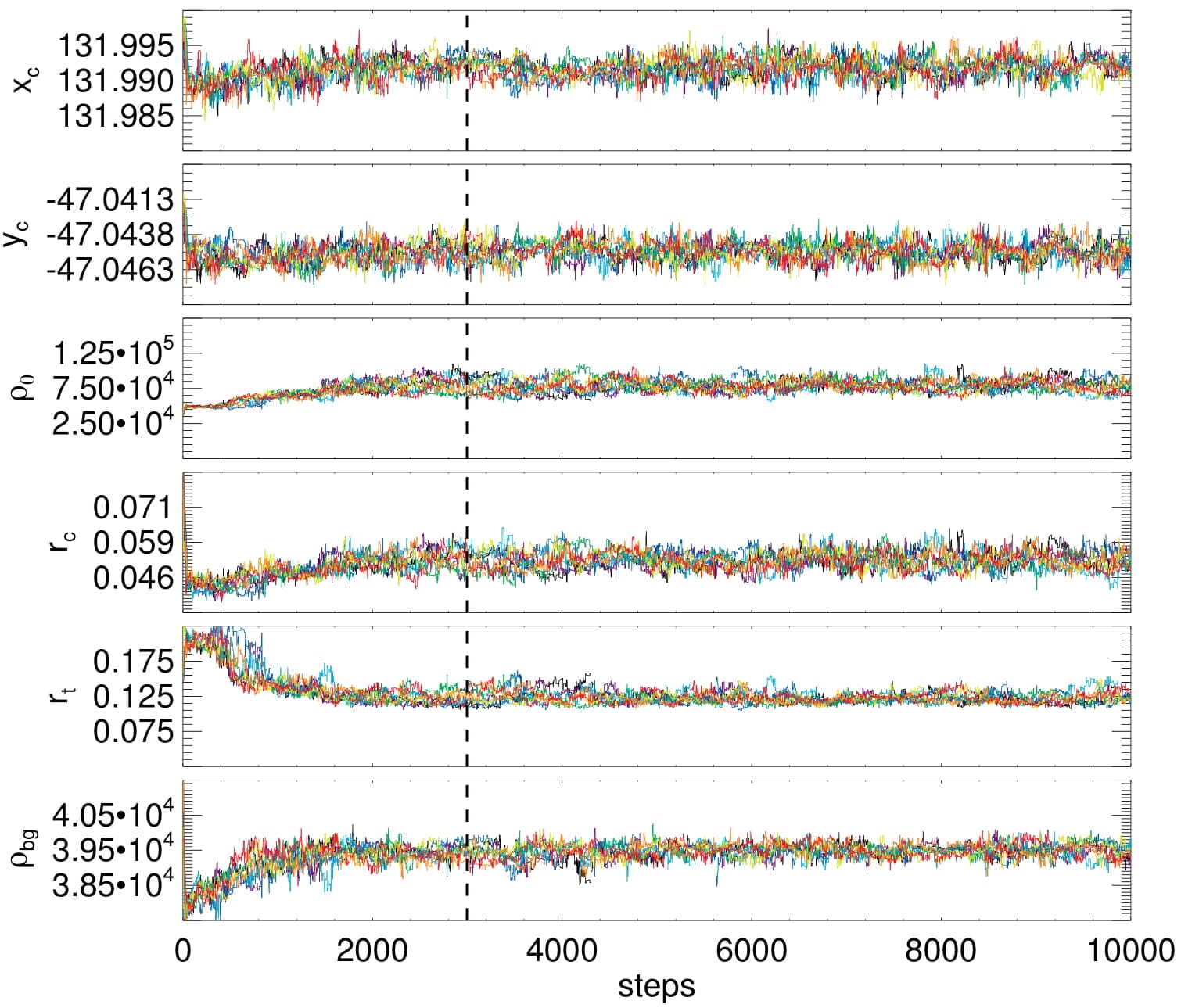}
   \includegraphics[width=0.5\linewidth]{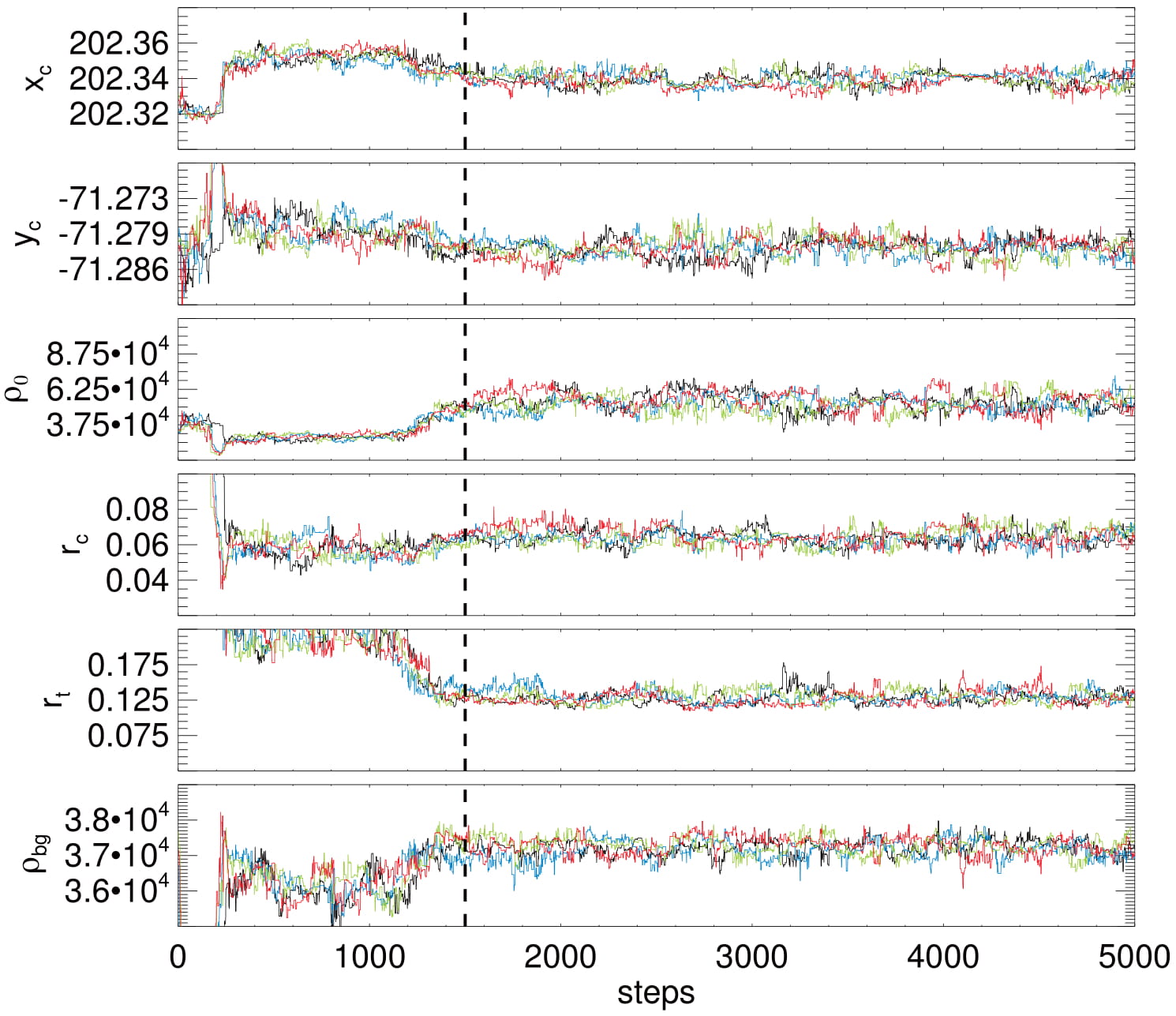}
}

\caption{Same of Figure \ref{fig:mctime}, but for the OCs ESO\,518-3 (top left panel), Ruprecht\,121 (top right panel), ESO\,134-12 (middle left panel), NGC\,6573 (middle right panel), ESO\,260-7 (bottom left panel) and ESO\,065-7 (bottom right panel).}
\label{fig:mctime_all_clusters}
\end{figure*}

\begin{figure*}

\parbox[c]{1.0\textwidth}
{

\includegraphics[width=0.5\linewidth]{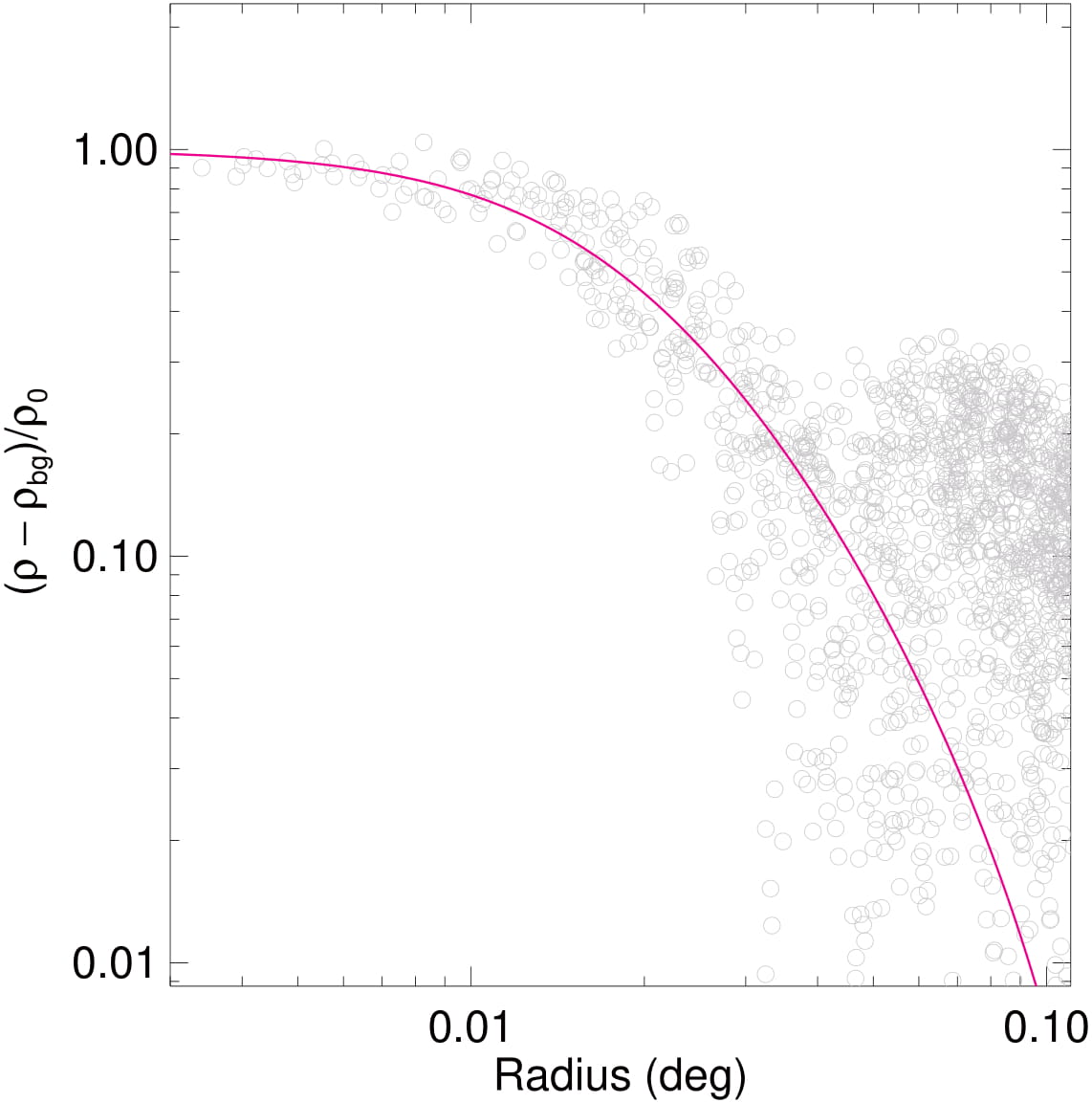}
\includegraphics[width=0.5\linewidth]{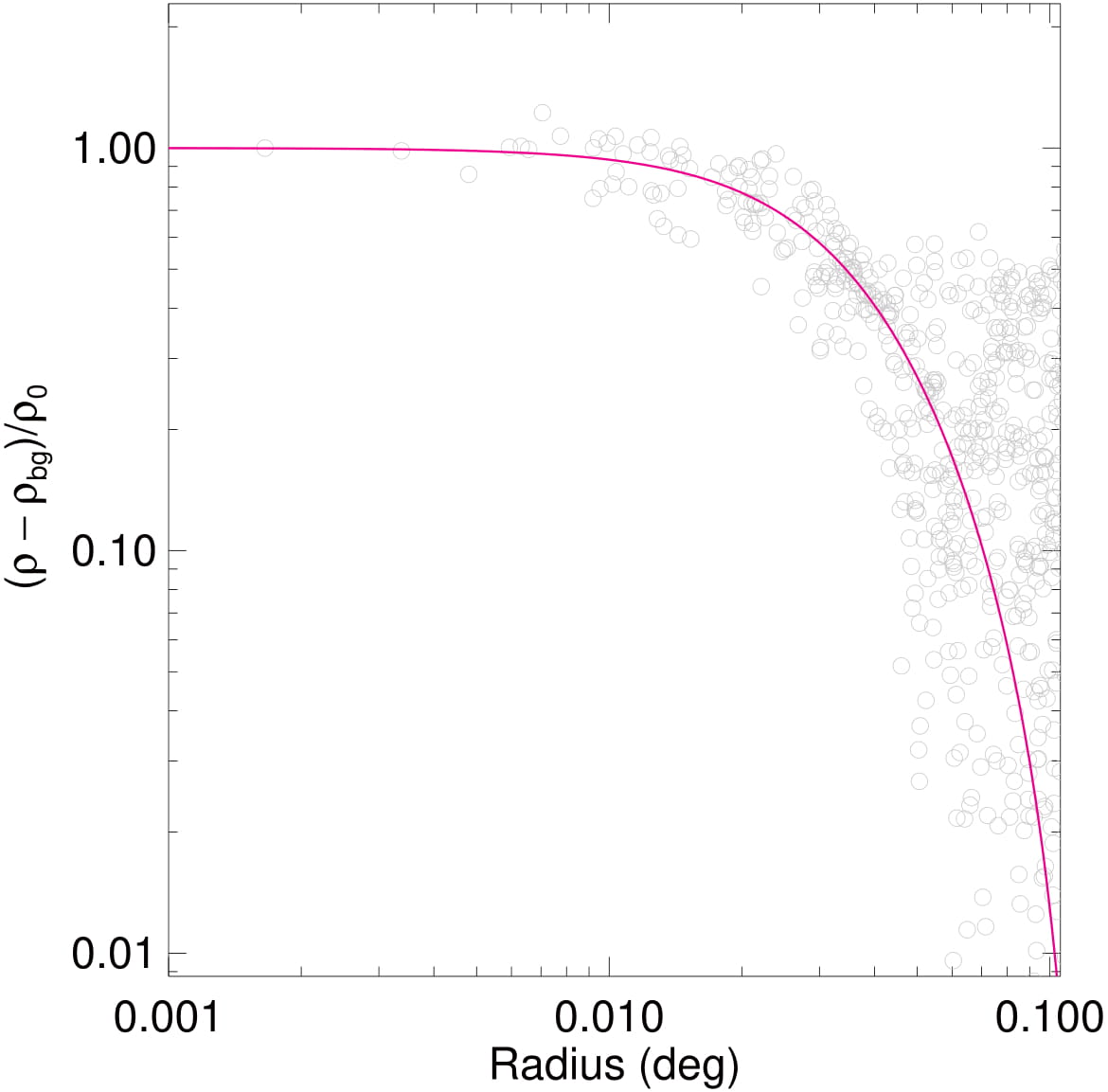}
\includegraphics[width=0.5\linewidth]{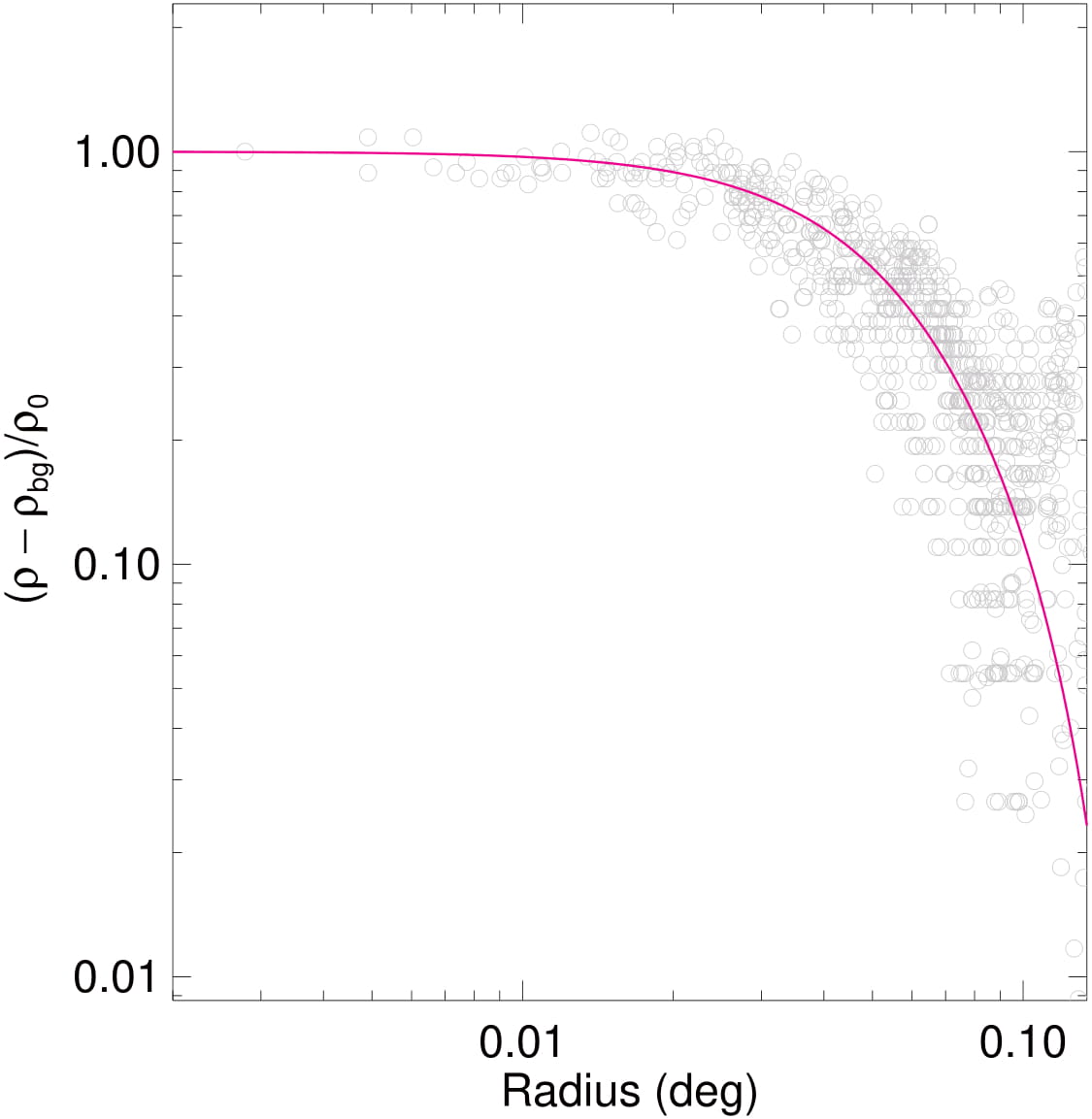}
\includegraphics[width=0.5\linewidth]{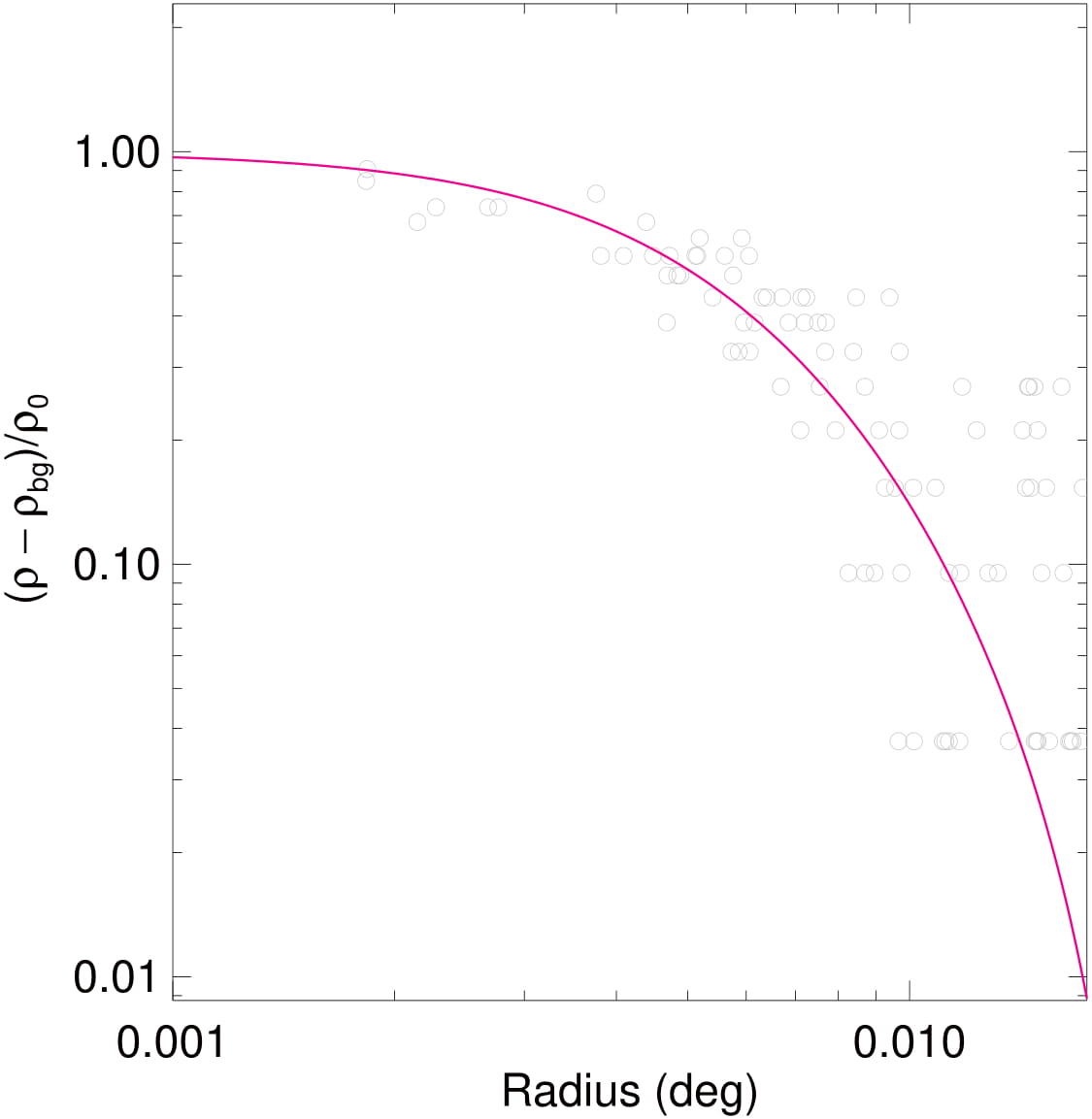}

}

\caption{\textbf{Normalised background subtracted} radial density profiles of ESO\,518-3 (top left panel), Ruprecht\,121 (top right panel), ESO\,134-12 (bottom left panel) and NGC\,6573 (bottom right panel), showing a K62 profile model function (solid line) fitted to the data.}
\label{fig:mcfit_all_clusters_part1}
\end{figure*}

\begin{figure*}

\parbox[c]{1.0\textwidth}
{

\includegraphics[width=0.5\linewidth]{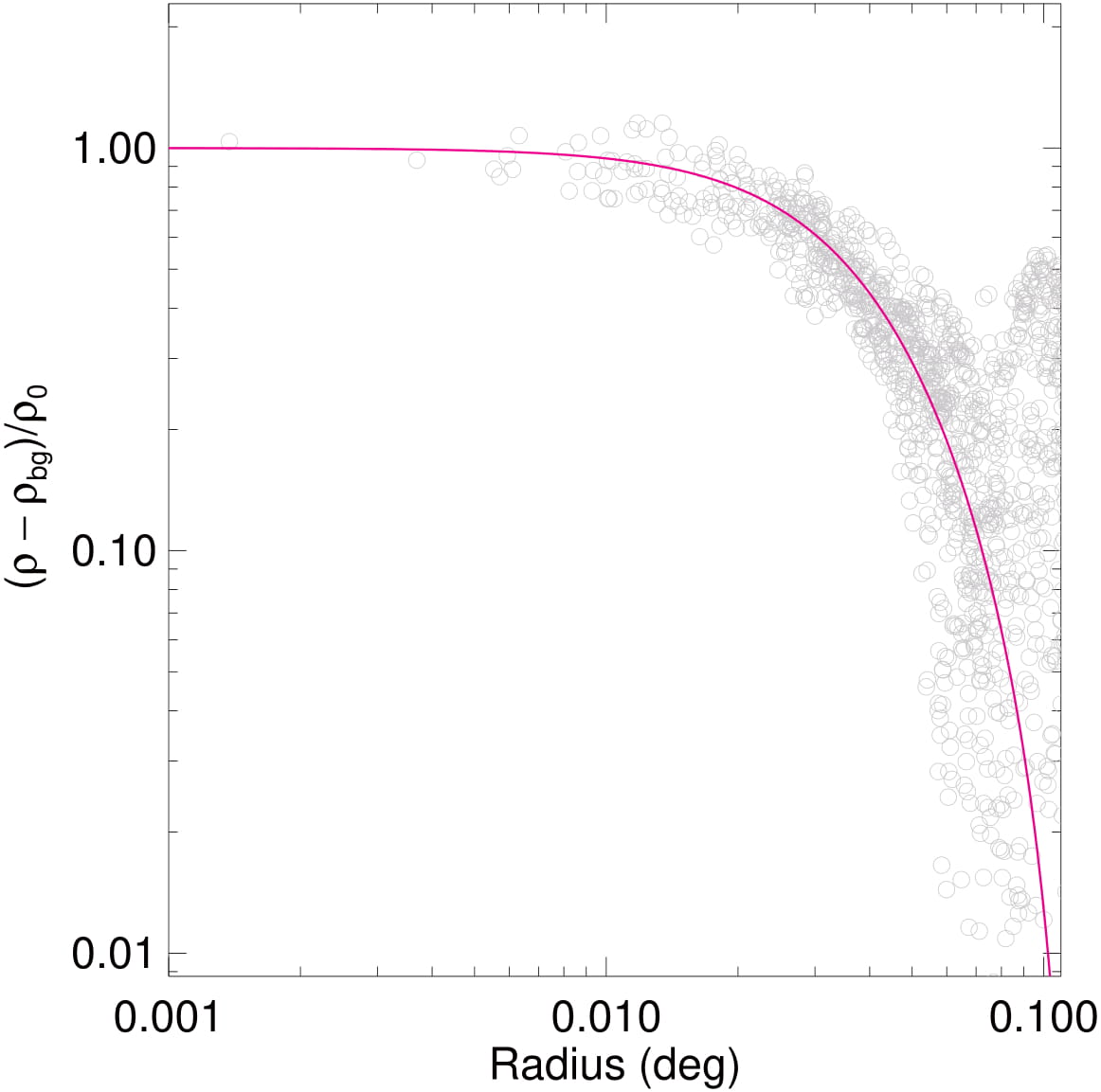}
\includegraphics[width=0.5\linewidth]{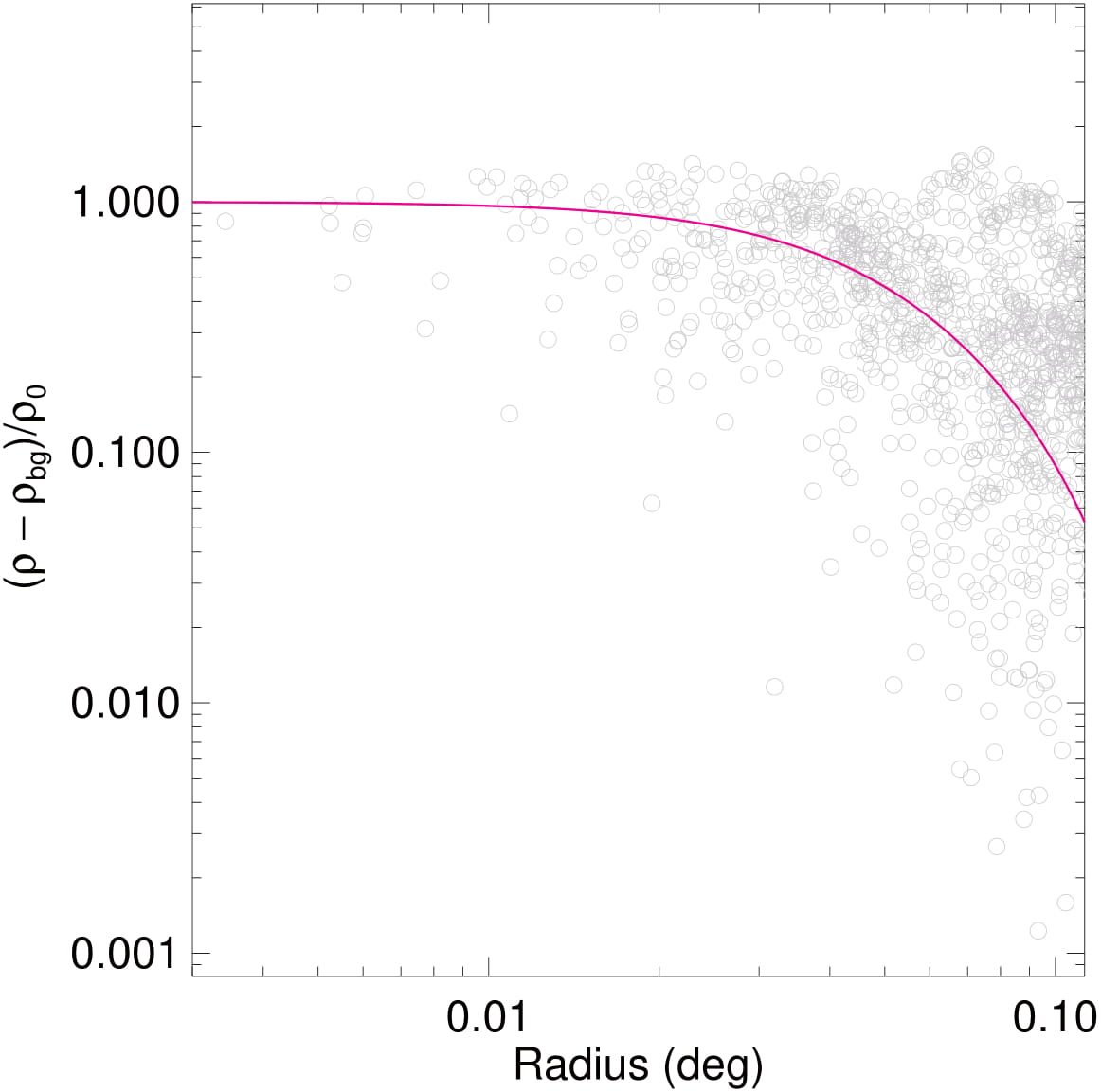}

}

\caption{Same of Figure \ref{fig:mcfit_all_clusters_part1}, but for the OCs ESO\,260-7 (left panel) and ESO\,065-7 (right panel).}
\label{fig:mcfit_all_clusters_part2}
\end{figure*}





\begin{figure*}
\begin{center}
\begin{minipage}{185mm}
\parbox[c]{1.0\textwidth}
  {
   
    \includegraphics[width=0.5\textwidth]{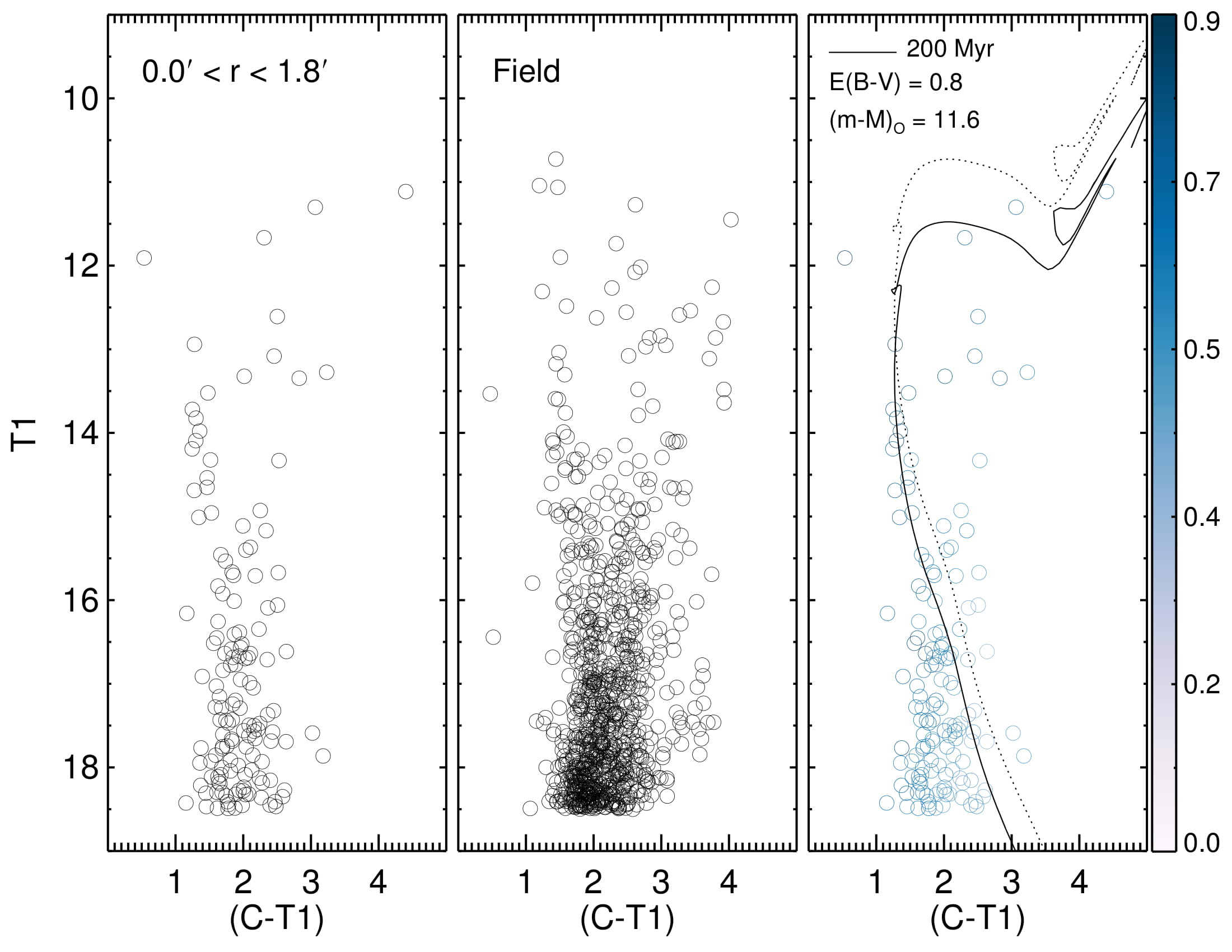}
    \includegraphics[width=0.5\textwidth]{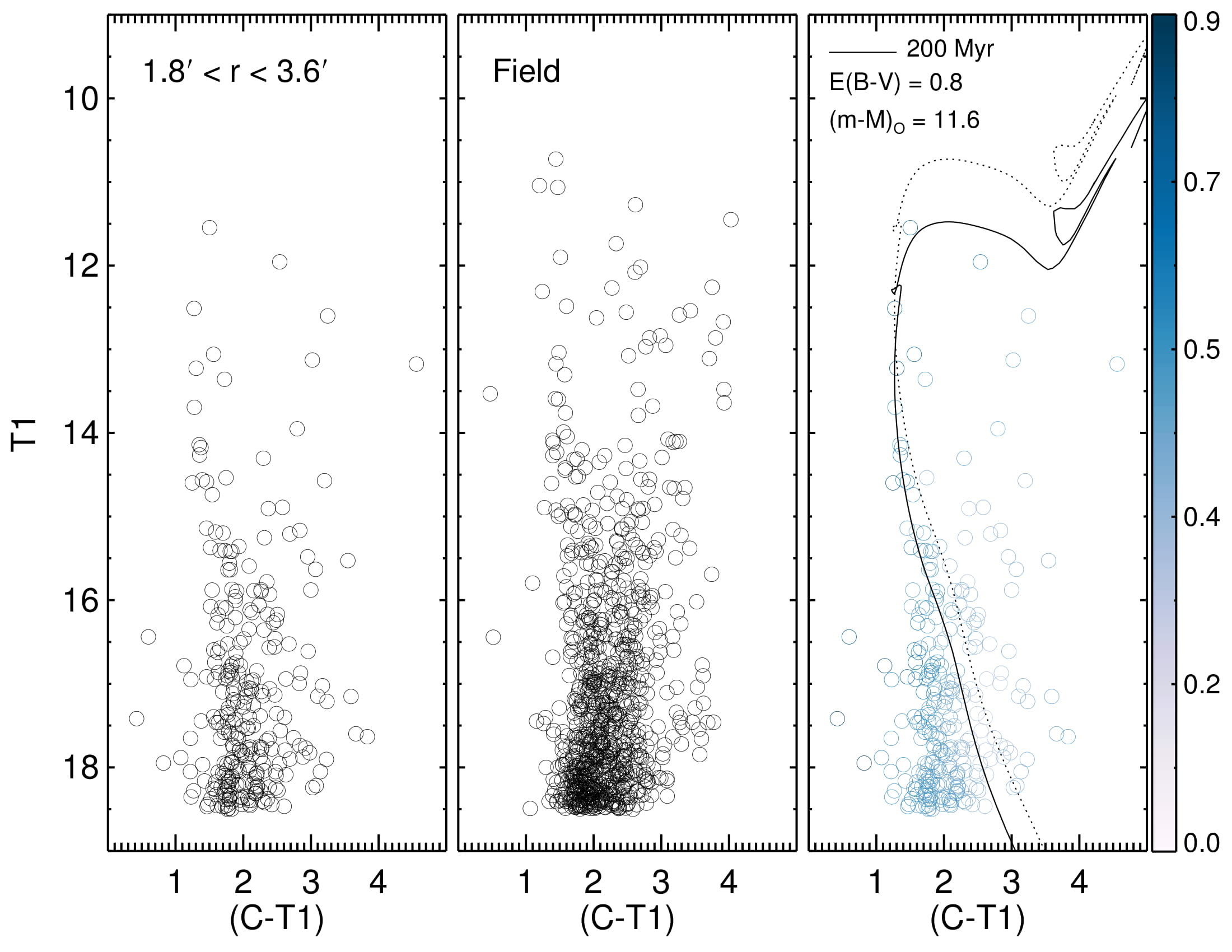}
    
    \begin{center}
      \includegraphics[width=0.5\textwidth]{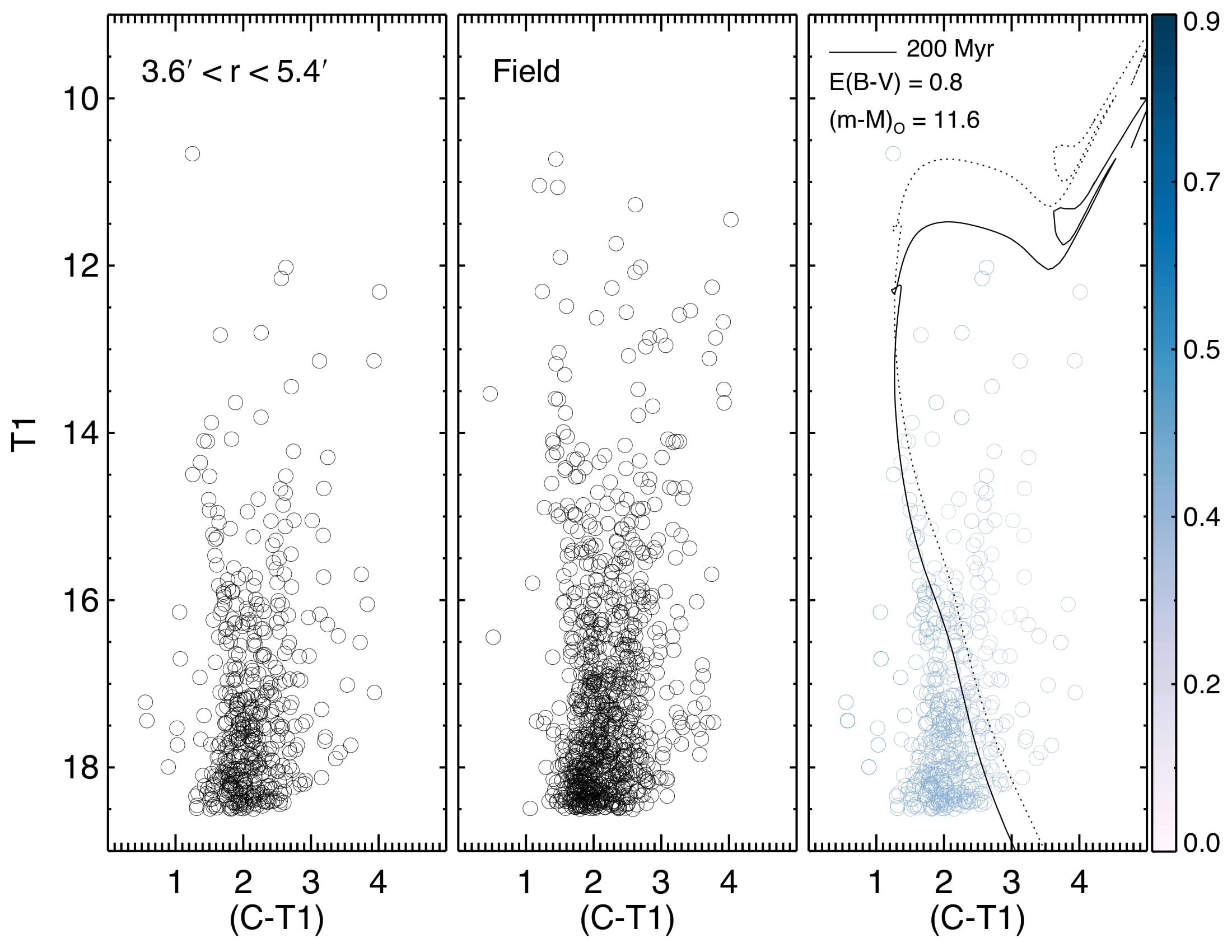}
    \end{center}
  }
  
\caption{  Same of Figure \ref{CMDs_decontam_cascas_exemplo}, but for ESO\,518-3.  }
\label{CMDs_decontam_cascas_ESO518-3}
\end{minipage}
\end{center}
\end{figure*}

\begin{figure*}
\begin{center}
\begin{minipage}{185mm}
\parbox[c]{1.0\textwidth}
  {
   
    \includegraphics[width=0.5\textwidth]{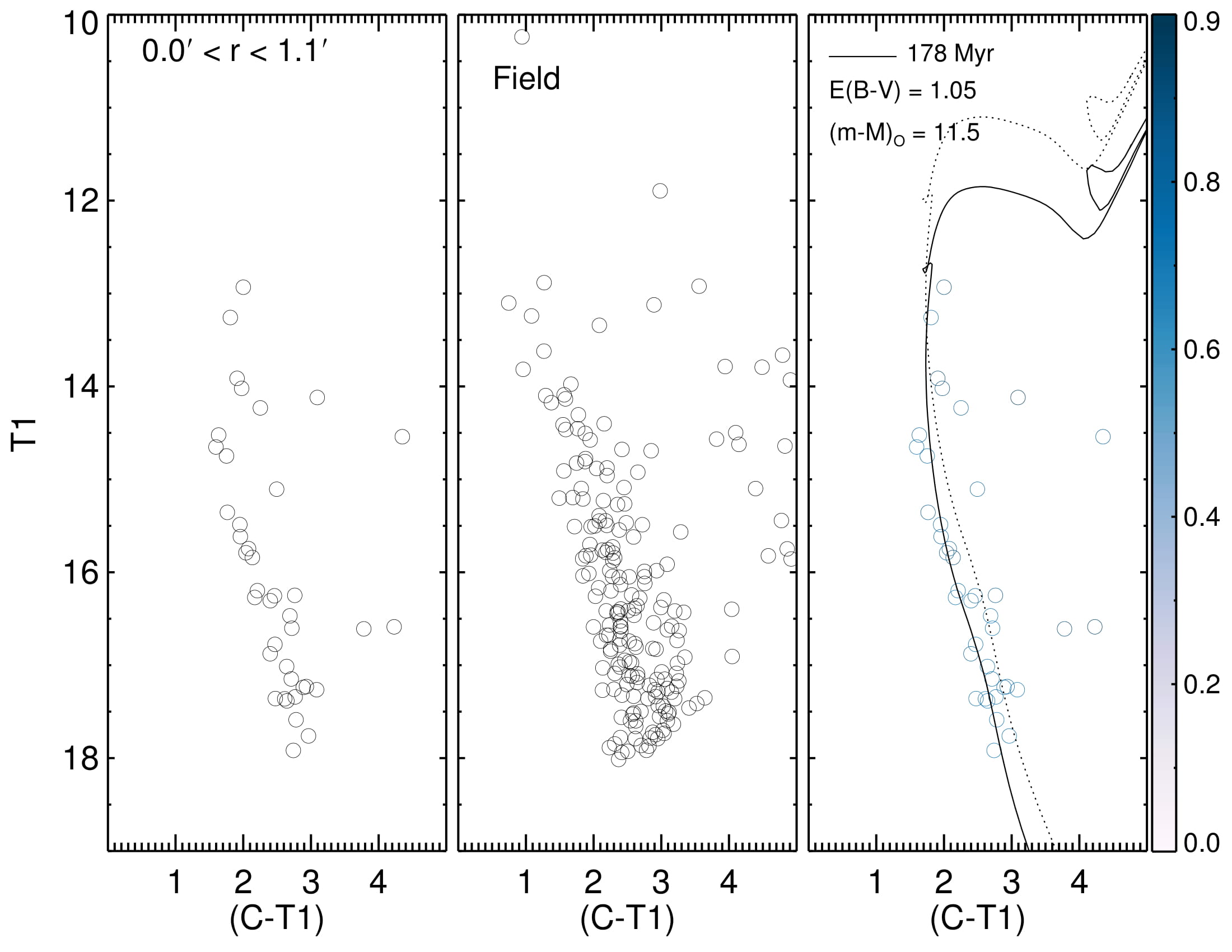}
    \includegraphics[width=0.5\textwidth]{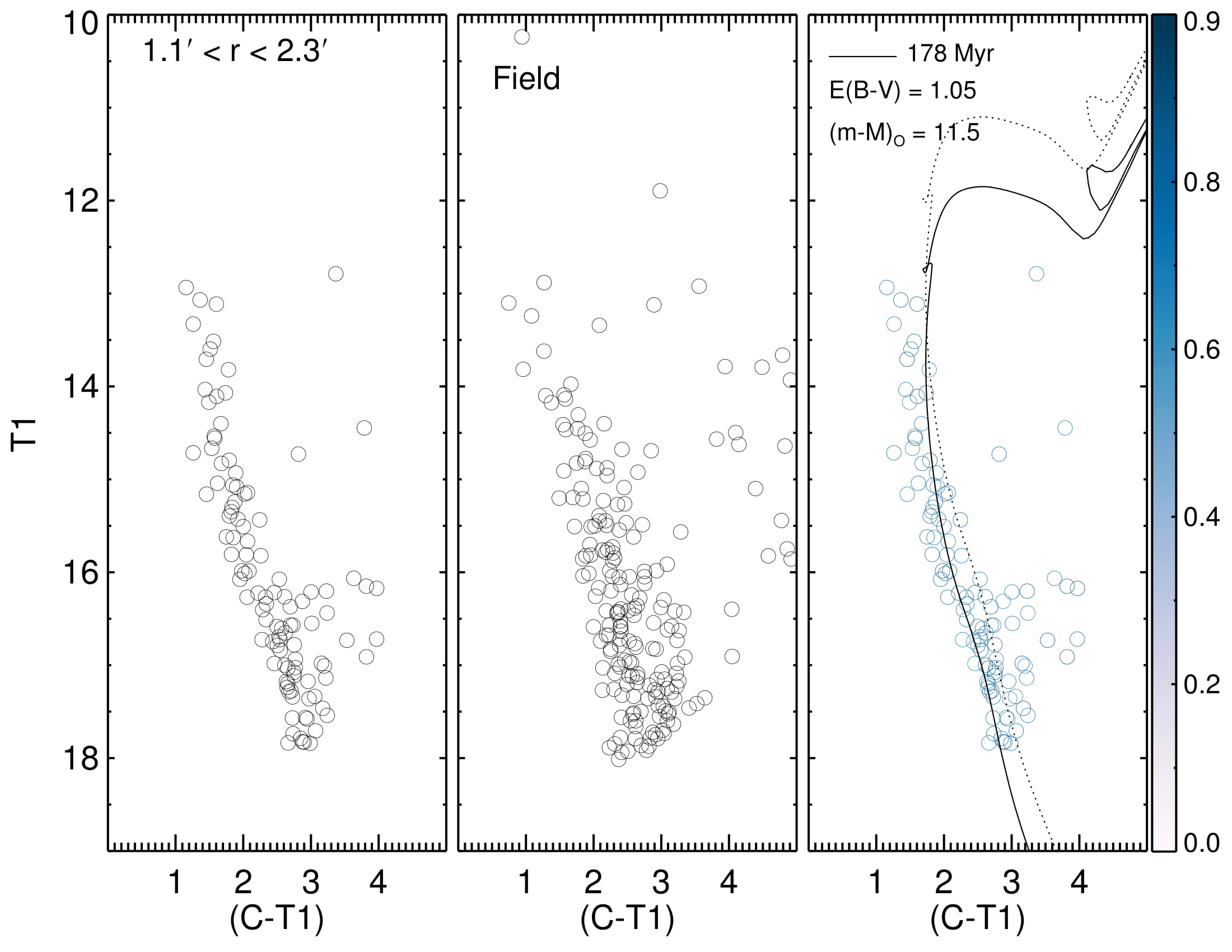}
    \includegraphics[width=0.5\textwidth]{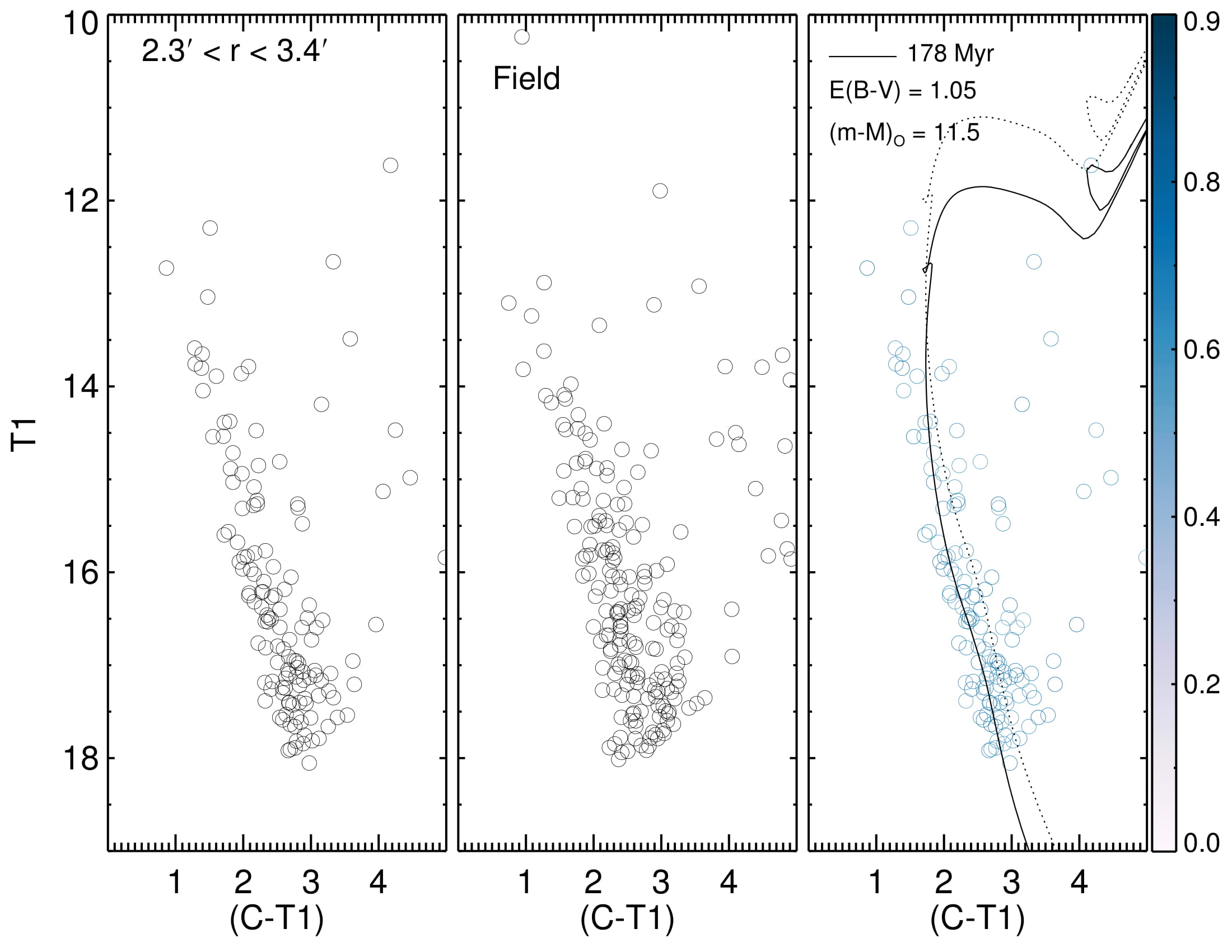}
    \includegraphics[width=0.5\textwidth]{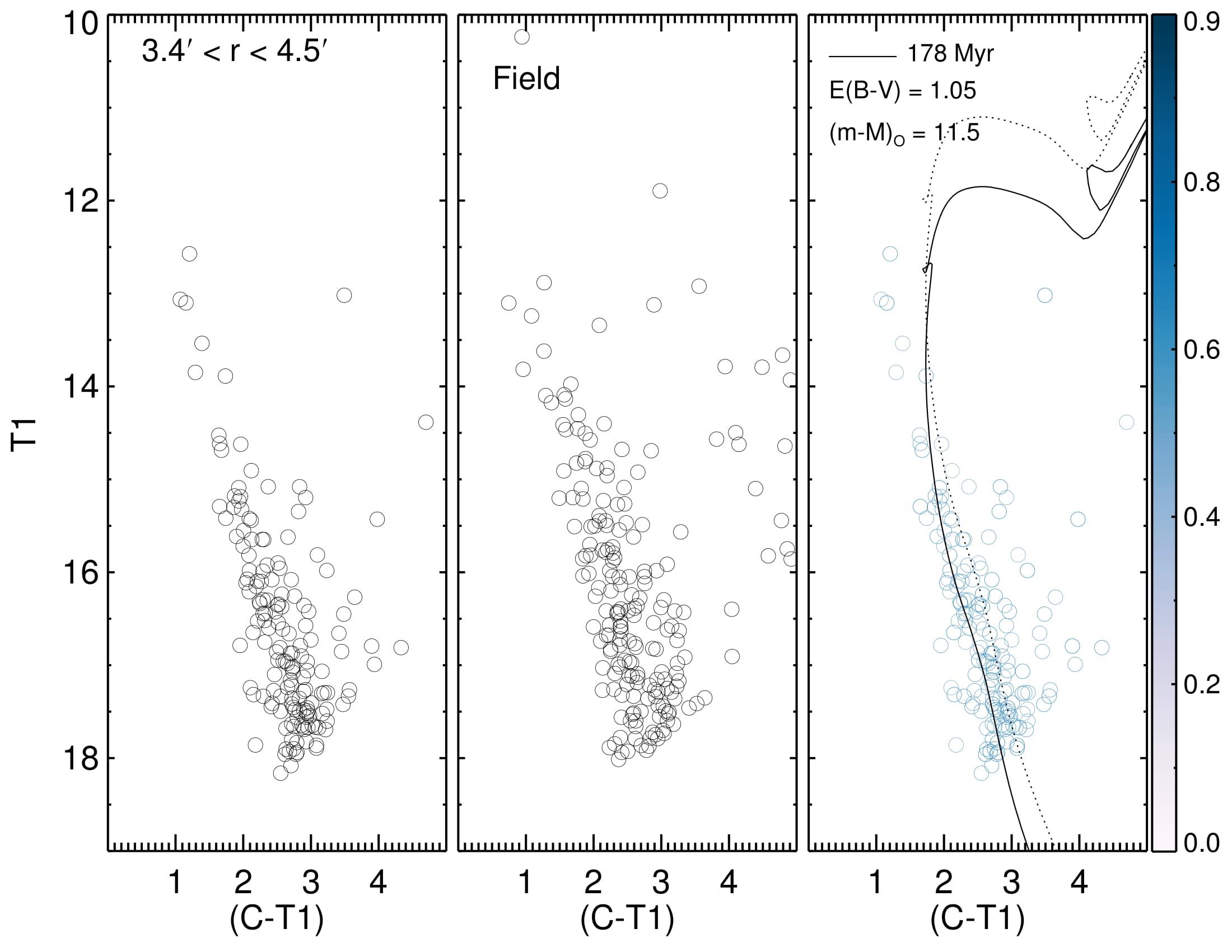}
    \includegraphics[width=0.5\textwidth]{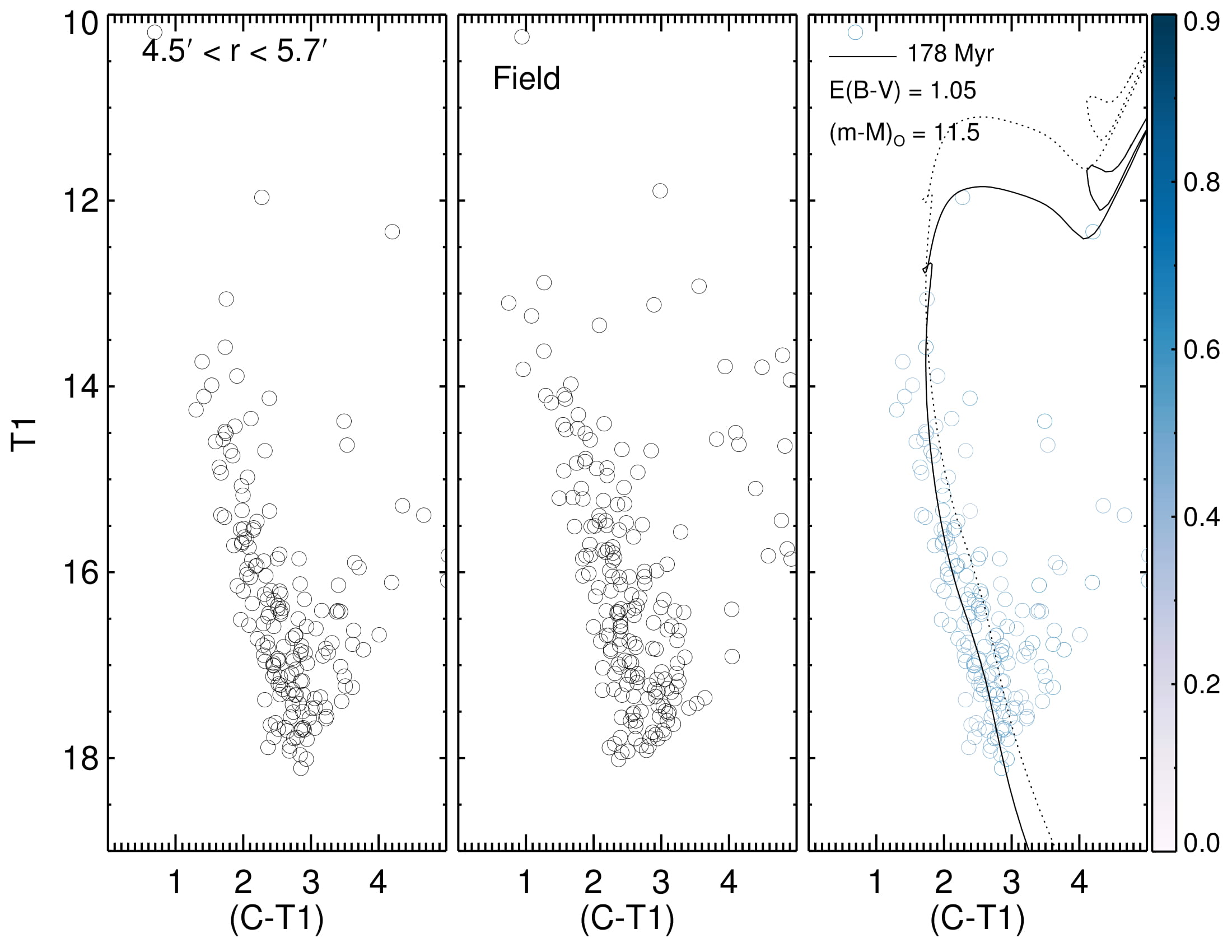}
    \includegraphics[width=0.5\textwidth]{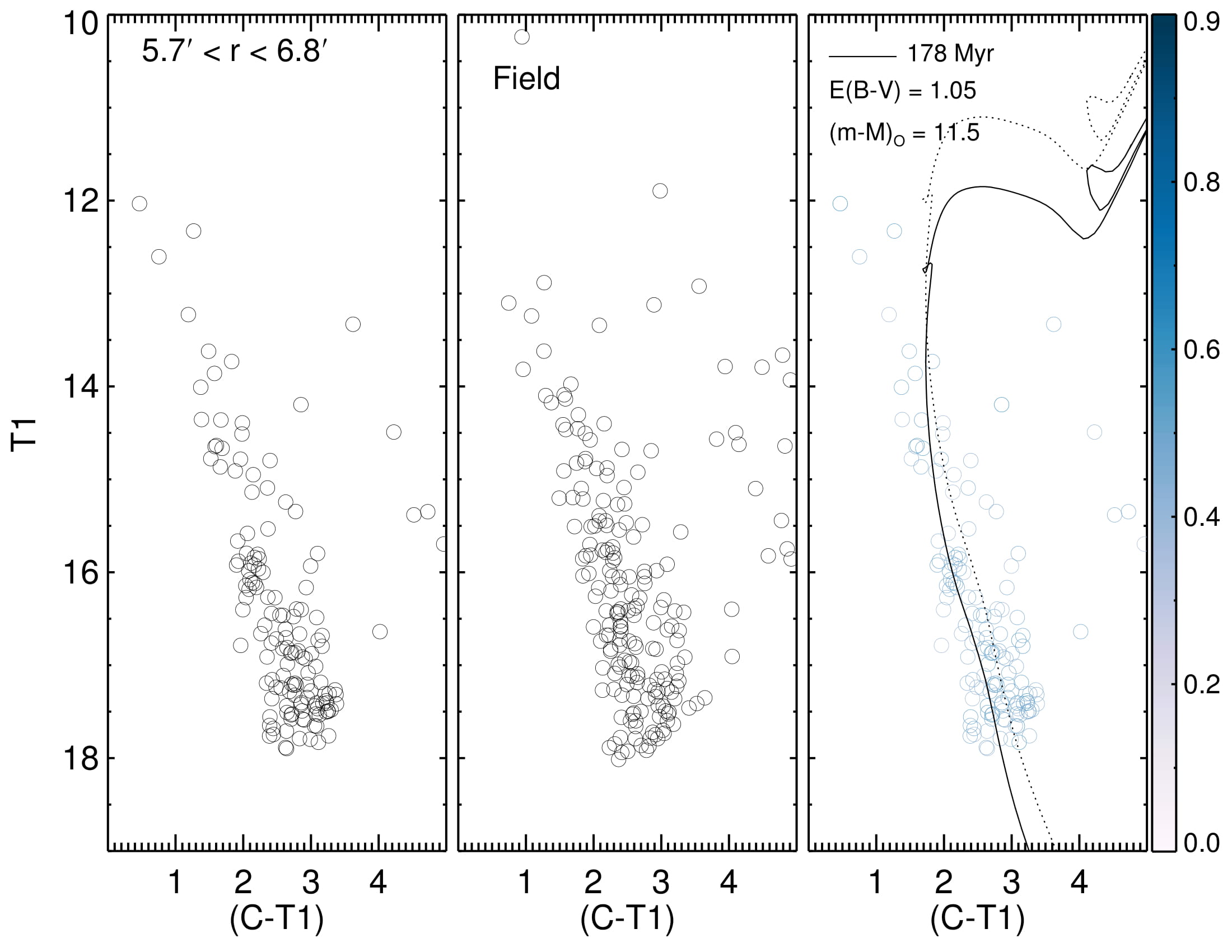}
  }
  
\caption{  Same of Figure \ref{CMDs_decontam_cascas_exemplo}, but for ESO\,134-12.  }
\label{CMDs_decontam_cascas_ESO134-12}
\end{minipage}
\end{center}
\end{figure*}

\begin{figure*}
\begin{center}
\begin{minipage}{185mm}
\parbox[c]{1.0\textwidth}
  {
   
    \includegraphics[width=0.5\textwidth]{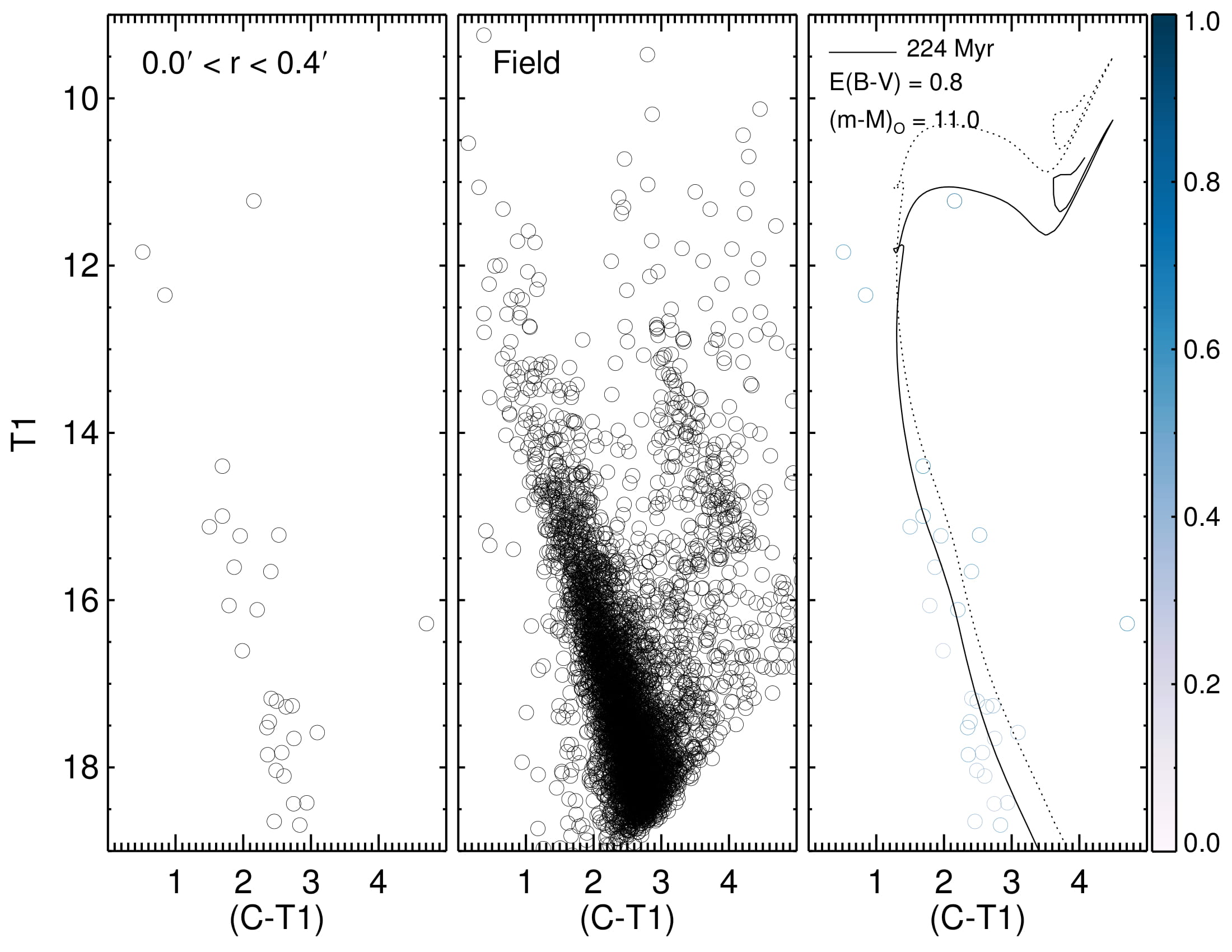}
    \includegraphics[width=0.5\textwidth]{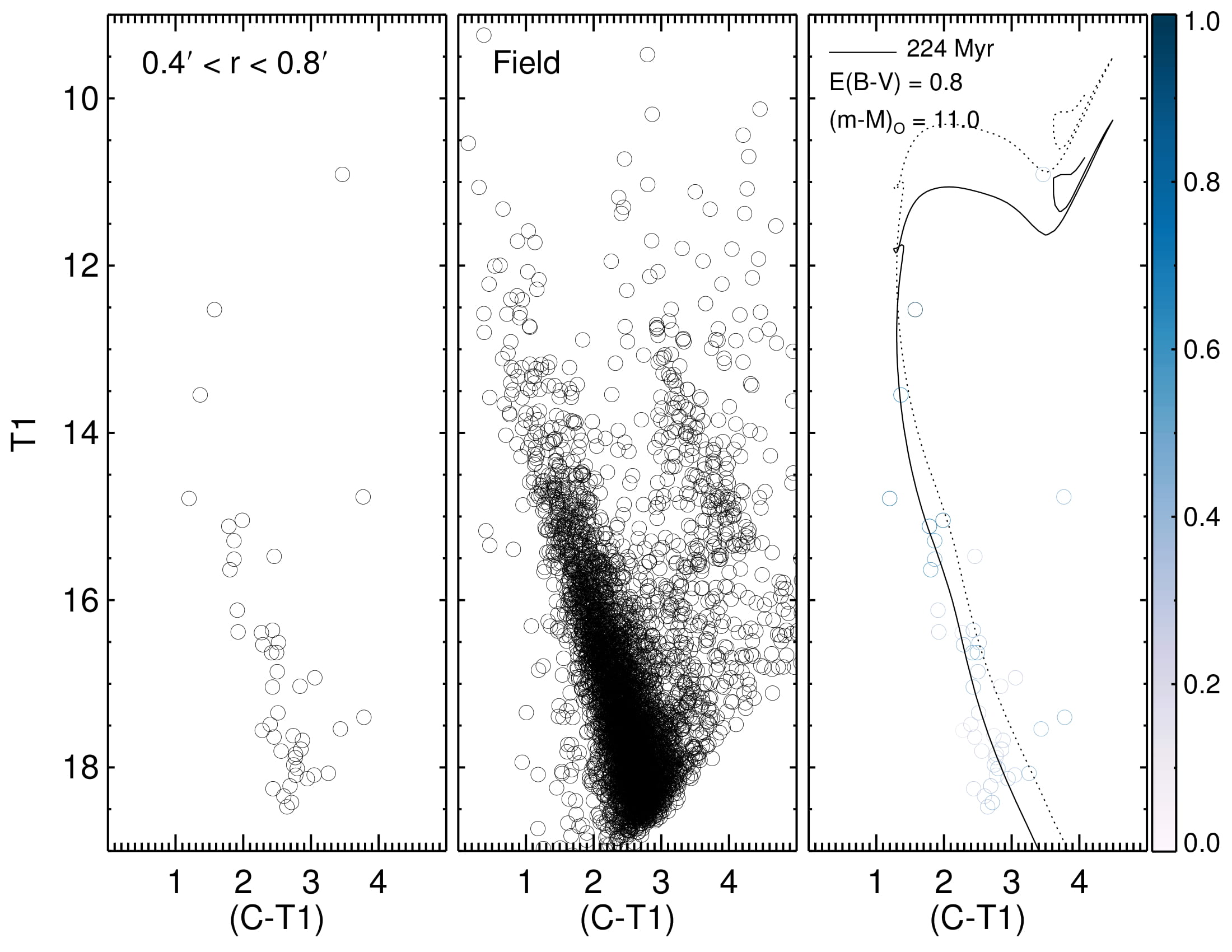}
    \includegraphics[width=0.5\textwidth]{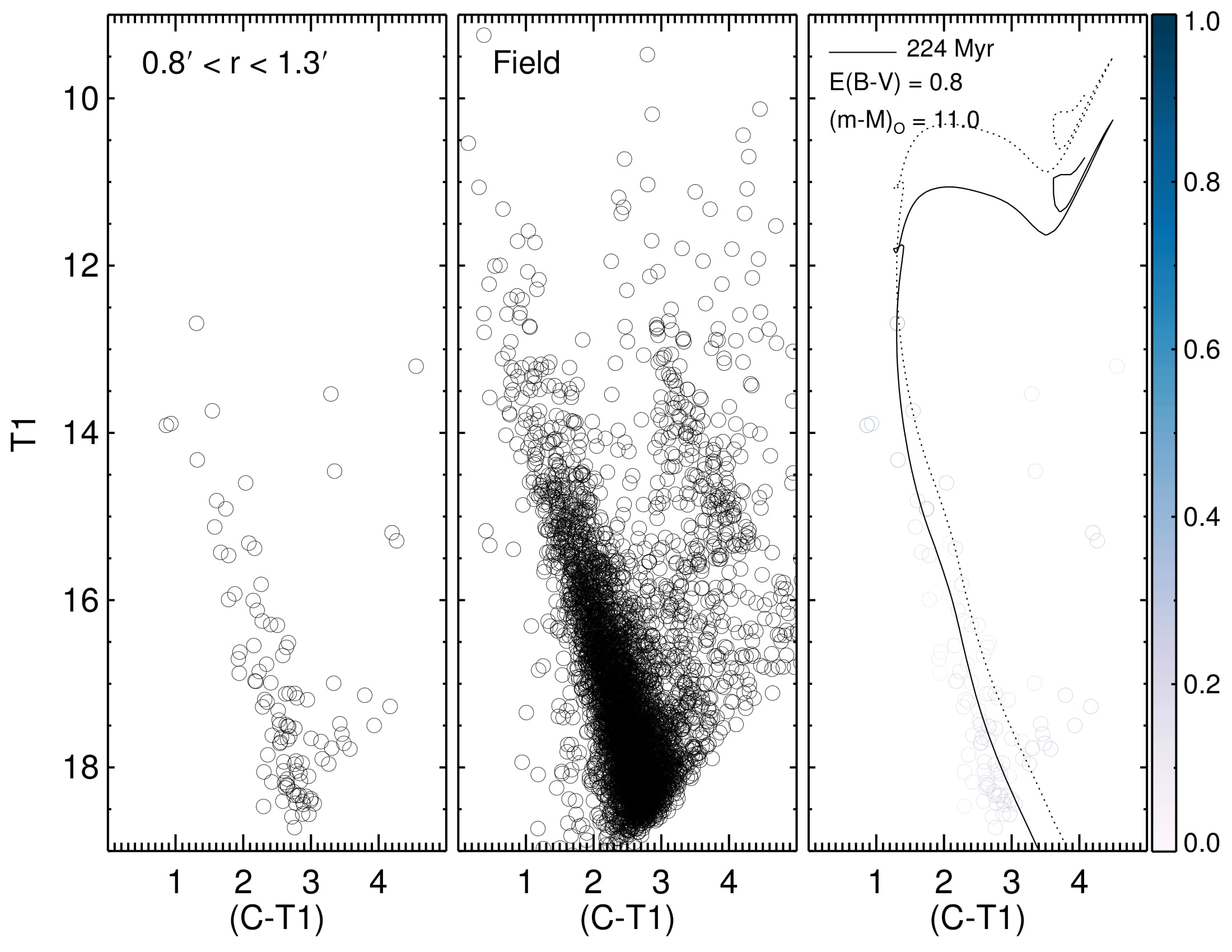}
    \includegraphics[width=0.5\textwidth]{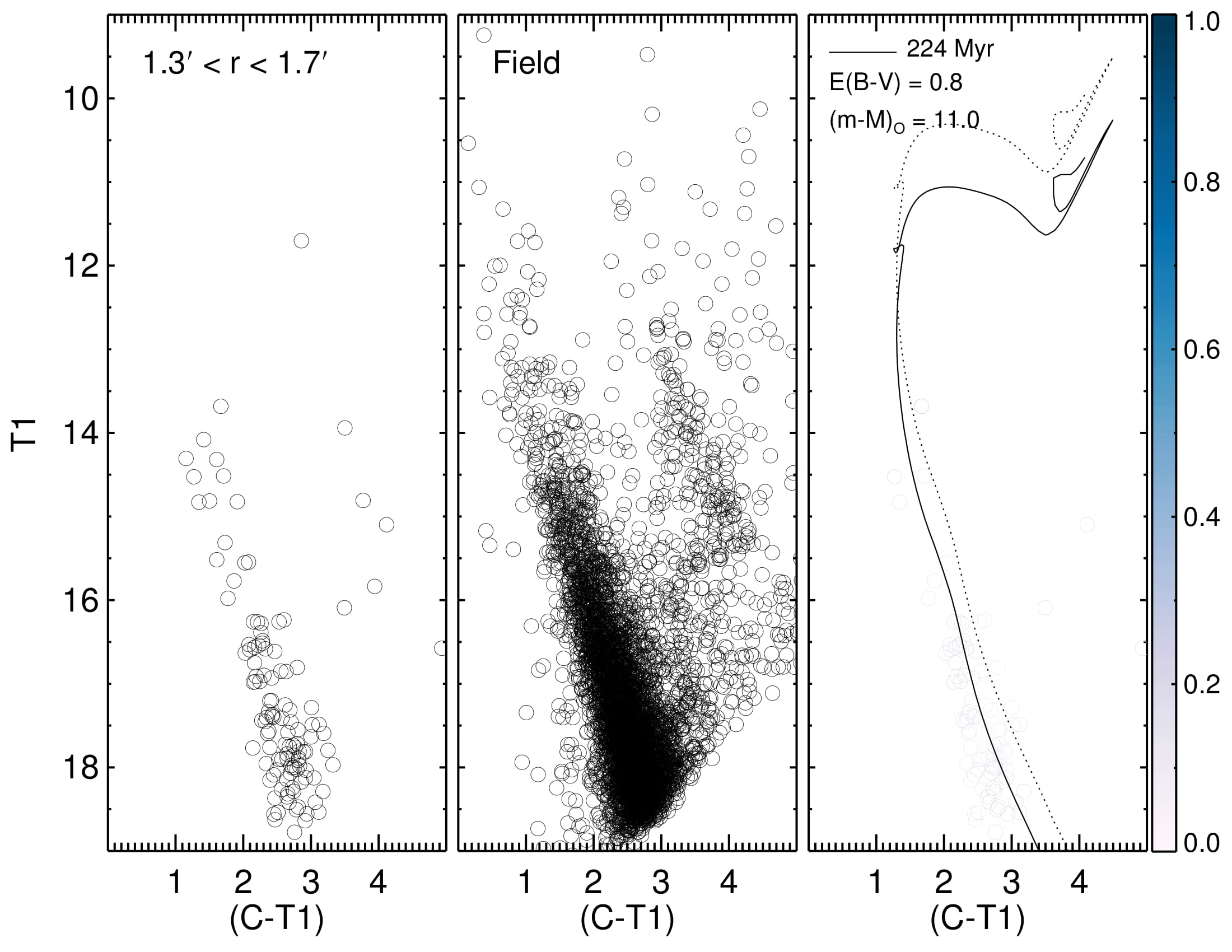}
  }
  
\caption{   Same of Figure \ref{CMDs_decontam_cascas_exemplo}, but for NGC\,6573.   }
\label{CMDs_decontam_cascas_NGC6573}
\end{minipage}
\end{center}
\end{figure*}

\begin{figure*}
\begin{center}
\begin{minipage}{185mm}
\parbox[c]{1.0\textwidth}
  {
   
    \includegraphics[width=0.5\textwidth]{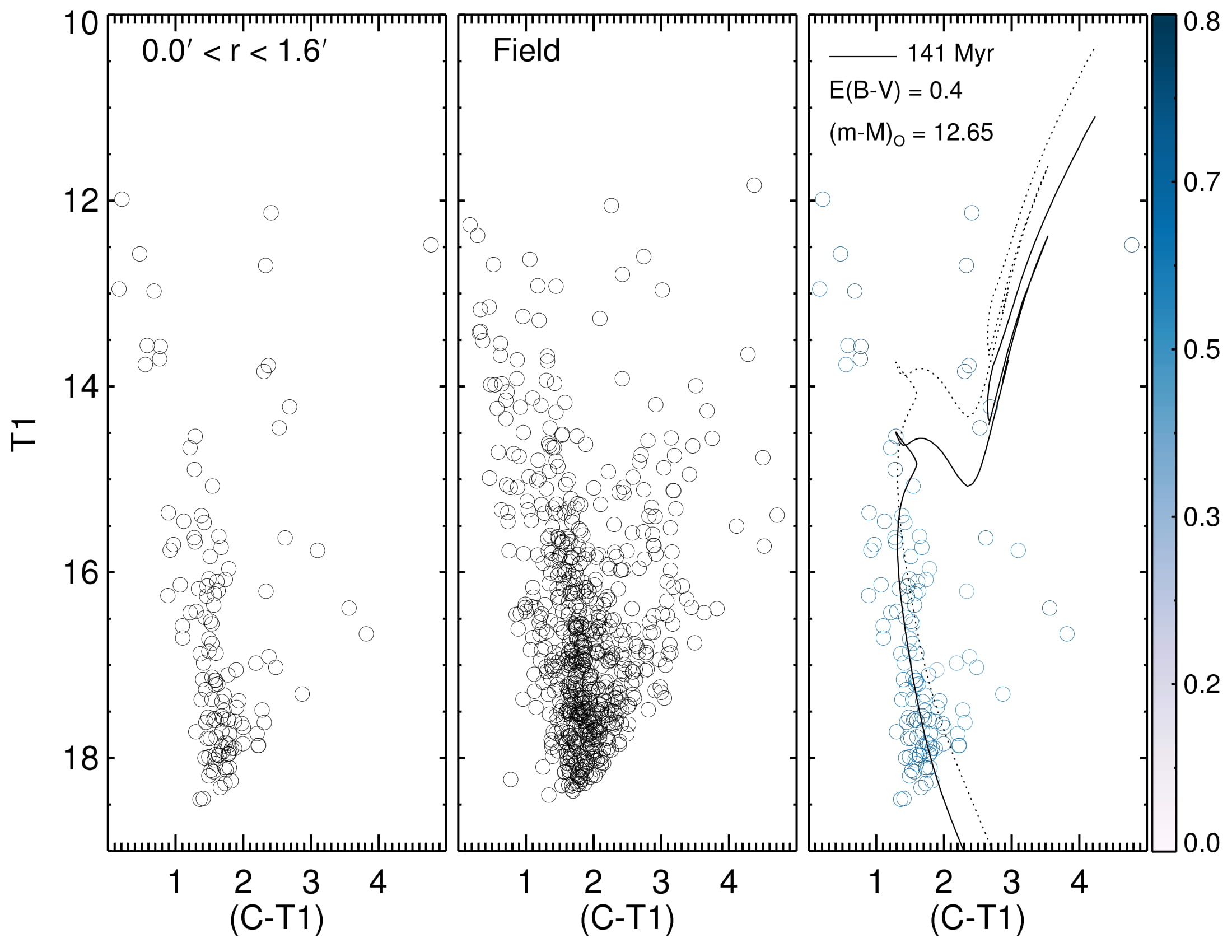}
    \includegraphics[width=0.5\textwidth]{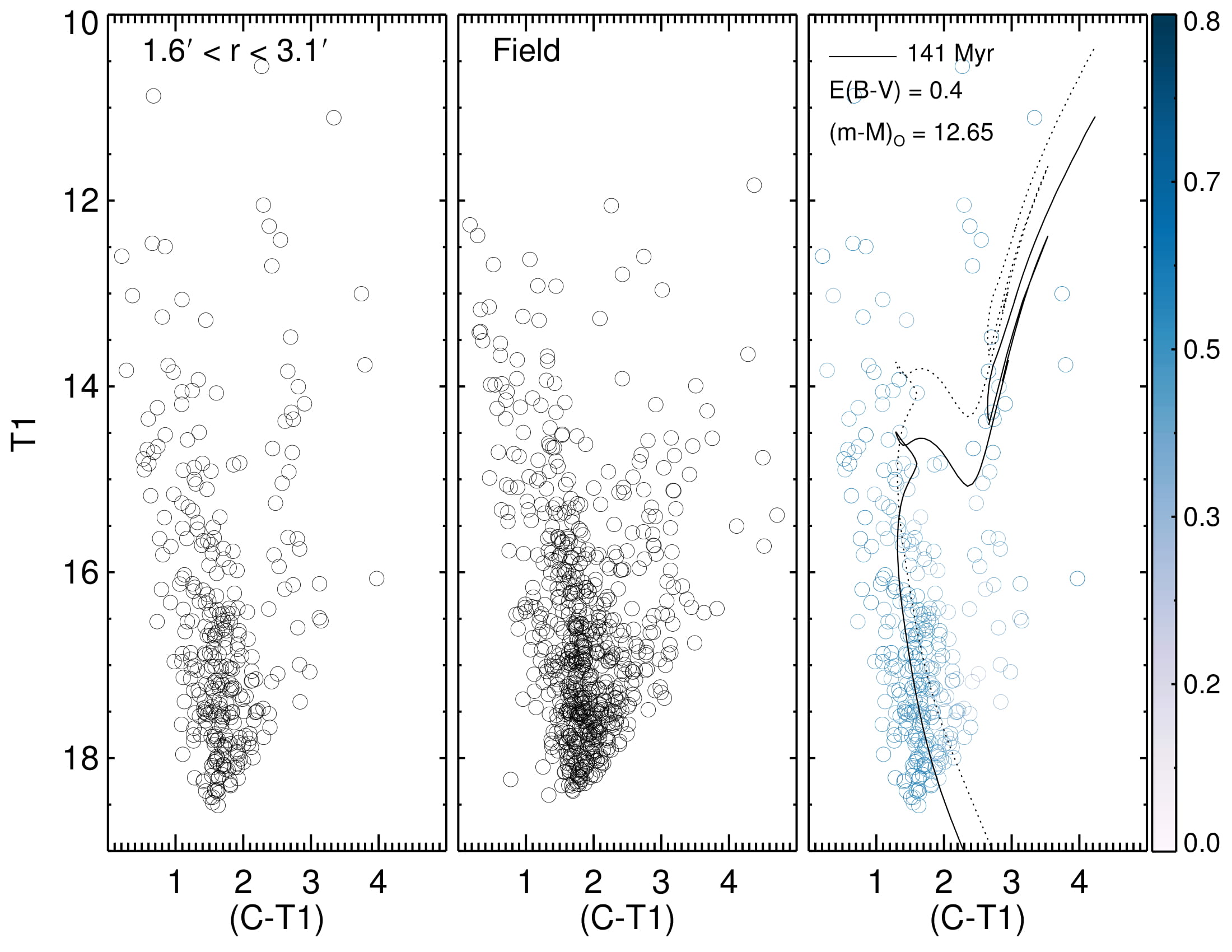}
    \includegraphics[width=0.5\textwidth]{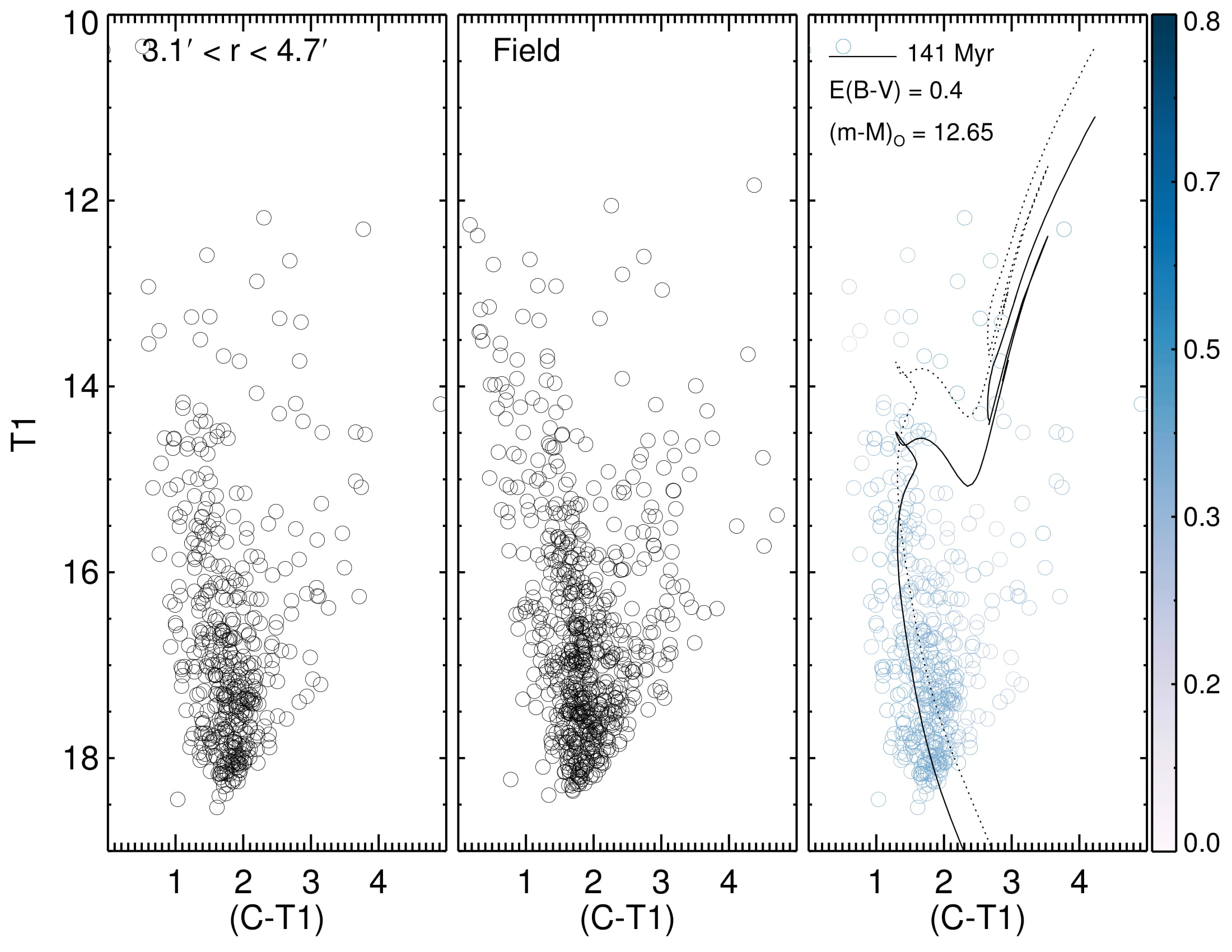}
    \includegraphics[width=0.5\textwidth]{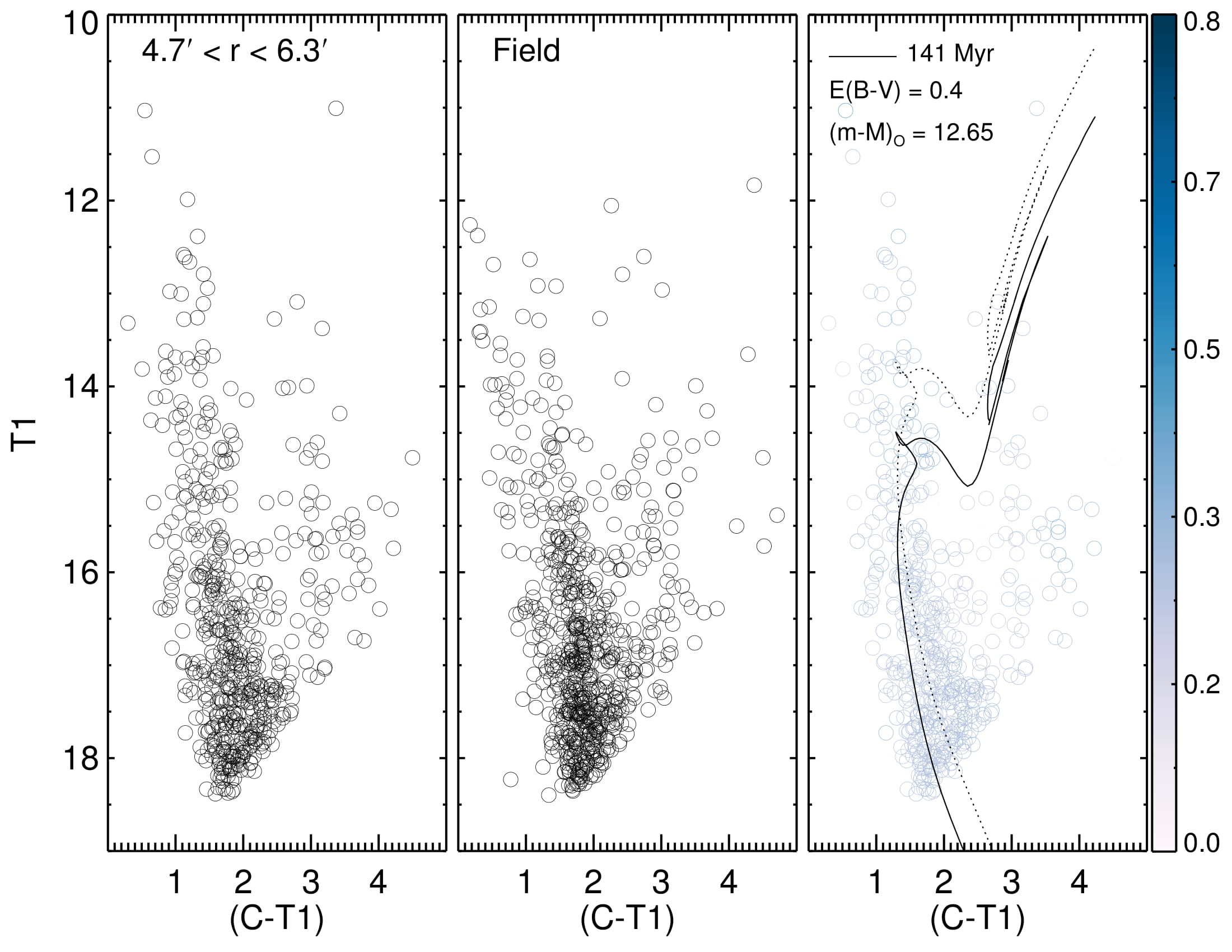}
  }
  
\caption{   Same of Figure \ref{CMDs_decontam_cascas_exemplo}, but for ESO\,260-7.   }
\label{CMDs_decontam_cascas_ESO260-7}
\end{minipage}
\end{center}
\end{figure*}

\begin{figure*}
\begin{center}
\begin{minipage}{185mm}
\parbox[c]{1.0\textwidth}
  {
   
    \includegraphics[width=0.5\textwidth]{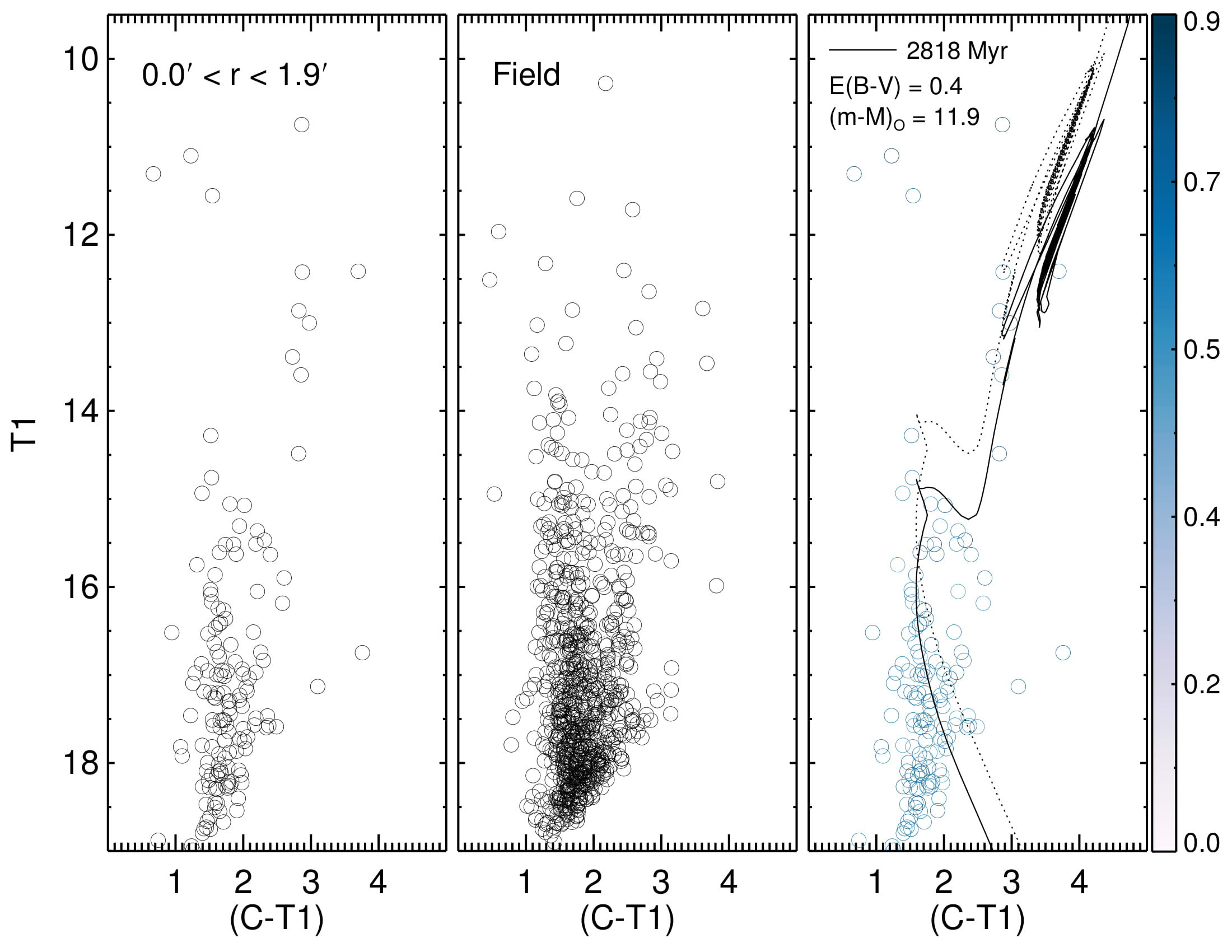}
    \includegraphics[width=0.5\textwidth]{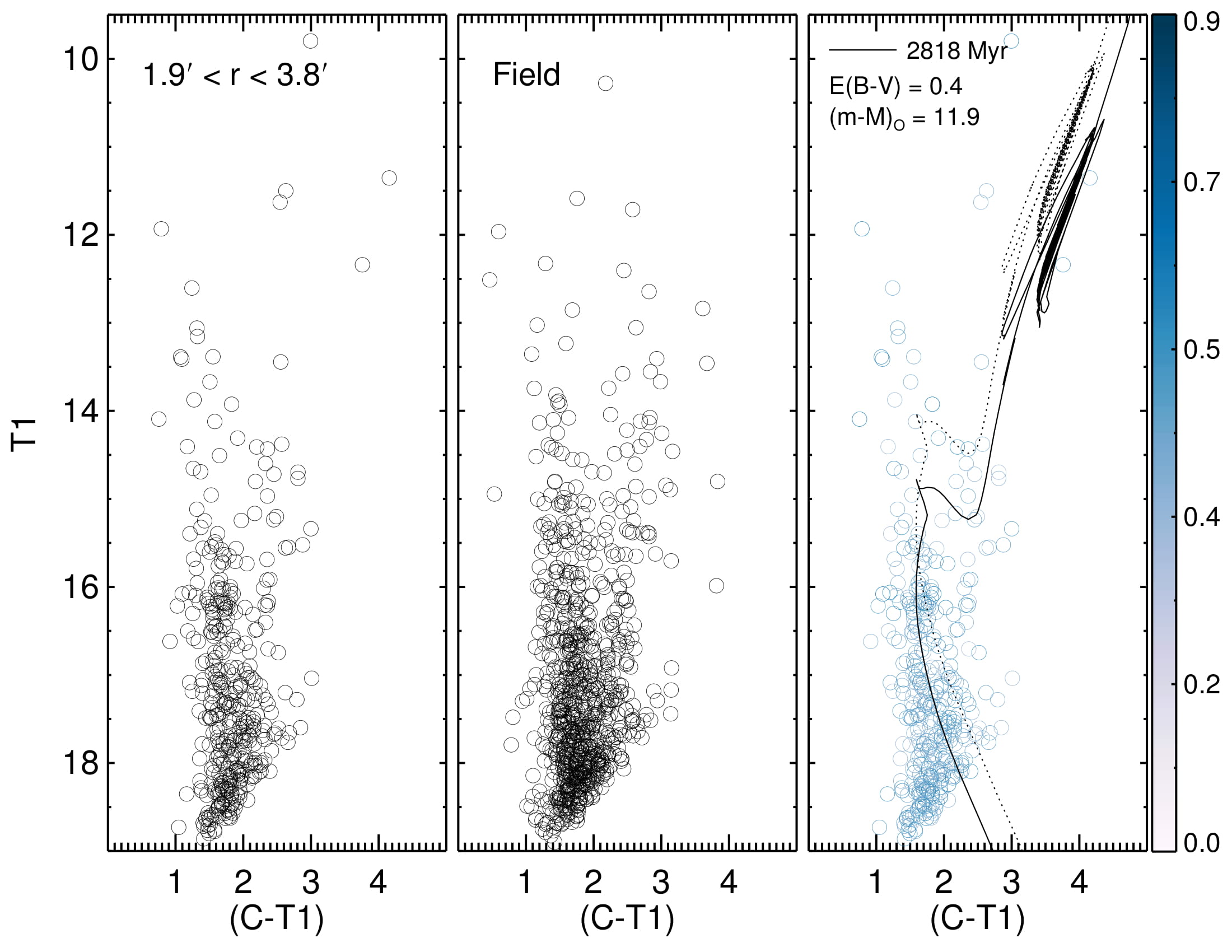}
    \begin{center}
       \includegraphics[width=0.5\textwidth]{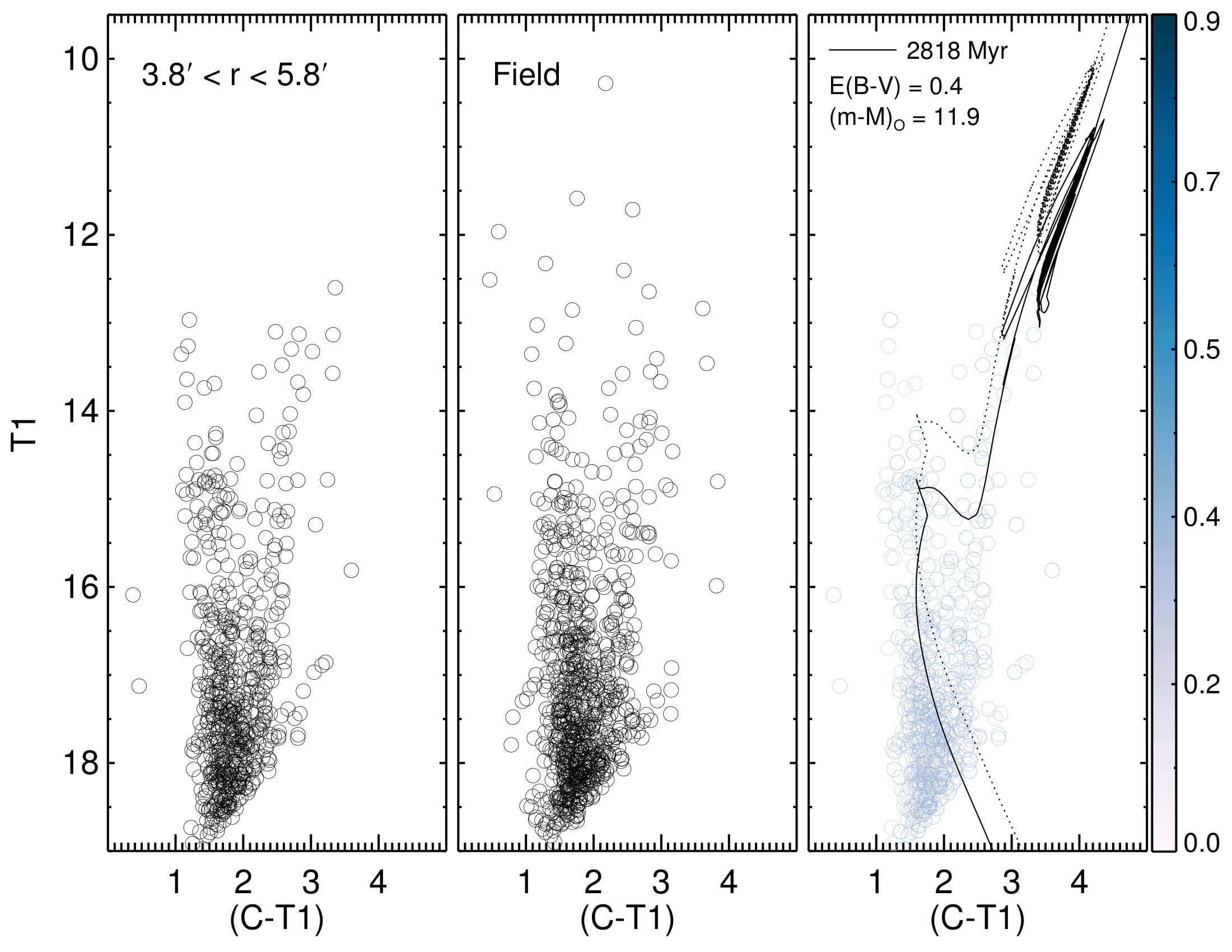}
    \end{center}
  }
  
\caption{   Same of Figure \ref{CMDs_decontam_cascas_exemplo}, but for ESO\,065-7.   }
\label{CMDs_decontam_cascas_ESO065-7}
\end{minipage}
\end{center}
\end{figure*}

\begin{figure*}


\parbox[c]{1.0\textwidth}
  {
   
    \includegraphics[width=0.5\textwidth]{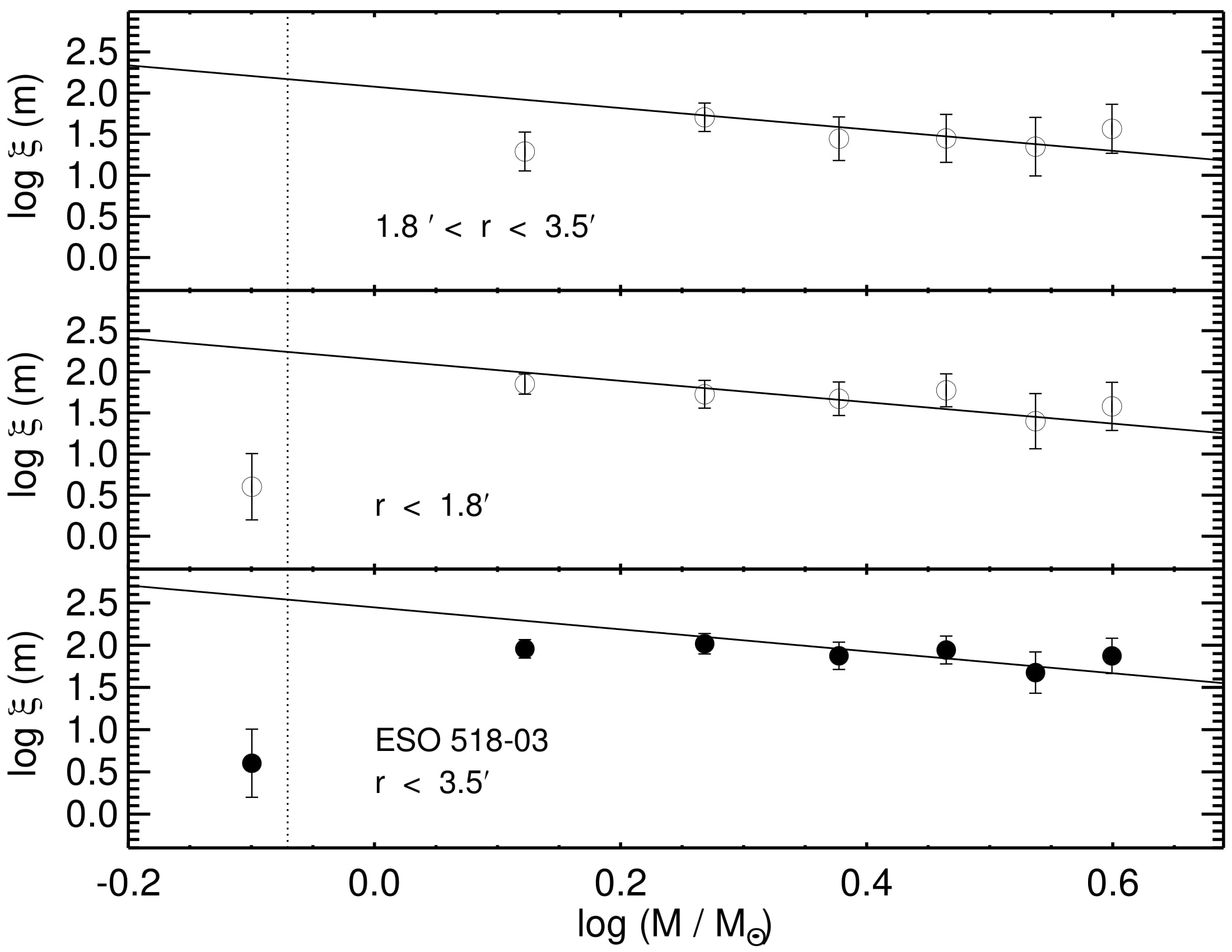}  
    \includegraphics[width=0.5\textwidth]{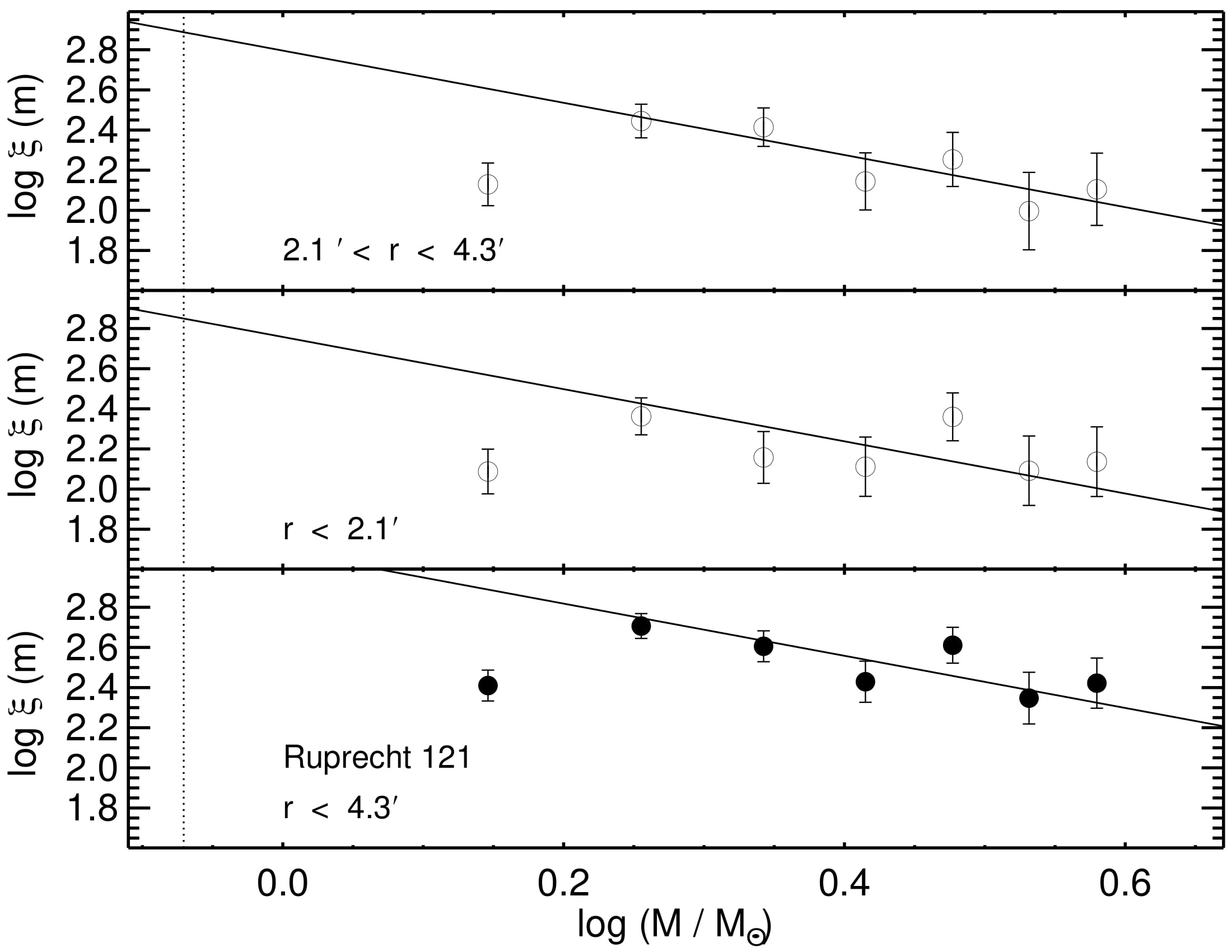} 
    \includegraphics[width=0.5\textwidth]{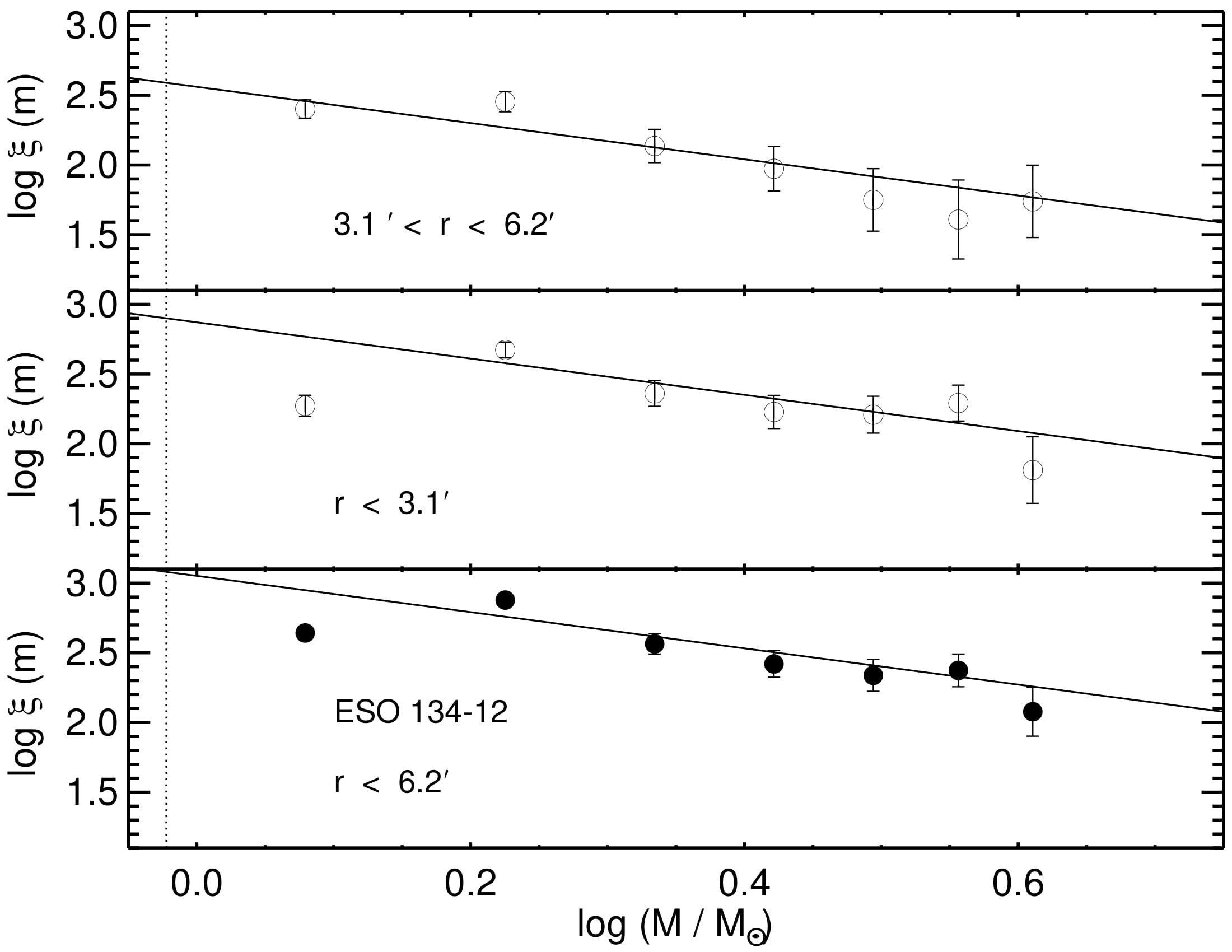} 
    \includegraphics[width=0.5\textwidth]{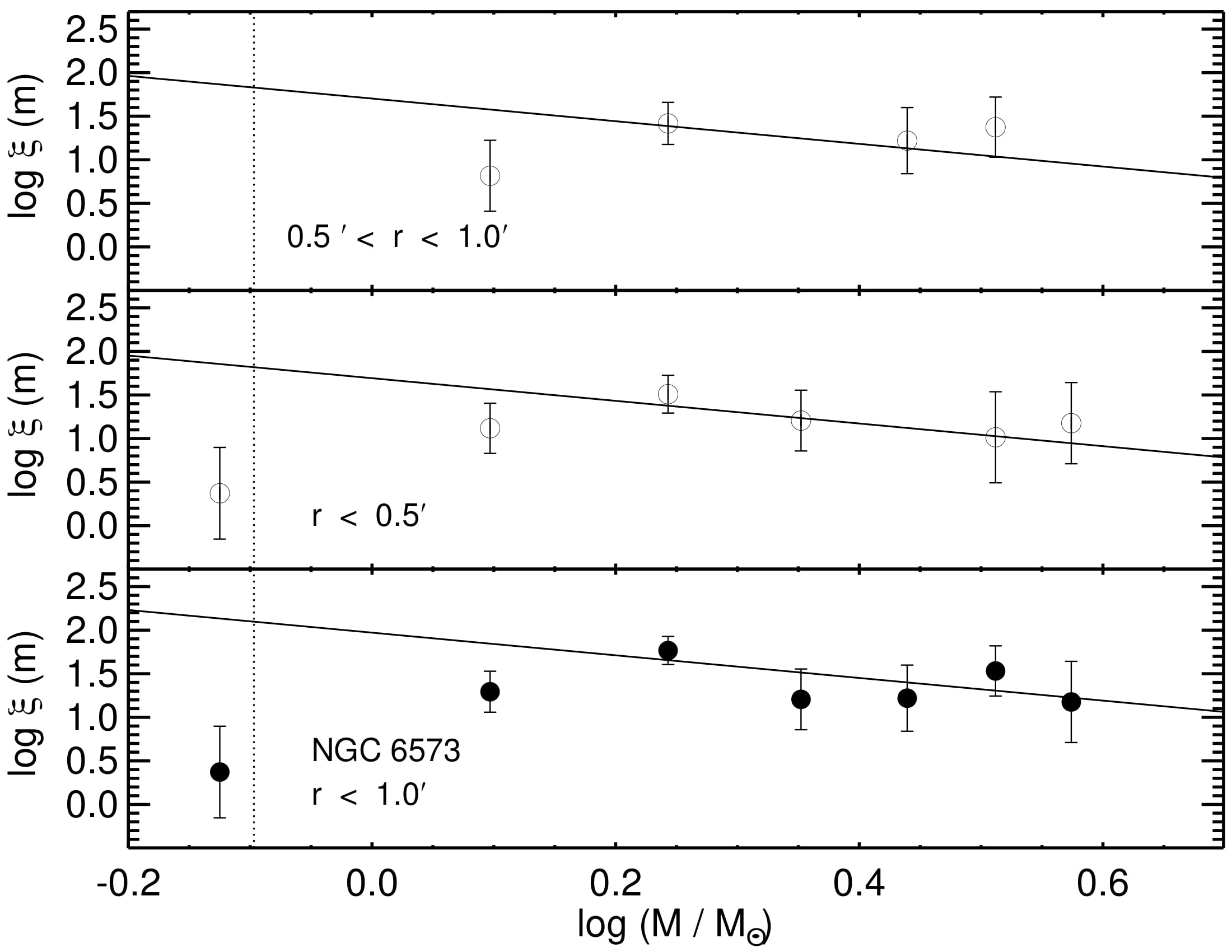} 
    
    \begin{center}
    \includegraphics[width=0.5\textwidth]{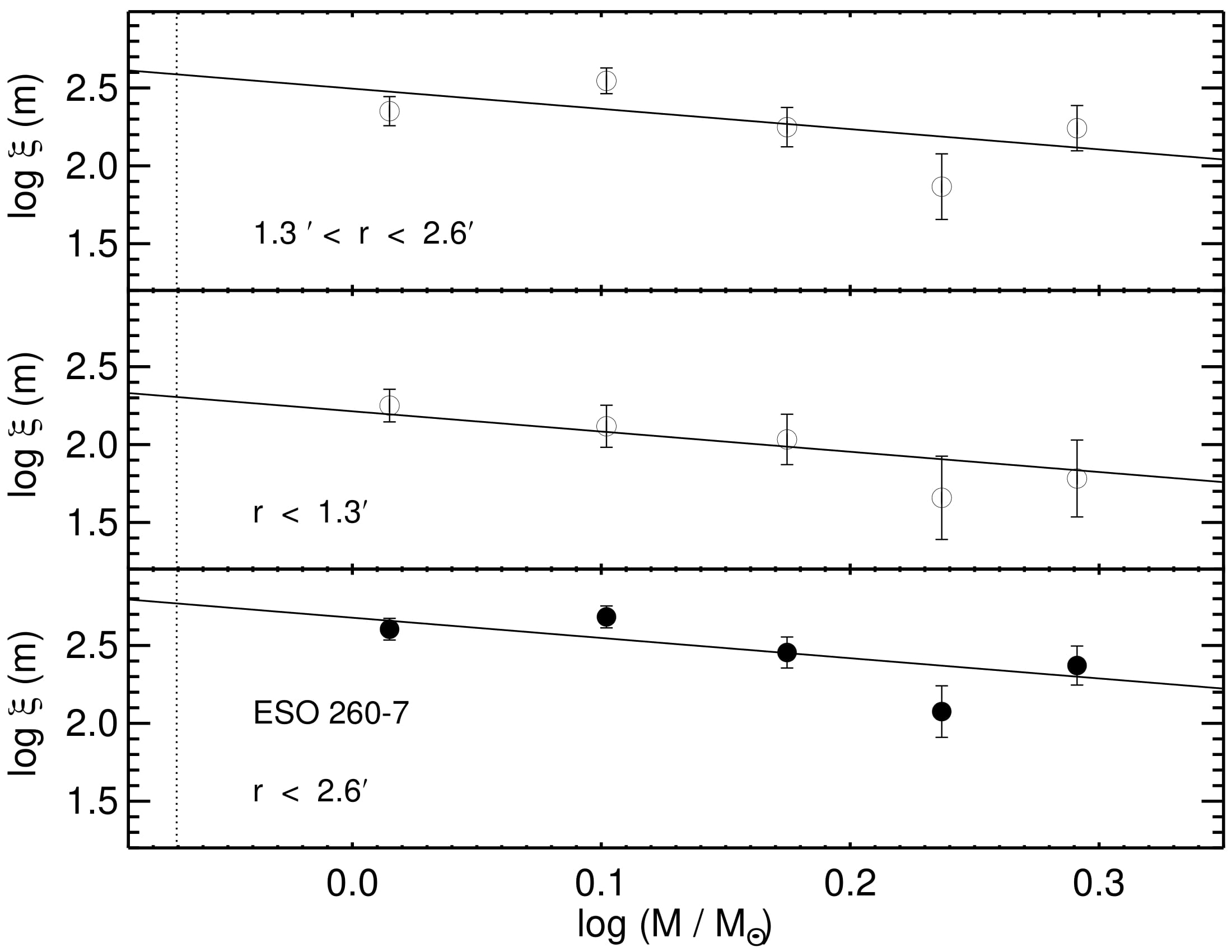}
    \end{center}
  }
\caption{   Same of Figure \ref{phi_m_allregions_ESO065-7}, but for the OCs ESO\,518-3 (top left panel), Ruprecht\,121 (top right panel), ESO\,134-12 (middle left panel), NGC\,6573 (middle right panel) and ESO\,260-7 (bottom panel).   }

\label{phi_m_allregions}


\end{figure*}


\bsp

\label{lastpage}

\end{document}